# Diffraction Effects in the Near Field

by

Marek Wlodzimierz Kowarz

Submitted in Partial Fulfillment

of the

Requirements for the Degree

Doctor of Philosophy

Supervised by

Professor Emil Wolf

The Institute of Optics

The College

School of Engineering and Applied Sciences

University of Rochester

Rochester, New York

1995

*To the memory of my grandfather,*

*Boleslaw Bursztyn,*

*who showed me how to seek the answers.*

# Curriculum Vitae

The author was born in Warsaw, Poland on November 9, 1967. He attented the University of Pennsylvania from 1985 to 1989 and graduated (Summa Cum Laude) with a Bachelor of Science degree in Electrical Engineering and a Bachelor of Arts degree in Physics. Since September of 1989, he has been a doctoral student at The Institute of Optics, University of Rochester. His Ph.D. research in physical optics has been supervised by Professor Emil Wolf.

# List of Publications

# Acknowledgments

I would like to thank my advisor, Professor Emil Wolf, for his guidance, support and friendship. I have learned a great deal more than optics from him.

I have also greatly benefited from interactions with the past and present members of Professor Wolf's group, in particular with Brian Cairns, Weijian Wang, David G. Fischer and Daniel F. V. James, and from discussions with various faculty members and graduate students in The Institute of Optics. I am also grateful to Professors Dwight Jaggard and Nader Engheta of the University of Pennsylvania for sparking my initial interest in electromagnetic theory and optics.

Many people have enhanced my life in Rochester: my friends from The Institute of Optics who increased my tolerance, from the Graduate Organizing Group who broadened my views of various academic and non-academic pursuits, and from Ultimate Frisbee who enlarged my lungs and my shoulders; and Jen who recently enriched my heart.

I gratefully acknowledge financial support from The Institute of Optics, the Department of Education, the Department of Energy, the Army Research Office, the National Science Foundation, and the New York State Science and Technology Foundation.

Finally, I would like to thank my mother, sister, grandmother and aunt for their continuous love and support throughout this journey.

# Abstract

This dissertation is concerned with understanding and analyzing some of the effects of diffraction in the near field. The near field has become of interest at optical wavelengths with the advent of promising applications such as near-field optical microscopy, near-field spectroscopy and high-density optical data storage.

First, a comprehensive review is presented of the scalar and the electromagnetic theories of diffraction by an aperture in a planar opaque screen. Two new theorems concerning the behavior of the phase near extrema of the amplitude are established that provide a novel insight into the relationship between the amplitude and the phase of the field in diffraction patterns.

The contributions of homogeneous and of evanescent waves to two-dimensional near-field diffraction patterns of scalar fields are then examined in detail. Exact relations are obtained for calculating these contributions for arbitrary propagation distances, along with approximate expressions for the near field. The behavior of the two contributions is illustrated for the case of a plane wave diffracted by a slit in an opaque screen.

The finite-difference time-domain (FD-TD) method is used examine the influence of exact boundary values on the near field for the case of a slit in a thin perfectly conducting screen. The FD-TD numerical results are displayed in color images that illustrate the intricate behavior of the amplitude and the phase of the field in the vicinity of the slit. These numerical results are compared with the predictions of approximate theories.

Some new methods for determining near-fields in rigorous diffraction problems involving thin screens are discussed. Specifically, new approximate theories of diffraction are introduced for both scalar and electromagnetic fields and an iterative Fourier-based algorithm is proposed for solving the rigorous boundary value problem.

In order to understand the effects of an optical vortex on diffraction, the field emerging from a spiral phase plate illuminated by a Gaussian beam is examined. It is shown that the amplitude profile of the emerging field changes appreciably over propagation distances that are much smaller than the Rayleigh range.

# Table of Contents

# List of Tables

# List of Figures

# List of Main Symbols

| Symbol | Quantity |
|--------|----------|
| $a(\mathbf{u}_\perp)$ | Angular spectrum amplitude |
| $a^{(\mathrm{inc})}(\mathbf{u}_\perp)$ | Angular spectrum amplitude of incident field |
| $a^{(\mathrm{rfl})}(\mathbf{u}_\perp)$ | Angular spectrum amplitude of reflected field |
| $a^{(\mathrm{trn})}(\mathbf{u}_\perp)$ | Angular spectrum amplitude of transmitted field |
| A | Amplitude of scalar field; |
| | *also* maximum amplitude of Gaussian beam |
| **A** | Aperture |
| $\mathrm{B}_{u_\perp}^{(n)}$ | $n$th-order Bessel beam with parameter $u_\perp$ |
| $c$ | Speed of light in vacuum |
| $c_n(u_\perp)$ | Bessel-beam expansion coefficients |
| $d$ | Width of slit |
| $\mathbf{e}(\mathbf{u}_\perp)$ | Vectorial angular spectrum amplitude of electric field |
| $\mathbf{e}^{(\mathrm{inc})}(\mathbf{u}_\perp)$ | Vectorial angular spectrum amplitude of incident electric field |
| $\mathbf{e}^{(\mathrm{rfl})}(\mathbf{u}_\perp)$ | Vectorial angular spectrum amplitude of reflected electric field |
| $\mathbf{e}^{(\mathrm{trn})}(\mathbf{u}_\perp)$ | Vectorial angular spectrum amplitude of transmitted electric field |
| $\mathbf{e}_\perp(\mathbf{u}_\perp)$ | Transverse vector component of $\mathbf{e}(\mathbf{u}_\perp)$ |
| $\mathbf{e}_\perp^{(\mathrm{inc})}(\mathbf{u}_\perp)$ | Transverse vector component of $\mathbf{e}^{(\mathrm{inc})}(\mathbf{u}_\perp)$ |
| $\mathbf{e}_\perp^{(\mathrm{rfl})}(\mathbf{u}_\perp)$ | Transverse vector component of $\mathbf{e}^{(\mathrm{rfl})}(\mathbf{u}_\perp)$ |
| $\mathbf{e}_\perp^{(\mathrm{trn})}(\mathbf{u}_\perp)$ | Transverse vector component of $\mathbf{e}^{(\mathrm{trn})}(\mathbf{u}_\perp)$ |
| $E$ | Magnitude of electric field |

$\left.\begin{array}{c} E_x \\ E_y \\ E_z \end{array}\right\}$ Cartesian components of electric field

$\mathbf{E}$ — Electric field

$\mathbf{E}_\perp$ — Transverse vector component of $\mathbf{E}$

$\mathbf{E}_\perp^{(0)}$ — $\mathbf{E}_\perp^{(\text{trn})}$ in aperture plane

$\mathbf{E}_\perp^{(\text{trn})}$ — Transverse vector component of $\mathbf{E}^{(\text{trn})}$

$\mathbf{E}^{(e)}$ — Electric field ($e$-theory)

$\mathbf{E}_{\text{EP}}$ — Electric field for E-polarization

$\mathbf{E}_{\text{HP}}$ — Electric field for H-polarization

$\mathbf{E}^{(\text{inc})}$ — Incident electric field

$\mathbf{E}^{(K)}$ — Electric field (Kottler-Kirchhoff theory)

$\mathbf{E}^{(m)}$ — Electric field ($m$-theory)

$\mathbf{E}^{(M1)}$ — Electric field (first modified electromagnetic theory)

$\mathbf{E}^{(M2)}$ — Electric field (second modified electromagnetic theory)

$\mathbf{E}^{(\text{rfl})}$ — Reflected electric field

$\mathbf{E}^{(\text{trn})}$ — Transmitted electric field

$F_{\text{tot}}$ — Total energy flux

$\mathbf{F}$ — Real energy flux vector for scalar field

$\tilde{\mathbf{F}}$ — Complex energy flux vector for scalar field

$G$ — Green's function

$\mathbf{h}(\mathbf{u}_\perp)$ — Vectorial angular spectrum amplitude of magnetic field

$\mathbf{h}^{(\text{inc})}(\mathbf{u}_\perp)$ — Vectorial angular spectrum amplitude of incident magnetic field

$\mathbf{h}^{(\text{rfl})}(\mathbf{u}_\perp)$ — Vectorial angular spectrum amplitude of reflected magnetic field

$\mathbf{h}^{(\text{trn})}(\mathbf{u}_\perp)$ — Vectorial angular spectrum amplitude of transmitted magnetic field

$\mathbf{h}_\perp(\mathbf{u}_\perp)$ — Transverse vector component of $\mathbf{h}(\mathbf{u}_\perp)$

| | |
|---|---|
| $I$ | Intensity of scalar field |
| $I_h$ | Intensity of homogeneous contribution |
| $I_{hi}$ | Interference term of homogeneous and evanescent contributions |
| $I_i$ | Intensity of evanescent contribution |
| $I_{\text{tot}}$ | Total intensity |
| $I_{\text{tot}}^{(h)}$ | Total homogeneous intensity |
| $I_{\text{tot}}^{(i)}$ | Total evanescent intensity |
| $k$ | Free-space wavenumber |
| $M_m$ | Gaussian beam modifying function |
| $\mathbf{n}_e$ | Polarization unit vector of linearly polarized electric field |
| $O$ | Origin of coordinates |
| $\mathbf{r}$ | Position vector |
| $\mathbf{r}_s$ | Position vector of amplitude extremum |
| $R$ | Radius of curvature of Gaussian beam |
| $\mathsf{S}$ | Surface of planar screen |
| $\mathbf{S}$ | Poynting's vector |
| $t$ | Time |
| $T$ | FD-TD total computation time |
| $\left.\begin{array}{l} u_x \\ u_y \\ u_z \end{array}\right\}$ | Cartesian components of $\mathbf{u}$ |
| $u_\perp$ | Magnitude of $\mathbf{u}_\perp$ |
| $\mathbf{u}$ | Unit vector (generally complex) |
| $\mathbf{u}_\perp$ | Transverse vector component of $\mathbf{u}$ (real) |
| $\mathbf{u}_e$ | Polarization unit vector of electric field (generally complex) |
| $\mathbf{u}_m$ | Polarization unit vector of magnetic field (generally complex) |
| $U$ | Scalar field |
| $U_0$ | Gaussian beam |
| $U^{(0)}$ | Transmitted scalar field in aperture plane |
| $U_z^{(0)}$ | $z$-derivative of transmitted scalar field in aperture plane |
| $U^{(I)}$ | Scalar field (Rayleigh-Sommerfeld theory of the first kind) |
| $U^{(II)}$ | Scalar field (Rayleigh-Sommerfeld theory of the second kind) |
| $U_D$ | Scalar field (Dirichlet-type screen) |
| $U_h$ | Homogeneous contribution |
| $U_i$ | Evanescent contribution |
| $U^{(inc)}$ | Incident scalar field |
| $U^{(K)}$ | Scalar field (Kirchhoff theory) |
| $U_m$ | Beam emerging from $m$th-order spiral phase plate |
| $U^{(M1)}$ | Scalar field (first modified theory) |
| $U^{(M2)}$ | Scalar field (second modified theory) |
| $U_N$ | Scalar field (Neumann-type screen) |
| $U^{(rfl)}$ | Reflected scalar field |
| $U^{(trn)}$ | Transmitted scalar field |
| $V$ | Time-dependent scalar field |
| $w$ | Width of Gaussian beam |
| $w_e$ | Electric energy density |
| $w_m$ | Magnetic energy density |
| $w_o$ | Waist of Gaussian beam |
| $W_{tot}$ | Total reactive energy |
| $x$ | Cartesian coordinate |
| $\hat{\mathbf{x}}$ | Unit vector in $x$-direction |
| $X$ | FD-TD grid size in $x$-direction |
| $y$ | Cartesian coordinate |
| $\hat{\mathbf{y}}$ | Unit vector in $y$-direction |
| $z$ | Cartesian or cylindrical coordinate |
| $z_r$ | Rayleigh range |
| $\hat{\mathbf{z}}$ | Unit vector in $z$-direction |
| $Z$ | FD-TD grid size in $z$-direction |
| $\Delta t$ | FD-TD time spacing |
| $\Delta x$ | FD-TD grid spacing in $x$-direction |
| $\Delta z$ | FD-TD grid spacing in $z$-direction |
| $\varphi$ | Cylindrical coordinate |
| $\hat{\boldsymbol{\varphi}}$ | Unit vector in $\varphi$-direction |
| $\phi$ | Phase of scalar field |
| $\phi_e$ | Phase of linearly polarized electric field |
| $\lambda$ | Free-space wavelength |
| $\rho$ | Cylindrical coordinate |
| $\boldsymbol{\rho}$ | Transverse position vector |
| $\omega$ | Frequency |
| $\psi$ | Angle of $\mathbf{u}$ in cylindrical coordinates; |
| | *also* axial phase factor of Gaussian beam |

## Errata

(Added in 2006)

Below are corrected versions of certain equations that appear in this thesis.

### Chapter 3

$$\mathrm{H}_h(x,z) \;=\; \frac{ikz}{2\sqrt{x^2+z^2}}\, J_1\!\left(k\sqrt{x^2+z^2}\right) \;+\; \frac{k}{\pi}\sum_{m=0}^{\infty}(-1)^m\,\frac{2^m\,m!}{(2m)!}\,(kz)^{2m}\,\frac{j_m(kx)}{(kx)^m}$$

$$(3.3.14a)$$

$$\mathrm{H}_i(x,z) \;=\; -\frac{kz}{2\sqrt{x^2+z^2}}\, Y_1\!\left(k\sqrt{x^2+z^2}\right) \;-\; \frac{k}{\pi}\sum_{m=0}^{\infty}(-1)^m\,\frac{2^m\,m!}{(2m)!}\,(kz)^{2m}\,\frac{j_m(kx)}{(kx)^m}$$

$$(3.3.14b)$$

$$\mathrm{H}_h(x,z) \;\approx\; \frac{k}{\pi}\frac{\sin(kx)}{kx} \;+\; ikz\,\frac{k}{2}\frac{J_1(kx)}{kx} \qquad (3.4.1a)$$

$$U_h(x,z) \;\approx\; U_h(x,0) \;+\; ikz\,\frac{k}{2}\int_{-\infty}^{\infty}\frac{J_1\!\left[k(x-x')\right]}{k(x-x')}\,U(x',0)\,dx'$$

$$(3.4.2a)$$

### Appendix C

$$\mathrm{H}_h(x,z) \;=\; \frac{ikz}{2\sqrt{x^2+z^2}}\, J_1\!\left(k\sqrt{x^2+z^2}\right) \;+\; \frac{k}{\pi}\sum_{m=0}^{\infty}(-1)^m\,\frac{2^m\,m!}{(2m)!}\,(kz)^{2m}\,\frac{j_m(kx)}{(kx)^m}\,.$$

$$(C.13)$$

CHAPTER 1

# INTRODUCTION

## 1.1  DIFFRACTION OF LIGHT

The early observations of diffraction of light by an obstacle were the primary pieces of evidence for the adoption of the wave theory of light.[1-3]  Despite the fact that diffraction phenomena were first described in detail more than three centuries ago, the theory of diffraction continues to be a subject of great interest today in many branches of both basic and applied physics.

The analysis of diffraction of light by an aperture in a planar opaque screen[*†] can be separated into the treatment of two distinct problems[‡]: the boundary value problem and the propagation problem.  The boundary value problem consists of determining the boundary value of the field in the plane immediately behind the screen for a given illuminating field, given material characteristics of the screen and a given aperture shape.  The propagation problem involves finding the field at some distance behind the screen from this boundary field.

---

[*] We take an opaque screen to be one for which the normal component of the energy flux vector is identically zero over the portion of the screen in the shadow region.  Therefore, in our terminology, a perfectly conducting screen is opaque even though it is highly reflecting.  However, other definitions of opaque screens are often used in the literature

[†] An extensive review of scalar and electromagnetic diffraction can be found in Ref. 4.  References to more recent work are given in Refs. 5-7.

[‡] As is stressed in Ref. 8, this approach can sometimes simplify the analysis of rigorous diffraction problems.

The approach one uses to solve the propagation problem is straightforward, although there can be some computational difficulties in evaluating the necessary integrals: one simply applies the appropriate propagator (diffraction formula) to the boundary field. However, the boundary field itself is very difficult to determine, especially if the optical field is treated by full electromagnetic theory. For the idealized case of an electromagnetic field incident upon an infinitely thin perfectly conducting screen, analytic solutions have been obtained for a small number of geometries[*], including the half-plane,[8-10] the slit[4,11] and the circular aperture[4,12,13] (see also Ref. 6, Chapters 4, 8 and 14), along with the complementary strip and circular disk geometries. But the analytic results for the slit and the circular aperture are not always useful in practice, because they contain infinite series that converge slowly when the linear dimensions of the aperture are larger than about one wavelength.

In optics, one usually treats the electromagnetic field by a scalar model and one simplifies the boundary value problem by approximating the (scalar) boundary field by the unperturbed incident field. The diffraction theory based on this approximation is sometimes called the Rayleigh-Sommerfeld theory of the first kind. Two other approximate diffraction theories are also occasionally used:[14-18] the Rayleigh-Sommerfeld theory of the second kind and the older classic Kirchhoff theory. Even though the boundary fields given by the three theories are very different, all three often yield nearly identical results, and agree very well with experimental observations, when the size of the aperture and the propagation distance are both much larger than the wavelength. These conditions are usually satisfied in conventional optics, but they are not satisfied in near-field optics.

---

[*] For these same geometries, analytic solutions exist for scalar fields interacting with Dirichlet-type or Neumann-type screens.

## 1.2  THE NEAR FIELD

For our purposes, the near zone of a radiating, scattering or diffracting object in free space will be considered to be the region of space where there are non-negligible contributions from evanescent waves associated with high spatial frequencies of the field. Specifically, if $\tilde{U}_o(f_x, f_y)$ is the Fourier transform of the field $U_o(x, y)$ in a plane in close proximity to the object,

$$\tilde{U}_o(f_x, f_y) = \iint U_o(x, y)\, e^{i(f_x x + f_y y)} dx\, dy \;, \qquad (1.2.1)$$

and if $\tilde{U}_d(f_x, f_y)$ is the corresponding Fourier transform after a propagation distance $d$, then those spatial frequencies $f_x$ and $f_y$ for which $\sqrt{f_x^2 + f_y^2} > k$ ($k$ being the free-space wavenumber) are exponentially attenuated so that

$$\tilde{U}_d(f_x, f_y) = \tilde{U}_o(f_x, f_y)\, \exp\left[-d\sqrt{f_x^2 + f_y^2 - k^2}\,\right] . \qquad (1.2.2)$$

Hence, for any evanescent waves to be present, the distance $d$ can be at most of the order of the wavelength. The resolution of conventional optical systems is therefore fundamentally limited by the wavelength $\lambda$, since the propagation distances involved are several orders of magnitude larger than $\lambda$ (see Fig. 1-1).

Optical systems that interact directly with the near field can overcome this resolution limit. Recently, there has been considerable interest in these systems[19-24] stemming from promising applications such as near-field optical microscopy, near-field spectroscopy and high-density optical data storage. The near-field interaction can be accomplished by a variety of schemes, for example, with a sub-wavelength fiber optic probe or a sub-wavelength aperture which is scanned at the distance from the object that is smaller than the wavelength (see Fig. 1-2). However, the interaction is very complicated and it is difficult to interpret experimental data. Consequently, there has been significant interest in using rigorous diffraction theories[25-28] and numerical simulations[29,30] to model the sub-wavelength near-field probe and the probe-object interaction.

It should be pointed out that near-field measurements of antennas have been used for several decades as a means of determining far-field radiation patterns.[31-33] However, these types of near-field measurements can be performed at distances larger than the wavelength, because evanescent waves are of no importance to the far field.

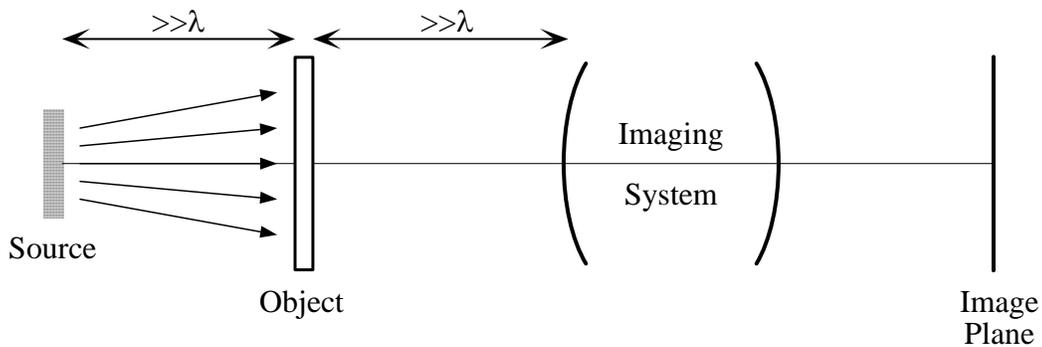

**Figure 1-1**  Diagram of a conventional optical microscope.

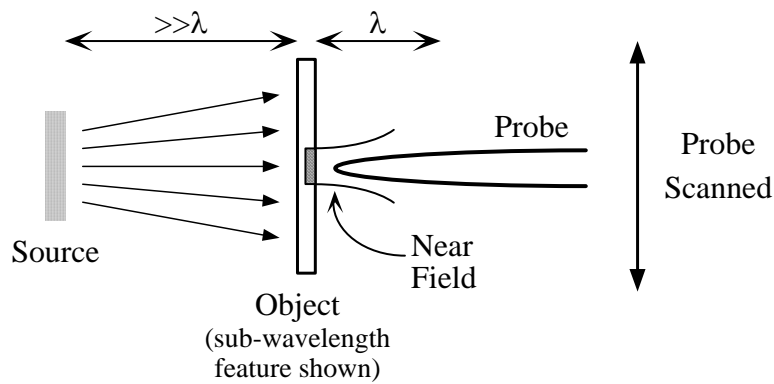

**Figure 1-2**  Diagram of a collection mode near-field scanning optical microscope.

## 1.3 OVERVIEW OF THE THESIS

This thesis is concerned with the effects of diffraction in the near field. Although the primary motivation for this work is near-field optics, the results of the research are also applicable to longer wavelengths of electromagnetic radiation and to acoustics. In fact, the new methods for aperture diffraction presented in Chapter 5 should be useful for aperture antenna analysis and design.

Chapter 2 contains an introductory discussion of the theory of diffraction by an aperture in a planar opaque screen for both the scalar and the electromagnetic cases. The discussion provides the theoretical background for the analysis in the subsequent chapters. It includes a review of boundary conditions, exact and approximate diffraction formulas, the angular spectrum representation, the Bessel-beam representation and energy flow. In addition, two new theorems are proven concerning the behavior of the phase of a scalar field near extrema of the amplitude.

Chapter 3 examines, in detail, the contributions of homogeneous and of evanescent waves to two-dimensional near-field diffraction patterns of scalar fields. Exact relations are derived for calculating these two contributions for arbitrary propagation distances, along with approximate expressions for the near field. The behavior of the two contributions is illustrated for the case of a plane wave diffracted by a slit in an opaque screen, with the use of approximate boundary conditions.

In Chapter 4, the near-field in the slit diffraction problem is then reexamined with rigorous boundary conditions appropriate to a thin perfectly conducting screen. Because it is not efficient to determine the near-field at many points in space from the exact series solution, the finite-difference time-domain method is used to numerically evaluate the near field. These numerical results are compared with the predictions of the approximate Rayleigh-Sommerfeld theories. The validity of two theorems of Chapter 2 on the behavior of the phase near extrema of the amplitude is readily observed in the near-field diffraction patterns.

Chapter 5 introduces some new, relatively untested methods for determining the near-field in rigorous aperture diffraction problems. First, in Section 5.1, two new approximate scalar theories of diffraction are described that are based on simple modifications of the Rayleigh-Sommerfeld theories. These new scalar theories are formulated for thin Dirichlet-type and Neumann-type screens. New approximate electromagnetic theories of diffraction, intended for thin perfectly conducting screens, are also discussed. Then, in Section 5.2, a new iterative Fourier-based algorithm for numerically solving the rigorous boundary value is proposed. The algorithm is obtained from dual integral equations for angular spectrum amplitude of the transmitted field and can be implemented for scalar fields incident upon either Dirichlet-type or Neumann-type screens, and also for electromagnetic fields incident upon perfectly conducting screens.

The last main chapter, Chapter 6, deals with a particular structure in the field that can only occur in three-dimensional diffraction, namely, the optical vortex or screw dislocation. The main aim of that chapter is to show, by a realistic example consisting of a spiral phase plate illuminated by a Gaussian beam, that the presence of a vortex can dramatically affect the diffraction of a field over very small propagation distances. Although the distances are larger than the wavelength, they are much smaller than the Rayleigh range. This result is rather surprising since, for most beams, the amplitude profile is unchanged for propagation distances considerably smaller than the Rayleigh range.

# CHAPTER 2

# APERTURE DIFFRACTION THEORY

In this Chapter, we discuss some fundamental aspects of the theory of diffraction of light by an aperture in a planar opaque screen. We consider the scalar and the electromagnetic cases separately and point out some of the basic differences between them. There are, in fact, a number of subtleties contained in the formulation of approximate electromagnetic theories of diffraction that are not encountered in the scalar case: a naive treatment can yield vector theories that do not satisfy Maxwell's equations.

Much of the material covered in this chapter is available in the published literature.[1-6] However, the Bessel-beam representation of the field, which we present in Section 2.1.4, is rarely discussed in any detail within the context of aperture diffraction. Furthermore, the analysis in Section 2.1.6 concerning the behavior of the phase near extrema of the amplitude appears to be new.

## 2.1  SCALAR THEORY OF DIFFRACTION

We shall first assume that the optical field may be described accurately by a scalar theory.[7-9] One situation where such a description is completely adequate is the case of two-dimensional diffraction, discussed in Section 2.2.4.

We consider a monochromatic scalar field $V^{(\mathrm{inc})}(\mathbf{r},t) = U^{(\mathrm{inc})}(\mathbf{r})\exp\{-i\omega t\}$, $\mathbf{r} = (x, y, z)$, of frequency $\omega$ that is incident from the half-space $z < 0$ upon an aperture

A in a thin, planar, opaque screen S (see Fig. 2-1). We take the screen to be located in the plane $z = 0$ and the rest of space to be free of matter. If we denote the total field by $U(\mathbf{r})$, then both $U^{(\text{inc})}(\mathbf{r})$ and $U(\mathbf{r})$ satisfy the Helmholtz equation in free-space,

$$\left(\nabla^2 + k^2\right)U^{(\text{inc})}(\mathbf{r}) = 0 , \tag{2.1.1}$$

$$\left(\nabla^2 + k^2\right)U(\mathbf{r}) = 0 , \tag{2.1.2}$$

where $\nabla^2 \equiv \partial^2/\partial x^2 + \partial^2/\partial y^2 + \partial^2/\partial z^2$ is the Laplacian and

$$k = \omega/c = 2\pi/\lambda \tag{2.1.3}$$

is the free-space wavenumber, $c$ being the speed of light in vacuum and $\lambda$ being the wavelength.

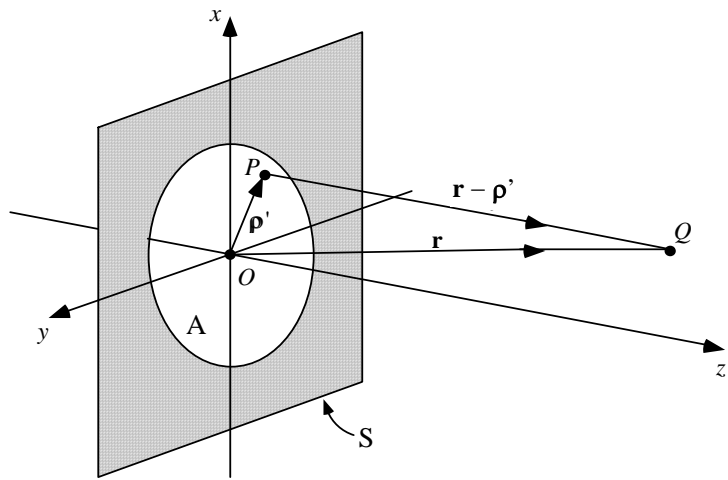

**Figure 2-1** An aperture A (of arbitrary shape) in a planar screen S. $P(\boldsymbol{\rho}')$ is a point in the aperture plane and $Q(\mathbf{r})$ is the observation point.

### 2.1.1 Boundary Conditions

To obtain a solution for the field $U(\mathbf{r})$, the boundary conditions in the aperture plane $z = 0$ need to be specified. Since in free space the field and its derivatives are continuous at every point, the field inside the aperture and its $z$-derivative must also be continuous. Hence

$$U(\boldsymbol{\rho}, 0^+) \;=\; U(\boldsymbol{\rho}, 0^-) \qquad\qquad \text{in A} \qquad\qquad (2.1.4)$$

and

$$\left.\frac{\partial U(\boldsymbol{\rho}, z)}{\partial z}\right|_{z=0^+} = \left.\frac{\partial U(\boldsymbol{\rho}, z)}{\partial z}\right|_{z=0^-} \quad \text{in A} \,, \qquad (2.1.5)$$

where $\boldsymbol{\rho} = (x, y)$ and the notation* $0^-$ $(0^+)$ indicates the limit as $z$ approaches zero from the negative (positive) $z$-direction. On the surface of an infinitesimally thin opaque screen, two different kinds of boundary conditions are usually applied to a scalar field:[1,4] Dirichlet conditions, for which the field is zero,

$$U_D(\boldsymbol{\rho}, 0) \;=\; 0 \qquad\qquad \text{on S} \,, \qquad\qquad (2.1.6)$$

or Neumann conditions, for which the normal derivative is zero,

$$\left.\frac{\partial U_N(\boldsymbol{\rho}, z)}{\partial z}\right|_{z=0} \;=\; 0 \qquad \text{on S} \,. \qquad (2.1.7)$$

A screen that satisfies Dirichlet boundary conditions is sometimes said to be soft, while a Neumann-type one is sometimes said to be hard. Both may be termed opaque because, on S, the normal component of the energy flux vector is identically zero in either case (see Section 2.1.6).

---

* We will also use this notation when we are referring to a specific side of the plane z = 0 where the screen is located. Whenever z = 0 appears without a superscript, the equations apply to both sides.

The boundary conditions on the screen together with the continuity conditions (2.1.4) and (2.1.5) completely describe what happens to the field as it traverses the plane $z = 0$. When the continuity conditions are applied to a Dirichlet-type screen, one finds that the $z$-derivative of the field in the aperture is equal to the $z$-derivative of the incident field,

$$\frac{\partial U_D(\boldsymbol{\rho},z)}{\partial z}\bigg|_{z=0} = \frac{\partial U^{(\text{inc})}(\boldsymbol{\rho},z)}{\partial z}\bigg|_{z=0} \qquad \text{in A .} \qquad (2.1.8)$$

However, when they are applied to a Neumann-type screen, the field in the aperture is simply the incident field,

$$U_N(\boldsymbol{\rho},0) = U^{(\text{inc})}(\boldsymbol{\rho},0) \qquad \text{in A .} \qquad (2.1.9)$$

Thus, the aperture diffraction problem is now posed as a mixed* boundary value problem.[10] For the Dirichlet-type screen, the field must satisfy the mixed boundary conditions given by Eqs. (2.1.6) and (2.1.8) whereas, for the Neumann-type screen, it must satisfy those given by Eqs. (2.1.7) and (2.1.9).

Although at first it might appear that, for a given incident field, the above mixed boundary conditions together with the radiation condition should uniquely specify the field $U(\mathbf{r})$ everywhere in space, there can, in general, exist an infinite number of solutions for $U(\mathbf{r})$ (see Ref. 4, Chapter 9, or Ref. 11). This non-uniqueness, which is not present in continuously varying media, can occur here because of the sharp edge of the aperture. However, only one of the solutions is physically sensible. The others contain singularities that act as primary sources of radiation. It is possible to choose the correct one by applying the so-called edge conditions.[4,11,12] We shall not discuss these edge conditions further here, but it should be pointed out that the numerical results given in Chapter 4 do, in fact, satisfy them.

---

* A boundary value problem is said to be mixed when the field is specified over some portion of a surface and its normal derivative is specified over the remaining portion.

### 2.1.2 The Rayleigh Diffraction Formulas

Let us suppose that the boundary value problem has been solved and that the field $U(\boldsymbol{\rho}, 0^+)$ behind the screen is known for all values of $\boldsymbol{\rho}$. Then the field $U(\boldsymbol{\rho}, z)$ in the half-space $z \geq 0$ can be determined from Rayleigh's first diffraction formula (see Ref. 3, Chapter 5, or Ref. 13),

$$U(\boldsymbol{\rho}, z) \;=\; -\frac{1}{2\pi} \int U(\boldsymbol{\rho}', 0^+)\; \frac{\partial G(\boldsymbol{\rho} - \boldsymbol{\rho}', z)}{\partial z}\, d^2\rho' \;. \qquad (2.1.10)$$

Here the integration is over the entire plane $z = 0^+$ and the function

$$G(\boldsymbol{\rho}, z) \;=\; \frac{e^{ik\sqrt{\rho^2 + z^2}}}{\sqrt{\rho^2 + z^2}} \;, \qquad (2.1.11)$$

$\rho = |\boldsymbol{\rho}| = \sqrt{x^2 + y^2}$, is the three-dimensional free-space Green function that obeys the radiation condition at infinity. Alternatively, if the boundary value of the derivative $\partial U(\boldsymbol{\rho}, z)/\partial z \big|_{z=0^+}$ is known for all values of $\boldsymbol{\rho}$, $U(\boldsymbol{\rho}, z \geq 0)$ can be determined from Rayleigh's second diffraction formula (see Ref. 3, Chapter 5, or Ref. 13),

$$U(\boldsymbol{\rho}, z) \;=\; -\frac{1}{2\pi} \int \frac{\partial U(\boldsymbol{\rho}', z')}{\partial z'} \bigg|_{z'=0^+} G(\boldsymbol{\rho} - \boldsymbol{\rho}', z)\, d^2\rho' \;. \qquad (2.1.12)$$

Obviously, if both $U(\boldsymbol{\rho},0^+)$ and $\partial U(\boldsymbol{\rho},z)/\partial z \big|_{z=0^+}$ are known exactly, the two Rayleigh formulas yield the same result and so does the linear superposition

$$U(\boldsymbol{\rho},z) = -\frac{C}{2\pi} \int U(\boldsymbol{\rho}',0^+) \frac{\partial G(\boldsymbol{\rho}-\boldsymbol{\rho}',z)}{\partial z} \, d^2\rho'$$
$$-\frac{(1-C)}{2\pi} \int \frac{\partial U(\boldsymbol{\rho}',z')}{\partial z'} \bigg|_{z'=0^+} G(\boldsymbol{\rho}-\boldsymbol{\rho}',z) \, d^2\rho' \, ,$$

$$(2.1.13)$$

where C is a real constant.

### 2.1.3    The Angular Spectrum Representation

A representation in terms of an angular spectrum of plane waves can often provide more insight into the propagation of the field than do the Rayleigh diffraction formulas.  With this representation, the field in the half-space $z \geq 0$ may be expressed in the form[13-15]

$$U(\boldsymbol{\rho},z) = \int a(\mathbf{u}_\perp) \, e^{ik\mathbf{u}_\perp \cdot \boldsymbol{\rho}} \, e^{iku_z z} \, d^2 u_\perp \, , \qquad (2.1.14)$$

where the integration extends over all real transverse vectors $\mathbf{u}_\perp \equiv (u_x, u_y)$.  The (generally complex) angular spectrum amplitude $a(\mathbf{u}_\perp)$ determines the amplitude of the plane wave specified by the unit vector $\mathbf{u} \equiv (u_x, u_y, u_z)$.  The longitudinal component of this vector is given by the formulas

$$u_z = \sqrt{1 - u_\perp^2} \qquad \text{for } u_\perp = \sqrt{u_x^2 + u_y^2} \leq 1 \, , \qquad (2.1.15a)$$

$$u_z = i\sqrt{u_\perp^2 - 1} \qquad \text{for } u_\perp = \sqrt{u_x^2 + u_y^2} > 1 \, . \qquad (2.1.15b)$$

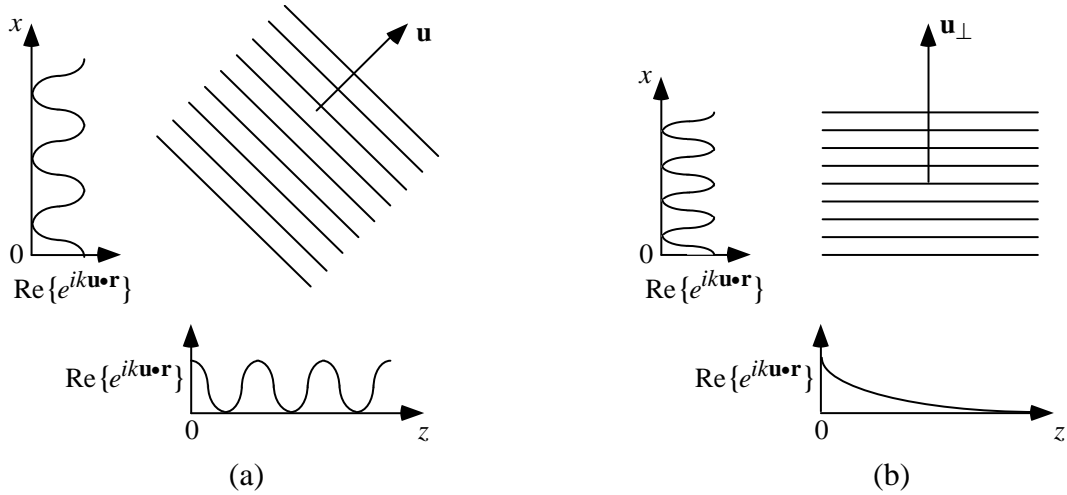

**Figure 2-2** Illustration of (a) a homogeneous wave propagating along the direction specified by the unit vector $\mathbf{u}$ and (b) an evanescent wave propagating along the $\mathbf{u}_\perp$-direction and decaying exponentially along the $z$-axis.

The waves associated with Eq. (2.1.15a) are the usual homogeneous plane waves, with phase fronts perpendicular to the vector $\mathbf{u}$ [see Fig. 2-2(a)]. Those associated with Eq. (2.1.15b) are evanescent (inhomogeneous) plane waves whose phase fronts are perpendicular to $\mathbf{u}_\perp$ and whose amplitudes decay exponentially with increasing distance $z$ from the aperture [see Fig. 2-2(b)].

By setting $z = 0^+$ in Eq. (2.1.14) and performing a Fourier transform transform with respect to $\boldsymbol{\rho}$, one finds that the angular spectrum amplitude $a(\mathbf{u}_\perp)$ may be determined from the field distribution $U(\boldsymbol{\rho}, 0^+)$ via the relation

$$a(\mathbf{u}_\perp) = \left(\frac{k}{2\pi}\right)^2 \int U(\boldsymbol{\rho}', 0^+)\, e^{-ik\mathbf{u}_\perp \cdot \boldsymbol{\rho}'} d^2 \rho' \ . \qquad (2.1.16)$$

Alternatively, $a(\mathbf{u}_\perp)$ may be determined from the derivative $\partial U(\boldsymbol{\rho}, z) / \partial z \big|_{z=0^+}$:

$$a(\mathbf{u}_\perp) \; = \; -\frac{i}{ku_z} \left(\frac{k}{2\pi}\right)^2 \int \frac{\partial U(\boldsymbol{\rho}',z')}{\partial z'} \bigg|_{z'=0^+} e^{-ik\mathbf{u}_\perp \cdot \boldsymbol{\rho}'} d^2\rho' \; . \quad (2.1.17)$$

The relationship between the angular spectrum representation and the Rayleigh diffraction formulas can be readily established with the help of Weyl's expansion for a spherical wave,[13]

$$G(\boldsymbol{\rho},z) \; = \; \frac{e^{ik\sqrt{\rho^2+z^2}}}{\sqrt{\rho^2+z^2}} \; = \; \frac{ik}{2\pi}\int \frac{1}{u_z} \, e^{ik\mathbf{u}_\perp \cdot \boldsymbol{\rho}} \, e^{iku_z z} \, d^2u_\perp \; . \quad (2.1.18)$$

By substituting either Eq. (2.1.16) or (2.1.17) into Eq. (2.1.14) and making use of this expansion, one obtains Rayleigh's first or second diffraction formula, respectively.

### 2.1.4   The Bessel-Beam Representation

It is possible to introduce a representation that is completely equivalent to the angular spectrum representation, but which contains nondiffracting and evanescent Bessel-beams instead of homogeneous and evanescent plane waves. This Bessel-beam representation[16,17] (see also Ref. 5, Section 4.11) is rarely used as a starting point in propagation and diffraction problems, but it is often applied to problems of scattering and waveguiding by circular cylinders, where it is sometimes called a cylindrical wave expansion.[18]

Nondiffracting Bessel beams have themselves been of considerable interest recently[19-23] because their intensity distribution does not change on propagation and, therefore, they have infinite depth of focus. Unfortunately, they also have infinite energy. Hence, in practice, such beams can only be approximated over a finite propagation distance.

To obtain the Bessel-beam representation from the angular spectrum representation, we expand the angular spectrum amplitude in a Fourier series over the angular domain of $\mathbf{u}_\perp$:

$$a(\mathbf{u}_\perp) \;=\; \frac{1}{2\pi u_\perp} \sum_{n=-\infty}^{\infty} c_n(u_\perp)\,(i)^{-n}\,e^{in\psi}\;, \qquad (2.1.19)$$

where $\cos\psi = u_x/u_\perp$, $\sin\psi = u_y/u_\perp$ and the expansion coefficients $c_n(u_\perp)$ are related to $a(\mathbf{u}_\perp)$ by

$$c_n(u_\perp) \;=\; i^n u_\perp \int_0^{2\pi} a(\mathbf{u}_\perp)\,e^{-in\psi}\,d\psi\;. \qquad (2.1.20)$$

If we now substitute from Eq. (2.1.19) into Eq. (2.1.14) and change to circular cylinder coordinates, $[\,\mathbf{r} = (\rho, \varphi, z),\ \rho = |\boldsymbol{\rho}| = \sqrt{x^2 + y^2}\,,\ \cos\varphi = x/\rho,\ \sin\varphi = y/\rho\,]$, we arrive at the expression

$$U(\rho, \varphi, z) \;=\; \frac{1}{2\pi} \sum_{n=-\infty}^{\infty} \int_0^{\infty} c_n(u_\perp)\,(i)^{-n}\,e^{iku_z z} \left\{ \int_0^{2\pi} e^{in\psi}\,e^{iku_\perp \rho \cos(\psi - \varphi)}\,d\psi \right\} du_\perp\;. $$

$$(2.1.21)$$

The term in the curly brackets may be rewritten in terms of a Bessel function of the first kind and $n$th order, with the help of the identity[24]

$$J_n(v) \;=\; \frac{(i)^{-n}}{\pi} \int_0^{\pi} \cos(n\psi)\,e^{iv\cos\psi}\,d\psi\;. \qquad (2.1.22)$$

We then obtain the Bessel-beam representation for a field propagating into the half-space $z > 0$,

$$U(\rho, \varphi, z) \;=\; \sum_{n=-\infty}^{\infty} \int_{0}^{\infty} c_n(u_\perp) \; \mathrm{B}_{u_\perp}^{(n)}(\rho, \varphi, z) \, du_\perp \,. \qquad (2.1.23)$$

The functions

$$\mathrm{B}_{u_\perp}^{(n)}(\rho, \varphi, z) \;\equiv\; J_n(ku_\perp \rho) \, e^{in\varphi} \, e^{iku_z z} \qquad (2.1.24)$$

are $n$th-order Bessel beams with parameter $u_\perp$ [$u_z$ is given by Eqs. (2.1.15a) and (2.1.15b)]. Bessel beams are solutions of the Helmholtz equation that are finite when $\rho = 0$ and that tend to zero as $|\rho| \to \infty$. For $u_\perp \leq 1$, the amplitude distribution of $\mathrm{B}_{u_\perp}^{(n)}(\mathbf{r})$ does not change on propagation along $z$,

$$\left| \mathrm{B}_{u_\perp}^{(n)}(\boldsymbol{\rho}, z) \right| \;=\; \left| \mathrm{B}_{u_\perp}^{(n)}(\boldsymbol{\rho}, 0) \right| \,, \qquad (2.1.25)$$

and $\mathrm{B}_{u_\perp}^{(n)}(\mathbf{r})$ may be termed a nondiffracting Bessel beam. These are the usual Bessel beams encountered in discussions of nondiffracting fields. [19-23] On the other hand, for $u_\perp > 1$, the amplitude distribution of $\mathrm{B}_{u_\perp}^{(n)}(\mathbf{r})$ decays exponentially along the positive $z$-direction,

$$\left| \mathrm{B}_{u_\perp}^{(n)}(\boldsymbol{\rho}, z) \right| \;=\; \left| \mathrm{B}_{u_\perp}^{(n)}(\boldsymbol{\rho}, 0) \right| e^{-kz\sqrt{u_\perp^2 - 1}} \,, \qquad (2.1.26)$$

and $\mathrm{B}_{u_\perp}^{(n)}(\mathbf{r})$ may be termed an evanescent Bessel beam.

It should be stressed that, in general, the Bessel-beam representation is not paraxial. In the paraxial regime, the expansion coefficients $c_n(u_\perp)$ are non-zero only for $u_\perp^2 \ll 1$ and $u_z$ can be approximated by $u_z = 1 - u_\perp^2/2$.

The completeness relation for Bessel functions,[25]

$$\int_0^\infty J_n(\alpha\rho)\, J_n(\alpha'\rho)\, d\rho \;=\; \frac{\delta(\alpha-\alpha')}{\alpha}\;, \qquad (2.1.27)$$

$\delta(\alpha)$ being the Dirac delta function, can be used to show that the expansion coefficients may be expressed directly in terms of the field distribution $U(\rho,0^+)$ in the plane $z = 0^+$ as

$$c_n(u_\perp) \;=\; \frac{k^2}{2\pi}\, u_\perp \int_0^\infty\!\!\int_0^{2\pi} U(\rho,\varphi,0^+)\, e^{-in\varphi}\, J_n(ku_\perp\rho)\, \rho\, d\rho\, d\varphi\;. \qquad (2.1.28)$$

They may also be determined from the derivative $\partial U(\rho,z)/\partial z\,\big|_{z=0^+}$ through the relationship

$$c_n(u_\perp) \;=\; -\frac{ik}{2\pi}\,\frac{u_\perp}{u_z}\int_0^\infty\!\!\int_0^{2\pi} \frac{\partial U(\rho,\varphi,z')}{\partial z'}\,\bigg|_{z'=0^+} e^{-in\varphi}\, J_n(ku_\perp\rho)\, \rho\, d\rho\, d\varphi\;.$$
$$(2.1.29)$$

As explained in the previous section, the Weyl expansion provides the connection between the angular spectrum representation and the two Rayleigh diffraction formulas. In the Bessel-beam representation, that connection can be established with the use of Eqs. (2.1.28) and (2.1.29) and the following expansion for the free-space Green function:

$$G(\mathbf{r} - \boldsymbol{\rho}') \;\; = \;\; \frac{e^{ik|\mathbf{r} - \boldsymbol{\rho}'|}}{|\mathbf{r} - \boldsymbol{\rho}'|}$$

$$= \; ik \sum_{n=-\infty}^{\infty} \int_{0}^{\infty} \frac{u_{\perp}}{u_z} \, J_n(ku_{\perp}\rho) \, J_n(ku_{\perp}\rho') \, e^{in(\varphi - \varphi')} \, e^{iku_z|z-z'|} \, du_{\perp} \; .$$

$$(2.1.30)$$

### 2.1.5 The Rayleigh-Sommerfeld Theories and the Kirchhoff Theory

The various diffraction formulas given in Sections (2.1.2)-(2.1.4) make it possible to calculate the field $U(\boldsymbol{\rho}, z)$ throughout the half-space $z \geq 0$ from the boundary value $U(\boldsymbol{\rho}, 0^+)$ of the field or from the boundary value $\partial U(\boldsymbol{\rho}, z)/\partial z \mid_{z=0^+}$ of its derivative. However, the rigorous diffraction problem is posed as a mixed boundary problem (see Section 2.1.1) and it is difficult to determine either $U(\boldsymbol{\rho}, 0^+)$ or $\partial U(\boldsymbol{\rho}, z)/\partial z \mid_{z=0^+}$ for all values of $\boldsymbol{\rho}$. Therefore, one often resorts to theories that contain approximate boundary values. We will discuss three of these approximate theories here: the Rayleigh-Sommerfeld theories of the first and the second kinds and the Kirchhoff theory.[26-28] These are the ones used most often in optics. Although they were not originally intended to be used as approximations for field diffracted by apertures in Dirichlet-type or Neumann-type screens, they are frequently used for such screens, especially in comparisons with exact results (see, for example, Refs. 26 and 29).

In the Rayleigh-Sommerfeld theory of the first kind, the field, denoted by $U^{(I)}(\boldsymbol{\rho}, z)$, is obtained by approximating the boundary values in Rayleigh's first diffraction formula (2.1.10) by

$$U^{(I)}(\boldsymbol{\rho}, 0^+) \; = \; 0 \qquad\qquad \text{on S} \qquad (2.1.31a)$$

and

$$U^{(I)}(\boldsymbol{\rho}, 0^+) \; = \; U^{(inc)}(\boldsymbol{\rho}, 0) \qquad \text{in A} \;, \qquad (2.1.31b)$$

i.e., it is assumed that the field on the screen is zero in the shadow region and that the field in the aperture is simply the unperturbed incident field. Therefore, $U^{(I)}(\boldsymbol{\rho}, z)$ in the half-space $z \geq 0$ is given by the expression

$$U^{(I)}(\boldsymbol{\rho}, z) \; = \; -\frac{1}{2\pi} \int_A U^{(inc)}(\boldsymbol{\rho}, 0) \; \frac{\partial G(\boldsymbol{\rho} - \boldsymbol{\rho}', z)}{\partial z} \; d^2\rho' \;, \qquad (2.1.32)$$

where the domain of integration extends over the aperture A.

The field in the Rayleigh-Sommerfeld theory of the second kind, denoted by $U^{(II)}(\boldsymbol{\rho}, z)$, involves a similar approximation to Rayleigh's second diffraction formula (2.1.12):

$$\left. \frac{\partial U^{(II)}(\boldsymbol{\rho}, z)}{\partial z} \right|_{z=0^+} \; = \; 0 \qquad \qquad \text{on S} \qquad (2.1.33a)$$

and

$$\left. \frac{\partial U^{(II)}(\boldsymbol{\rho}, z)}{\partial z} \right|_{z=0^+} = \left. \frac{\partial U^{(inc)}(\boldsymbol{\rho}, z)}{\partial z} \right|_{z=0} \qquad \text{in A} \;. \qquad (2.1.33b)$$

Hence,

$$U^{(II)}(\boldsymbol{\rho}, z) \; = \; -\frac{1}{2\pi} \int_A \left. \frac{\partial U^{(inc)}(\boldsymbol{\rho}', z')}{\partial z'} \right|_{z'=0} G(\boldsymbol{\rho} - \boldsymbol{\rho}', z) \; d^2\rho' \;, \quad \text{for } z \geq 0.$$

$$(2.1.34)$$

Lastly, the Kirchhoff theory is obtained from Eq. (2.1.13) with $C = 1/2$, viz.,

$$U(\boldsymbol{\rho},z) \;=\; -\frac{1}{4\pi}\int U(\boldsymbol{\rho}',0^{+})\;\frac{\partial G(\boldsymbol{\rho}-\boldsymbol{\rho}',z)}{\partial z}\,d^{2}\rho'$$

$$-\frac{1}{4\pi}\int \left.\frac{\partial U(\boldsymbol{\rho}',z')}{\partial z'}\right|_{z'=0^{+}} G(\boldsymbol{\rho}-\boldsymbol{\rho}',z)\,d^{2}\rho'\;,$$

$$(2.1.35)$$

and the following values for $U(\boldsymbol{\rho},0^{+})$ and $\partial U(\boldsymbol{\rho},z)/\partial z\,|_{z=0^{+}}$ :

$$U(\boldsymbol{\rho},0^{+}) \;=\; 0\;,\;\; \left.\frac{\partial U(\boldsymbol{\rho},z)}{\partial z}\right|_{z=0^{+}} \;=\; 0 \qquad\qquad\text{on S}$$

$$(2.1.36a)$$

and

$$U(\boldsymbol{\rho},0^{+}) \;=\; U^{(\text{inc})}(\boldsymbol{\rho},0)\;,\;\; \left.\frac{\partial U(\boldsymbol{\rho},z)}{\partial z}\right|_{z=0^{+}} \;=\; \left.\frac{\partial U^{(\text{inc})}(\boldsymbol{\rho},z)}{\partial z}\right|_{z=0} \qquad\text{in A}\;.$$

$$(2.1.36b)$$

The resulting field $U^{(K)}(\boldsymbol{\rho},z)$ is the average of $U^{(I)}(\boldsymbol{\rho},z)$ and $U^{(II)}(\boldsymbol{\rho},z)$:

$$U^{(K)}(\boldsymbol{\rho},z) \;=\; \frac{1}{2}\left[U^{(I)}(\boldsymbol{\rho},z)\;+\;U^{(II)}(\boldsymbol{\rho},z)\right].\qquad(2.1.37)$$

It is well known that the limit of $U^{(K)}(\boldsymbol{\rho},z)$ as $z\to 0^{+}$ does not yield the values of $U(\boldsymbol{\rho},0^{+})$ and $\partial U(\boldsymbol{\rho},z)/\partial z\,|_{z=0^{+}}$ given by Eqs. (2.1.36a) and (2.1.36b). However, as was stressed by Kottler,[30,31] there are no logical inconsistencies if the Kirchhoff theory is viewed as a solution to a saltus problem, which involves a discontinuity in the field and its $z$-derivative, rather than a boundary value problem.

It should be pointed out that, unlike the Eqs. (2.1.10), (2.1.12) and (2.1.13), which all produce the same result for the field distribution in the half-space z ≥ 0 if the boundary values $U(\boldsymbol{\rho},0^{+})$ and $\partial U(\boldsymbol{\rho},z)/\partial z\,|_{z=0^{+}}$ are known exactly for all values of $\boldsymbol{\rho}$, Eqs. (2.1.32), (2.1.34) and (2.1.37) each produce a different result because of the approximations that are made to these boundary values.

### 2.1.6 Energy Flux through the Aperture

The time-averaged, real, energy flux vector $\mathbf{F(r)}$ associated with a monochromatic, complex, scalar field $U(\mathbf{r})$ can be defined by the expression[32-34]

$$\mathbf{F(r)} \equiv i\omega\alpha\left[U(\mathbf{r})\nabla U^*(\mathbf{r}) - U^*(\mathbf{r})\nabla U(\mathbf{r})\right], \qquad (2.1.38)$$

which is similar to the definition of the probability current density in quantum mechanics. Here the asterisk denotes the complex conjugate and $\alpha$ is a real, positive constant that depends on the choice of units. From the Helmholtz equation, it can readily be shown that, in free space, $F(\mathbf{r})$ obeys the energy conservation law

$$\nabla \cdot \mathbf{F(r)} = 0 . \qquad (2.1.39)$$

In integral form, this conservation law may be written as

$$\int_{\Sigma} \mathbf{F(r)} \cdot \hat{\mathbf{n}}\, d\sigma = 0 , \qquad (2.1.40)$$

where $\Sigma$ is a closed surface with outward unit normal $\hat{\mathbf{n}}$ and surface area element $d\sigma$.

If we now consider the total energy flux $F_{\text{tot}}$ through the aperture A,

$$F_{\text{tot}} = \int_{A} \mathbf{F}(\boldsymbol{\rho}, 0^+) \cdot \hat{\mathbf{z}}\, d^2\rho , \qquad (2.1.41)$$

$\hat{\mathbf{z}}$ being a unit vector in the $z$-direction, and if we choose the surface $\Sigma$ to consist of the aperture A, the opaque screen S and an infinite hemisphere located in the half-space z > 0, then by virtue of Eq. (2.1.40), $F_{\text{tot}}$ is equal to the total energy radiated into the far zone[*]. With the help of Eqs. (2.1.14) and (2.1.38), we can express $F_{\text{tot}}$ in terms of angular spectrum amplitude $a(\mathbf{u}_\perp)$ as

$$F_{\text{tot}} = 8\pi^2 \alpha c \int\limits_{u_\perp \leq 1} u_z \left|a(\mathbf{u}_\perp)\right|^2 d^2 u_\perp .$$  (2.1.42)

Since the integration in this equation is limited by $u_\perp \leq 1$, the total energy flux contains only contributions from homogeneous plane waves. Equivalently, if Eq. (2.1.19) is used to expand the angular spectrum amplitude in terms of the Bessel-beam coefficients $c_n(u_\perp)$, $F_{\text{tot}}$ may also be written in the form[†]

$$F_{\text{tot}} = 4\pi \alpha c \sum_{n=-\infty}^{\infty} \int\limits_{u_\perp \leq 1} \frac{u_z}{u_\perp^2} \left|c_n(u_\perp)\right|^2 du_\perp ,$$  (2.1.43)

which contains only contributions from nondiffracting Bessel beams.

### 2.1.7 Coupled Amplitude-Phase Equations and the Behavior of the Phase Near Extrema of the Amplitude

We shall digress briefly from our discussion of aperture diffraction to derive coupled amplitude-phase equations and to establish two new theorems concerning the behavior of the phase near extrema of the amplitude. These theorems are included here because they provide a novel insight into the relationship between the amplitude and the phase of diffracted fields. They can be used to explain a variety of interesting phase phenomena including, for example, the so-called phase anomaly that occurs near a focus (see Ref. 2, Section 8.8.4).

---

[*] It can verified that, for both Dirichlet and Neumann-type screens, the integration over the screen $\mathcal{S}$ does not yield any contribution, because the $z$-component of the energy flux vector is identically zero on $\mathcal{S}$.

[†] Although the factor $u_\perp$ appears in the denominator of the integrand in Eq. (2.1.43) , the integrand is usually not singular because the coefficients $c_n(u_\perp)$ contain a multiplicative factor of $u_\perp$ [see Eq. (2.1.20)].

We can express the field in terms of its amplitude $A(\mathbf{r})$ and its phase $\phi(\mathbf{r})$ as

$$U(\mathbf{r}) \;=\; A(\mathbf{r}) \; e^{i\phi(\mathbf{r})} \;, \tag{2.1.44}$$

where $A(\mathbf{r})$ and $\phi(\mathbf{r})$ are real functions of position, $A = |U| \ge 0$, $\cos\phi = \mathrm{Re}\{U\}/|U|$ and $\sin\phi = \mathrm{Im}\{U\}/|U|$. Because the field and its derivatives are continuous throughout free space, $A$ is continuous at all points and $\nabla A$, $\phi$ and $\nabla\phi$ are continuous at points where $A$ is non-zero. If we substitute from Eq. (2.1.44) into the Helmholtz equation, we obtain at once the relation

$$\nabla^2 A \;+\; 2i\nabla A \cdot \nabla\phi + iA\nabla^2\phi - A(\nabla\phi)^2 + k^2 A \;=\; 0 \;. \tag{2.1.45}$$

The real and imaginary parts of Eq. (2.1.45) yield the following coupled amplitude-phase partial differential equations:

$$(\nabla\phi)^2 \;=\; k^2 + \frac{\nabla^2 A}{A} \tag{2.1.46}$$

and

$$\nabla \cdot \left[ A^2 \nabla\phi \right] \;=\; 0 \;. \tag{2.1.47}$$

Equations (2.1.46) and (2.1.47) are completely equivalent to the Helmholtz equation. The same equations appear in the derivation of geometrical optics from wave theory,[35,36] in which case the short-wavelength limit $k \to \infty$ is of interest. In this limit, Eq. (2.1.46) becomes the usual eikonal equation of geometrical optics* (see Ref. 2, Chapter 3 or Ref. 5, Chapter 2). However for finite $k$, if the amplitude $A$ is known, Eq. (2.1.46) may be viewed as a generalized eikonal equation that includes diffraction effects. It should be mentioned that equations analogous to (2.1.46) and (2.1.47) were used in an early attempt to obtain a hydrodynamical model for quantum mechanics.[37]

Let us substitute from Eq. (2.1.44) into expression (2.1.38) for the energy flux vector $\mathbf{F}$. $\mathbf{F}$ then takes the simple form†

$$\mathbf{F} \;=\; 2\omega\alpha\, A^2 \nabla\phi \;, \tag{2.1.48}$$

and Eqs. (2.1.46) and (2.1.47) may be rewritten as

$$F^2 \;=\; 4\omega^2\alpha^2 \left[ k^2 A^4 + A^3 \nabla^2 A \right] \tag{2.1.49}$$

and

$$\nabla \cdot \mathbf{F} \;=\; 0 \,, \tag{2.1.50}$$

respectively. Therefore, since Eq. (2.1.50) is the usual energy conservation law (see Section 2.1.6), Eq. (2.1.47) also expresses conservation of energy. If we write the energy flux vector as

$$\mathbf{F} \;=\; F\, \frac{\nabla\phi}{|\nabla\phi|} \;, \tag{2.1.51}$$

we see that the magnitude of $\mathbf{F}$ depends only on the amplitude of the field [see Eq. (2.1.49)] and its direction depends only on the phase.

---

* In free space, the eikonal equation is $\left(\nabla\phi\right)^2 = k^2$.

† Equation (2.1.48) displays a very simple relationship between the energy flux vector $\mathbf{F}$, the intensity $I = A^2$ and the gradient of the phase $\nabla\phi$. One might wonder whether an analogous expression can be obtained for an electromagnetic field. This issue is discussed in Appendix A.

We now consider a point $P(\mathbf{r}_s)$ where the amplitude $A$ has a non-zero extremum. If $A$ has a maximum at $P(\mathbf{r}_s)$, then $A(\mathbf{r}_s) > 0$, $\nabla A(\mathbf{r}_s) = 0$ and $\nabla^2 A(\mathbf{r}_s) < 0$ (this notation has the obvious meaning: $\nabla A(\mathbf{r}_s) = \nabla A(\mathbf{r})\big|_{\mathbf{r}=\mathbf{r}_s}$, etc.). Hence, according to Eqs. (2.1.46) and (2.1.47),

$$\left|\nabla\phi(\mathbf{r}_s)\right| \; < \; k \quad \text{and} \quad \nabla^2\phi(\mathbf{r}_s) \; = \; 0 \; . \tag{2.1.52}$$

However, if $A$ has a non-zero minimum, then $A(\mathbf{r}_s) > 0$, $\nabla A(\mathbf{r}_s) = 0$ and $\nabla^2 A(\mathbf{r}_s) > 0$ and, consequently,

$$\left|\nabla\phi(\mathbf{r}_s)\right| \; > \; k \quad \text{and} \quad \nabla^2\phi(\mathbf{r}_s) \; = \; 0 \; . \tag{2.1.53}$$

In the unusual case that the amplitude has a non-zero extremum with $\nabla^2 A(\mathbf{r}_s) = 0$, one has $\left|\nabla\phi(\mathbf{r}_s)\right| \; = \; k$.

Because the above results are very general, i.e., they apply to any field that satisfies the Helmholtz equation, we shall restate them in the following two theorems.

Theorem 1: *Near a point in space where the amplitude has a* **maximum***, the Laplacian of the phase is zero and the surfaces of constant phase are spaced* **further apart** *than the corresponding surfaces for a plane wave with wavenumber $k = \omega / c$.*

Theorem 2: *Near a point in space where the amplitude has a non-zero* **minimum***, the Laplacian of the phase is zero and the surfaces of constant phase are spaced* **closer together** *than the corresponding surfaces for a plane wave with wavenumber $k = \omega / c$.*

The condition $\nabla^2 \phi(\mathbf{r}_s) = 0$ ensures that the wavefront has a certain "flatness" in the vicinity $P(\mathbf{r}_s)$, but this does not imply that the wavefront is necessarily planar in that region. For example, if we choose our coordinate system so that the origin coincides with the point $P(\mathbf{r}_s)$ and so that the $z$-axis is along the direction of $\nabla \phi(\mathbf{r}_s)$, the phase could be of the form

$$\phi(\mathbf{r}) = bz + d\left(x^2 - y^2\right), \qquad (2.1.54)$$

where $b > 0$ and $d$ are constants and

$$b^2 = k^2 + \frac{\nabla^2 A(0)}{A(0)} . \qquad (2.1.55)$$

In this case, the surfaces of constant phase have saddle points along the $z$-axis. However, the phase could not be of the form

$$\phi(\mathbf{r}) = b'z + d'\left(x^2 + y^2\right), \qquad (2.1.56)$$

i.e., the field near an extemum of the amplitude cannot have a (paraxial) spherical wavefront.

## 2.2  ELECTROMAGNETIC THEORY OF DIFFRACTION

We now turn to the electromagnetic theory of diffraction.[1,2,38]  We consider the case of a monochromatic electromagnetic field of frequency $\omega$ incident from the half-space $z < 0$ upon an aperture $\mathcal{A}$ in an opaque planar screen $\mathcal{S}$, which is located in the plane $z = 0$ (see Fig. 2-1).  Throughout free-space, the total electric field $\mathbf{E}(\mathbf{r})\exp\{-i\omega t\}$ and the total magnetic field $\mathbf{H}(\mathbf{r})\exp\{-i\omega t\}$ obey the time-harmonic, source-free Maxwell equations (Gaussian units),

$$\nabla \times \mathbf{E}(\mathbf{r}) \ = \ ik\mathbf{H}(\mathbf{r}) \ , \tag{2.2.1a}$$

$$\nabla \times \mathbf{H}(\mathbf{r}) \ = \ -ik\mathbf{E}(\mathbf{r}) \ , \tag{2.2.1b}$$

$$\nabla \cdot \mathbf{E}(\mathbf{r}) \ = \ 0 \ , \tag{2.2.1c}$$

$$\nabla \cdot \mathbf{H}(\mathbf{r}) \ = \ 0 \ . \tag{2.2.1d}$$

Consequently, the electric and magnetic fields each also satisfy the Helmholtz equation,

$$\left(\nabla^2 + \ k^2\right)\mathbf{E}(\mathbf{r}) \ = \ 0, \tag{2.2.2a}$$

$$\left(\nabla^2 + \ k^2\right)\mathbf{H}(\mathbf{r}) \ = \ 0 \ . \tag{2.2.2b}$$

### 2.2.1 Boundary Conditions

As in the scalar theory, the boundary conditions for the plane $z = 0$ are needed. The continuity of the electric and magnetic fields in the aperture simply requires that

$$\mathbf{E}(\boldsymbol{\rho},0^+) \ = \ \mathbf{E}(\boldsymbol{\rho},0^-) \qquad \text{in A} \tag{2.2.3a}$$

and

$$\mathbf{H}(\boldsymbol{\rho},0^+) \ = \ \mathbf{H}(\boldsymbol{\rho},0^-) \qquad \text{in A} \ . \tag{2.2.3b}$$

We will only consider the case of an electromagnetically opaque screen $S$ that is a perfect electric conductor. The tangential components of the electric field and the normal component of the magnetic field are then identically zero on the screen:

$$\hat{\mathbf{z}} \times \mathbf{E}(\boldsymbol{\rho},0) \ = \ 0 \qquad\qquad \text{on } \mathrm{S} \ , \qquad\qquad (2.2.4)$$

$$\hat{\mathbf{z}} \cdot \mathbf{H}(\boldsymbol{\rho},0) \ = \ 0 \qquad\qquad \text{on } \mathrm{S} \ . \qquad\qquad (2.2.5)$$

When Eqs. (2.2.3a) and (2.2.3b) are used in conjunction with Eqs. (2.2.4) and (2.2.5), one finds that, in the aperture $\mathrm{A}$, the $z$-component of the electric field and the $x$- and $y$-components of the magnetic field are unaltered from their values in the absence of the screen.[1,38]  Hence, if we denote the incident electric and magnetic fields by $\mathbf{E}^{(\text{inc})}$ and $\mathbf{H}^{(\text{inc})}$,

$$\hat{\mathbf{z}} \cdot \mathbf{E}(\boldsymbol{\rho},0) \ = \ \hat{\mathbf{z}} \cdot \mathbf{E}^{(\text{inc})}(\boldsymbol{\rho},0) \qquad\qquad \text{in } \mathrm{A} \ , \qquad\qquad (2.2.6)$$

$$\hat{\mathbf{z}} \times \mathbf{H}(\boldsymbol{\rho},0) \ = \ \hat{\mathbf{z}} \times \mathbf{H}^{(\text{inc})}(\boldsymbol{\rho},0) \qquad\qquad \text{in } \mathrm{A} \ . \qquad\qquad (2.2.7)$$

Since the electromagnetic field components specified on the screen [Eqs. (2.2.4) and (2.2.5)] are different from those specified in the aperture [Eqs. (2.2.6) and (2.2.7)], the electromagnetic aperture diffraction problem is a mixed boundary value problem.  However, the vector nature of this mixed boundary value problem makes it more complicated than the one that occurs in scalar diffraction.

It should be pointed out that here, as in the scalar case, the boundary conditions together with the radiation condition do not ensure a unique solution for the electromagnetic field.  Fortunately, edge conditions can again be invoked to determine which solution is physically realizable.[4,11,12]

### 2.2.2  Exact Diffraction Formulas

If all of the components of electromagnetic field were known in the plane $z = 0^+$ for all values of $\boldsymbol{\rho}$, we could apply Rayleigh's first diffraction formula (2.1.10) to each to compute the field in the half-space $z \geq 0$. The corresponding vector diffraction formulas would then be of the form

$$\mathbf{E}(\boldsymbol{\rho}, z) \;=\; -\frac{1}{2\pi} \int \mathbf{E}(\boldsymbol{\rho}', 0^+) \; \frac{\partial G(\boldsymbol{\rho} - \boldsymbol{\rho}', z)}{\partial z} \, d^2\rho' \;, \qquad (2.2.8a)$$

$$\mathbf{H}(\boldsymbol{\rho}, z) \;=\; -\frac{1}{2\pi} \int \mathbf{H}(\boldsymbol{\rho}', 0^+) \; \frac{\partial G(\boldsymbol{\rho} - \boldsymbol{\rho}', z)}{\partial z} \, d^2\rho' \;, \qquad (2.2.8b)$$

where here, as in Eq. (2.1.10), the integration is over the entire plane $z = 0^+$ and $G(\boldsymbol{\rho}, z)$ is the free-space Green function. We could also use Rayleigh's second diffraction formula (2.1.12) to determine $\mathbf{E}$ and $\mathbf{H}$:

$$\mathbf{E}(\boldsymbol{\rho}, z) \;=\; -\frac{1}{2\pi} \int \left. \frac{\partial \mathbf{E}(\boldsymbol{\rho}', z')}{\partial z'} \right|_{z'=0^+} G(\boldsymbol{\rho} - \boldsymbol{\rho}', z) \, d^2\rho' \;, \qquad (2.2.9a)$$

$$\mathbf{H}(\boldsymbol{\rho}, z) \;=\; -\frac{1}{2\pi} \int \left. \frac{\partial \mathbf{H}(\boldsymbol{\rho}', z')}{\partial z'} \right|_{z'=0^+} G(\boldsymbol{\rho} - \boldsymbol{\rho}', z) \, d^2\rho' \;. \qquad (2.2.9b)$$

These two sets of diffraction formulas, Eqs. (2.2.8a)-(2.2.9b), are formally correct because, in the half-space $z \geq 0$, all six components of the electromagnetic field satisfy the Helmholtz equation. However, as a consequence of Maxwell's equations, only two components are actually independent. Equations (2.2.8a)-(2.2.9b) therefore make use of more information than is necessary. Furthermore, a complication arises if one naively attempts to obtain approximate theories starting from Eqs. (2.2.8a)-(2.2.9b): the resulting expressions can violate Maxwell's equations. For example, if in Eq. (2.2.8a) $\mathbf{E}(\boldsymbol{\rho}, 0^+)$ is approximated by

$$\mathbf{E}(\boldsymbol{\rho}, 0^+) \;=\; 0 \qquad\qquad \text{on } \mathrm{S} \qquad\qquad (2.2.10a)$$

and

$$\mathbf{E}(\boldsymbol{\rho}, 0^+) \; = \; \mathbf{E}^{(\text{inc})}(\boldsymbol{\rho}, 0) \qquad \text{in A} , \qquad (2.2.10b)$$

the resulting electric field is not divergence-free[*].

There are several different electromagnetic diffraction formulas that overcome the inadequacies of Eqs. (2.2.8a)-(2.2.9b), but they are all equivalent to each other when no approximations are made to the boundary values of the field. We will only discuss some of these diffraction formulas here.

It is possible to express the electric and magnetic fields in the half-space $z \geq 0$ in terms of the tangential electric field in the plane $z = 0^+$ by the use of the relations[42,43]

$$\mathbf{E}(\boldsymbol{\rho}, z) \; = \; \frac{1}{2\pi} \nabla \times \int \left[ \hat{\mathbf{z}} \times \mathbf{E}(\boldsymbol{\rho}', 0^+) \right] G(\boldsymbol{\rho} - \boldsymbol{\rho}', z) \, d^2\rho' , \qquad (2.2.11a)$$

$$\mathbf{H}(\boldsymbol{\rho}, z) \; = \; -\frac{i}{2\pi k} \nabla \times \nabla \times \int \left[ \hat{\mathbf{z}} \times \mathbf{E}(\boldsymbol{\rho}', 0^+) \right] G(\boldsymbol{\rho} - \boldsymbol{\rho}', z) \, d^2\rho' , \qquad (2.2.11b)$$

or in terms of the tangential magnetic field by the use of the analogous relations[42,43]

$$\mathbf{H}(\boldsymbol{\rho}, z) \; = \; \frac{1}{2\pi} \nabla \times \int \left[ \hat{\mathbf{z}} \times \mathbf{H}(\boldsymbol{\rho}', 0^+) \right] G(\boldsymbol{\rho} - \boldsymbol{\rho}', z) \, d^2\rho' , \qquad (2.2.12a)$$

$$\mathbf{E}(\boldsymbol{\rho}, z) \; = \; \frac{i}{2\pi k} \nabla \times \nabla \times \int \left[ \hat{\mathbf{z}} \times \mathbf{H}(\boldsymbol{\rho}', 0^+) \right] G(\boldsymbol{\rho} - \boldsymbol{\rho}', z) \, d^2\rho' . \qquad (2.2.12b)$$

It is obvious from the form of Eqs. (2.2.11a)-(2.2.12b) that the fields given by these formulas satisfy Maxwell's equations. The integral that appears in Eqs. (2.2.11a) and (2.2.11b) is proportional to the magnetic Hertz vector, whereas the one in Eqs.

---

[*] Although the flaws of such an approximate theory have been known for a long time, it has been used recently in the study of focused fields (see Ref. 6, Section 15.4.1) along with another theory[39-41] that does not satisfy $\nabla \cdot \mathbf{E} = 0$ .

(2.2.12a) and (2.2.12b) is proportional to the electric Hertz vector. Linear combinations of Eqs. (2.2.11a) and (2.2.12b) and of Eqs. (2.2.11b) and (2.2.12a) can also be used when both the tangential electric and the tangential magnetic fields at $z = 0^+$ are known exactly.

Alternatively, instead of Green's functions, an angular spectrum representation can be employed, in which case the electric and magnetic fields in the half-space $z \geq 0$ may be written as[44-46]

$$\mathbf{E}(\boldsymbol{\rho}, z) \;=\; \int \mathbf{e}(\mathbf{u}_\perp) \, e^{ik\mathbf{u}_\perp \cdot \boldsymbol{\rho}} \, e^{iku_z z} \, d^2 u_\perp \;, \qquad (2.2.13a)$$

$$\mathbf{H}(\boldsymbol{\rho}, z) \;=\; \int \mathbf{h}(\mathbf{u}_\perp) \, e^{ik\mathbf{u}_\perp \cdot \boldsymbol{\rho}} \, e^{iku_z z} \, d^2 u_\perp \;. \qquad (2.2.13b)$$

As in Section 2.1.3, $\mathbf{u} = (u_x, u_y, u_z)$ is a unit vector, the integration is over all $\mathbf{u}_\perp = (u_x, u_y)$ and $u_z$ is given by Eqs. (2.1.15a) and (2.1.15b). Because of Maxwell's equations, the (vectorial) angular spectrum amplitudes $\mathbf{e}(\mathbf{u}_\perp)$ and $\mathbf{h}(\mathbf{u}_\perp)$ satisfy the relations

$$\mathbf{u} \times \mathbf{e}(\mathbf{u}_\perp) \;=\; \mathbf{h}(\mathbf{u}_\perp) \;, \qquad (2.2.14a)$$

$$\mathbf{u} \times \mathbf{h}(\mathbf{u}_\perp) \;=\; -\mathbf{e}(\mathbf{u}_\perp) \;, \qquad (2.2.14b)$$

$$\mathbf{u} \cdot \mathbf{e}(\mathbf{u}_\perp) \;=\; 0 \;, \qquad (2.2.14c)$$

$$\mathbf{u} \cdot \mathbf{h}(\mathbf{u}_\perp) \;=\; 0 \;. \qquad (2.2.14d)$$

Hence, only two of the six components of $\mathbf{e}(\mathbf{u}_\perp)$ and $\mathbf{h}(\mathbf{u}_\perp)$ are independent and, consequently, Eqs. (2.2.13a) and (2.2.13b) can be rewritten as

$$\mathbf{E}(\boldsymbol{\rho}, z) \;=\; \int \left[ \mathbf{e}_\perp(\mathbf{u}_\perp) \;-\; \hat{\mathbf{z}} \, \frac{\mathbf{u}_\perp \cdot \mathbf{e}_\perp(\mathbf{u}_\perp)}{u_z} \right] e^{ik\mathbf{u}_\perp \cdot \boldsymbol{\rho}} \, e^{iku_z z} \, d^2 u_\perp \;, \qquad (2.2.15a)$$

$$\mathbf{H}(\boldsymbol{\rho}, z) = -\frac{i}{k} \nabla \times \int \left[ \mathbf{e}_\perp(\mathbf{u}_\perp) - \hat{\mathbf{z}} \frac{\mathbf{u}_\perp \cdot \mathbf{e}_\perp(\mathbf{u}_\perp)}{u_z} \right] e^{ik\mathbf{u}_\perp \cdot \boldsymbol{\rho}} e^{iku_z z} d^2 u_\perp .$$

$$(2.2.15b)$$

The vector $\mathbf{e}_\perp(\mathbf{u}_\perp)$ is the component of $\mathbf{e}(\mathbf{u}_\perp)$ transverse to the $z$-direction, i.e., $\mathbf{e}_\perp(\mathbf{u}_\perp) = \mathbf{e}(\mathbf{u}_\perp) - \hat{\mathbf{z}} \cdot \mathbf{e}(\mathbf{u}_\perp)$, which is related to the tangential electric field in the plane $z = 0^+$ by the Fourier transform

$$\mathbf{e}_\perp(\mathbf{u}_\perp) = \left(\frac{k}{2\pi}\right)^2 \int \mathbf{E}_\perp(\boldsymbol{\rho}', 0^+) e^{-ik\mathbf{u}_\perp \cdot \boldsymbol{\rho}'} d^2 \rho' , \qquad (2.2.16)$$

$$\mathbf{E}_\perp = \mathbf{E} - \hat{\mathbf{z}} \cdot \mathbf{E} .$$

From the Weyl expansion (2.1.18) of a spherical wave and Eq. (2.2.16), it can readily be verified that Eqs. (2.2.15a) and (2.2.15b) are equivalent to Eqs. (2.2.11a) and (2.2.11b). It should be mentioned that the angular spectrum equivalent of Eqs. (2.2.12a) and (2.2.12b) is similar to Eqs. (2.2.15a) and (2.2.15b), except that it involves $\mathbf{h}_\perp(\mathbf{u}_\perp) = \mathbf{h}(\mathbf{u}_\perp) - \hat{\mathbf{z}} \cdot \mathbf{h}(\mathbf{u}_\perp)$ instead of $\mathbf{e}_\perp(\mathbf{u}_\perp)$.

### 2.2.3 Approximate Theories

Since the electromagnetic aperture diffraction problem is a mixed boundary value problem, it is difficult to determine either the electric field or the magnetic field, or any of their components, in the plane $z = 0^+$ for all values of $\boldsymbol{\rho}$. Hence approximate theories, similar to those found in the scalar theory (see Section 2.1.5), are often employed.

One commonly used theory is obtained by approximating the tangential electric field at $z = 0^+$ by the values

$$\hat{\mathbf{z}} \times \mathbf{E}^{(m)}(\boldsymbol{\rho}, 0^+) \;=\; 0 \qquad\qquad \text{on } \mathrm{S} \qquad (2.2.17\text{a})$$

and

$$\hat{\mathbf{z}} \times \mathbf{E}^{(m)}(\boldsymbol{\rho}, 0^+) \;=\; \hat{\mathbf{z}} \times \mathbf{E}^{(\text{inc})}(\boldsymbol{\rho}, 0) \qquad\qquad \text{in } \mathrm{A} \; . \qquad (2.2.17\text{b})$$

Equations (2.2.11a) and (2.2.12b) then yield the following formulas for the electric and magnetic fields in the half-space $z \geq 0$: [1,42,47]

$$\mathbf{E}^{(m)}(\boldsymbol{\rho}, z) \;=\; \frac{1}{2\pi} \nabla \times \int_{\mathrm{A}} \left[ \hat{\mathbf{z}} \times \mathbf{E}^{(\text{inc})}(\boldsymbol{\rho}', 0) \right] G(\boldsymbol{\rho} - \boldsymbol{\rho}', z) \, d^2\rho' \; , \qquad (2.2.18\text{a})$$

$$\mathbf{H}^{(m)}(\boldsymbol{\rho}, z) \;=\; -\frac{i}{2\pi k} \nabla \times \nabla \times \int_{\mathrm{A}} \left[ \hat{\mathbf{z}} \times \mathbf{E}^{(\text{inc})}(\boldsymbol{\rho}', 0) \right] G(\boldsymbol{\rho} - \boldsymbol{\rho}', z) \, d^2\rho' \; .$$

$$(2.2.18\text{b})$$

This theory is sometimes known as the *m*-theory because it involves an approximation to the magnetic Hertz vector. We will adopt the same terminology. There is also an *e*-theory that contains the approximate boundary values

$$\hat{\mathbf{z}} \times \mathbf{H}^{(e)}(\boldsymbol{\rho}, 0^+) \;=\; 0 \qquad\qquad \text{on } \mathrm{S} \qquad (2.2.19\text{a})$$

and

$$\hat{\mathbf{z}} \times \mathbf{H}^{(e)}(\boldsymbol{\rho}, 0^+) \;=\; \hat{\mathbf{z}} \times \mathbf{H}^{(\text{inc})}(\boldsymbol{\rho}, 0) \qquad\qquad \text{in } \mathrm{A} \; . \qquad (2.2.19\text{b})$$

With these boundary values for the tangential magnetic field in the plane $z = 0^+$, Eqs. (2.2.12a) and (2.2.12b) become[1,42,47]

$$\mathbf{H}^{(e)}(\boldsymbol{\rho}, z) \;=\; \frac{1}{2\pi} \nabla \times \int_{\mathrm{A}} \left[ \hat{\mathbf{z}} \times \mathbf{H}^{(\text{inc})}(\boldsymbol{\rho}', 0) \right] G(\boldsymbol{\rho} - \boldsymbol{\rho}', z) \, d^2\rho' \; , \qquad (2.2.20\text{a})$$

$$\mathbf{E}^{(e)}(\boldsymbol{\rho},z) \;=\; \frac{i}{2\pi k}\,\nabla\times\nabla\times\int_{A}\Big[\hat{\mathbf{z}}\times\mathbf{H}^{(\mathrm{inc})}(\boldsymbol{\rho}',0)\Big]G(\boldsymbol{\rho}-\boldsymbol{\rho}',z)\,d^{2}\rho' \;. \quad (2.2.20\mathrm{b})$$

The average of the above two theories yields yet another well-known approximate theory,[1,31,42,47]

$$\mathbf{E}^{(K)}(\boldsymbol{\rho},z) \;=\; \frac{1}{2}\Big[\mathbf{E}^{(m)}(\boldsymbol{\rho},z)\;+\;\mathbf{E}^{(e)}(\boldsymbol{\rho},z)\Big], \qquad (2.2.21\mathrm{a})$$

$$\mathbf{H}^{(K)}(\boldsymbol{\rho},z) \;=\; \frac{1}{2}\Big[\mathbf{H}^{(m)}(\boldsymbol{\rho},z)\;+\;\mathbf{H}^{(e)}(\boldsymbol{\rho},z)\Big], \qquad (2.2.21\mathrm{b})$$

which is sometimes called the Kirchhoff-Kottler theory and which is similar to the usual scalar Kirchhoff theory (2.1.37).

### 2.2.4   Two-Dimensional Electromagnetic Diffraction

If the incident field is a function of only two coordinates, say $x$ and $z$, and if the aperture is a function of only the $x$-coordinate, so it is a slit (or several parallel slits), the electromagnetic diffraction problem is two dimensional.[1,48]   Then, by decomposing the field into E and H-polarizations,

$$\mathbf{E}(x,z) \;=\; \mathbf{E}_{\mathrm{EP}}(x,z)\;+\;\mathbf{E}_{\mathrm{HP}}(x,z)\;, \qquad (2.2.22\mathrm{a})$$

$$\mathbf{H}(x,z) \;=\; \mathbf{H}_{\mathrm{EP}}(x,z)\;+\;\mathbf{H}_{\mathrm{HP}}(x,z)\;, \qquad (2.2.22\mathrm{b})$$

the problem may be treated by two independent scalar theories.

For the E-polarized contribution, the electric field is parallel to the $y$-axis and the magnetic field only has $x$ and $z$-components, i.e.,

$$\mathbf{E}_{\text{EP}}(x,z) \,=\, \hat{\mathbf{y}}\, E_y(x,z) \;, \tag{2.2.23a}$$

$$\mathbf{H}_{\text{EP}}(x,z) \,=\, \hat{\mathbf{x}}\, H_x(x,z) \;+\; \hat{\mathbf{z}}\, H_z(x,z) \;, \tag{2.2.23b}$$

$\hat{\mathbf{x}}$, $\hat{\mathbf{y}}$ and $\hat{\mathbf{z}}$ being unit vectors in the $x$, $y$, and $z$-directions. Furthermore, from Maxwell's equations it follows that $E_y(x,z)$ satisfies the two-dimensional Helmholtz equation,

$$\left(\nabla^2 + k^2\right) E_y(x,z) \,=\, 0 \;, \tag{2.2.24}$$

$\nabla^2 \equiv \partial^2/\partial x^2 + \partial^2/\partial y^2$, and that the components of the magnetic field can be computed from $E_y(x,z)$ with the use of the formulas

$$H_x(x,z) \,=\, \frac{i}{k}\, \frac{\partial E_y(x,z)}{\partial z} \;, \quad H_z(x,z) \,=\, -\frac{i}{k}\, \frac{\partial E_y(x,z)}{\partial x} \;. \tag{2.2.25}$$

For a perfectly conducting screen, the boundary conditions (2.2.4) and (2.2.7) in the plane $z = 0$ then reduce to

$$E_y(x,z) \,=\, 0 \qquad\qquad\qquad \text{on S} \;, \tag{2.2.26a}$$

$$\frac{\partial E_y(x,z)}{\partial z}\bigg|_{z=0} = \frac{\partial E_y^{(\text{inc})}(x,z)}{\partial z}\bigg|_{z=0} \qquad\qquad \text{in A} \;. \tag{2.2.26b}$$

These boundary conditions are exactly the same as those for a Dirichlet-type screen in the scalar theory [see Eqs. (2.1.6) and (2.1.8)].

For the H-polarized contribution, on the other hand, the magnetic field is parallel to the $y$-axis and the electric field only has $x$ and $z$-components:

$$\mathbf{H}_{\text{HP}}(x,z) \ = \ \hat{\mathbf{y}} \, H_y(x,z) \tag{2.2.27a}$$

$$\mathbf{E}_{\text{HP}}(x,z) \ = \ \hat{\mathbf{x}} \, E_x(x,z) \ + \ \hat{\mathbf{z}} \, E_z(x,z) \ . \tag{2.2.27b}$$

In this case, $H_y(x,z)$ obeys the Helmholtz equation,

$$\left(\nabla^2 + \, k^2\right) H_y(x,z) \ = \ 0 \tag{2.2.28}$$

and the formulas

$$E_x(x,z) \ = \ -\frac{i}{k} \frac{\partial H_y(x,z)}{\partial z} \ , \quad E_z(x,z) \ = \ \frac{i}{k} \frac{\partial H_y(x,z)}{\partial x} \tag{2.2.29}$$

can be used to determine the components of the electric field from $H_y(x,z)$. For a perfectly conducting screen, the boundary conditions (2.2.4) and (2.2.7) in the plane $z = 0$ now become identical to those for a Neumann-type screen in the scalar theory [see Eqs. (2.1.7) and (2.1.9)]:

$$\frac{\partial H_y(x,z)}{\partial z} \ \bigg|_{z=0} = \ 0 \qquad \text{on } \mathbf{S} \ , \tag{2.2.30a}$$

$$H_y(x,0) \ = \ H_y^{(\text{inc})}(x,0) \qquad \text{in } \mathbf{A} \ . \tag{2.2.30b}$$

Therefore, two-dimensional electromagnetic diffraction can be analyzed by two independent scalar theories. The scalar theory for E-polarization is based on the *y*-component of the electric field and boundary conditions appropriate to a Dirichlet-type screen, whereas the one for H-polarization is based on the *y*-component of the magnetic field and boundary conditions appropriate to a Neumann-type screen.

CHAPTER 3

## CONTRIBUTIONS OF HOMOGENEOUS AND EVANESCENT WAVES IN THE NEAR-FIELD

In order to elucidate the effects of diffraction in the near-field, we shall now examine the separate contributions of homogeneous and inhomogeneous (more precisely evanescent) waves to two-dimensional (2-D) near-field diffraction patterns of scalar fields. Some discussion of near-field diffraction based on such a decomposition can also be found in papers by Harvey[1] and by Massey.[2] The consequences of neglecting evanescent waves in certain near-field distributions have been discussed by Carter.[3,4] For the sake of mathematical simplicity and also because, in the 2-D case, the exact electromagnetic diffraction problem can be described by a scalar theory (see Section 2.2.4), we restrict our analysis to 2-D diffraction.

First, in Section 3.1 we discuss some basic relations for calculating the homogeneous and the evanescent contributions to an arbitrary field. We then introduce the concepts of total homogeneous intensity and total evanescent intensity in Section 3.2 as convenient measures of the relative importance of the two contributions. In Section 3.3 we show how to convert certain integrals used in calculating the homogeneous and the evanescent contributions into expressions involving infinite series and in Section 3.4, from some of these expressions, we obtain approximate relations that are valid in the near field. Finally, in Section 3.5 we examine the case of a plane wave diffracted by a slit in an opaque screen as an example.

### 3.1 HOMOGENEOUS AND EVANESCENT CONTRIBUTIONS

We consider a 2-D monochromatic scalar optical field $V(x,z,t) = U(x,z)e^{-i\omega t}$ that obeys the 2-D Helmholtz equation

$$\left(\nabla^2 + k^2\right)U(x,z) = 0 \ , \tag{3.1.1}$$

where $\nabla^2 \equiv \partial^2/\partial x^2 + \partial^2/\partial z^2$ and $k = \omega/c = 2\pi/\lambda$ is the free-space wave number. We are interested in free-space propagation of the field $U(x,z)$ from the plane $z = 0$ into the half-space $z > 0$. We therefore assume that all sources, scatterers, diffracting apertures, etc., are located in the half-space $z \leq 0$. Using the angular spectrum representation (see Section 2.1.3),[5-10] we can express $U(x, z \geq 0)$ as the sum of a homogeneous contribution $U_h(x,z)$ and an evanescent (inhomogeneous) one $U_i(x,z)$,

$$U(x,z) = U_h(x,z) + U_i(x,z) \ . \tag{3.1.2}$$

$U_h(x,z)$ is a superposition of homogeneous plane waves that propagate into the half-space $z > 0$,

$$U_h(x,z) = \int\limits_{|u_x| \leq 1} a(u_x) \, e^{iku_x x} \, e^{ikz\sqrt{1 - u_x^2}} \, du_x \ , \tag{3.1.3a}$$

whereas $U_i(x,z)$ is a superposition of evanescent (inhomogeneous) plane waves that decay exponentially along the positive $z$-direction,

$$U_i(x,z) \;=\; \int\limits_{|u_x| > 1} a(u_x)\, e^{iku_x x}\, e^{-kz\sqrt{u_x^2 - 1}}\, du_x \;. \qquad (3.1.3b)$$

It should be noted that the homogeneous and the evanescent contributions each separately satisfy the Helmholtz equation.

As in the 3-D case (see Section 2.2.4), the angular spectrum amplitude $a(u_x)$ is the Fourier transform of the field distribution $U(x,0)$ in the plane $z = 0$:

$$a(u_x) \;=\; \frac{k}{2\pi} \int\limits_{-\infty}^{\infty} U(x',0)\, e^{-iku_x x'}\, dx' \;. \qquad (3.1.4)$$

For the near-field geometries of interest here, we can restrict our analysis to well-behaved functions $U(x,0)$ that are of finite support, i.e., well-behaved functions that vanish outside some finite $x$-range. $U(x,0)$ is then square-integrable; furthermore, $a(u_x)$ is the boundary value on the real $u_x$-axis of an entire analytic function.[11]

With the use of relation (3.1.4), Eqs. (3.1.3a) and (3.1.3b) may be rewritten as the convolutions

$$U_h(x,z) \;=\; \int\limits_{-\infty}^{\infty} \mathsf{H}_h(x - x',z)\, U(x',0)\, dx' \qquad (3.1.5a)$$

and

$$U_i(x,z) \;=\; \int\limits_{-\infty}^{\infty} \mathsf{H}_i(x - x',z)\, U(x',0)\, dx' \;, \qquad (3.1.5b)$$

with the kernels $H_h(x,z)$ and $H_i(x,z)$ given by the formulas

$$H_h(x,z) \equiv \frac{k}{\pi} \int_0^1 \cos(ku_x x)\, e^{ikz\sqrt{1-u_x^2}}\, du_x \qquad (3.1.6a)$$

and

$$H_i(x,z) \equiv \frac{k}{\pi} \int_1^\infty \cos(ku_x x)\, e^{-kz\sqrt{u_x^2-1}}\, du_x \ . \qquad (3.1.6b)$$

It can readily be shown that

$$\begin{aligned} H(x,z) &\equiv H_h(x,z) + H_i(x,z) \\ &= \frac{ikz}{2\sqrt{x^2+z^2}}\, H_1^{(1)}\!\left(k\sqrt{x^2+z^2}\right), \end{aligned} \qquad (3.1.7)$$

where $H_1^{(1)}$ is a Hankel function of the first kind and first order. Expression (3.1.7) is the usual 2-D free-space wave propagator.[12]

If we let $z = 0$ in Eqs. (3.1.6a) and (3.1.6b), the integrations with respect to $u_x$ can be performed at once and yield

$$H_h(x,0) = \frac{1}{\pi} \frac{\sin(kx)}{x} \qquad (3.1.8a)$$

and

$$H_i(x,0) = \delta(x) - \frac{1}{\pi} \frac{\sin(kx)}{x} \ , \qquad (3.1.8b)$$

where $\delta(x)$ is the Dirac delta function. After substituting these relations into Eqs. (3.1.5a) and (3.1.5b), we find that

$$U_h(x,0) \;=\; \frac{1}{\pi} \int\limits_{-\infty}^{\infty} \frac{\sin\!\big[k(x-x')\big]}{(x-x')} \; U(x',0) \, dx' \qquad (3.1.9a)$$

and

$$U_i(x,0) \;=\; U(x,0) \;-\; \frac{1}{\pi} \int\limits_{-\infty}^{\infty} \frac{\sin\!\big[k(x-x')\big]}{(x-x')} \; U(x',0) \, dx' \; . \qquad (3.1.9b)$$

Because the location of the plane $z = 0$ is essentially arbitrary, analogous expressions must also apply to any plane $z = \text{constant} \ge 0$:

$$U_h(x,z) \;=\; \frac{1}{\pi} \int\limits_{-\infty}^{\infty} \frac{\sin\!\big[k(x-x')\big]}{(x-x')} \; U(x',z) \, dx' \qquad (3.1.10a)$$

and

$$U_i(x,z) \;=\; U(x,z) \;-\; \frac{1}{\pi} \int\limits_{-\infty}^{\infty} \frac{\sin\!\big[k(x-x')\big]}{(x-x')} \; U(x',z) \, dx' \; . \qquad (3.1.10b)$$

We see from the preceding discussion that there are three alternative pairs of expressions that can be used to determine the homogeneous and the evanescent contributions to the field in any plane $z = \text{constant} \ge 0$:  Eqs. (3.1.3a) and (3.1.3b); Eqs. (3.1.5a) and (3.1.5b); Eqs. (3.1.10a) and (3.1.10b).   The first pair, Eqs. (3.1.3a) and (3.1.3b), expresses $U_h(x,z)$ and $U_i(x,z)$ in terms of the angular spectrum amplitude $a(u_x)$, which itself can be computed by taking a Fourier transform of the field distribution $U(x,0)$ in the plane $z = 0$ [see Eq. (3.1.4)].  In the near field, these equations are well suited for numerical implementation because all the integrations can be performed with fast Fourier transforms.

The second pair of equations, Eqs. (3.1.5a) and (3.1.5b), determines $U_h(x,z)$ and $U_i(x,z)$ directly from the field distribution $U(x,0)$ in the plane $z = 0$.  However for arbitrary $z$, the kernels $H_h(x,z)$ and $H_i(x,z)$, given by Eqs. (3.1.6a) and (3.1.6b), cannot readily be expressed in closed form. Nevertheless, as shown in Section 3.3, the integral expressions (3.1.6a) and (3.1.6b) can be converted into expressions that contain infinite series. Furthermore, as shown in Section 3.4, using these expressions one can obtain approximate closed-form relations for both $H_h(x,z)$ and $H_i(x,z)$ that are valid in the near field where $kz \ll 1$.

Lastly, Eqs. (3.1.10a) and (3.1.10b) can be used to determine $U_h(x,z)$ and $U_i(x,z)$ in any plane $z = \text{constant} \geq 0$ from knowledge of the field distribution $U(x,z)$ in that plane. If $U(x,z)$ is not known, but either $U_h(x,z)$ or $U_i(x,z)$ is known, Eqs. (3.1.10a) and (3.1.10b) are then Fredholm integral equations of the first and second kind, respectively, for the unknown $U(x,z)$. The integral operator that appears in Eq. (3.1.10a) has been studied extensively and occurs in a variety of contexts.[13-16]

## 3.2  TOTAL INTENSITIES

We shall now introduce the concepts of total (plane-integrated) homogeneous intensity and total (plane-integrated) evanescent intensity as rough convenient measures of the relative importance of the contributions $U_h(x,z)$ and $U_i(x,z)$ to the field distribution $U(x,z)$.

Using Eq. (3.1.2), we can write the intensity of the field $I(x,z) \equiv |U(x,z)|^2$ in the form

$$I(x,z) = I_h(x,z) + I_i(x,z) + I_{hi}(x,z) , \qquad (3.2.1)$$

where

$$I_h(x,z) \equiv |U_h(x,z)|^2 \qquad (3.2.2a)$$

and

$$I_i(x,z) \equiv |U_i(x,z)|^2 \qquad (3.2.2b)$$

are the intensities of the homogeneous and of the evanescent contributions, respectively, and the term

$$I_{hi}(x,z) \equiv U_h^*(x,z)U_i(x,z) + U_h(x,z)U_i^*(x,z) \qquad (3.2.2c)$$

arises from interference between the two contributions. We now define the total intensity $I_{\text{tot}}(z)$, the total homogeneous intensity $I_{\text{tot}}^{(h)}(z)$ and the total evanescent (inhomogeneous) intensity $I_{\text{tot}}^{(i)}(z)$ by the expressions

$$I_{\text{tot}}(z) \equiv \int_{-\infty}^{\infty} I(x,z)\,dx \ , \qquad (3.2.3)$$

$$I_{\text{tot}}^{(h)}(z) \equiv \int_{-\infty}^{\infty} I_h(x,z)\,dx \qquad (3.2.4a)$$

and

$$I_{\text{tot}}^{(i)}(z) \equiv \int_{-\infty}^{\infty} I_i(x,z)\,dx \ , \qquad (3.2.4b)$$

respectively.

From Eqs. (3.1.3a), (3.1.3b) and (3.2.1)-(3.2.4b), it follows that the total intensity is just the sum

$$I_{\text{tot}}(z) = I_{\text{tot}}^{(h)}(z) + I_{\text{tot}}^{(i)}(z) \ , \qquad (3.2.5)$$

where

$$I_{\text{tot}}^{(h)}(z) = \frac{2\pi}{k} \int_{|u_x| \leq 1} |a(u_x)|^2 \, du_x \qquad (3.2.6a)$$

and

$$I_{\text{tot}}^{(i)}(z) = \frac{2\pi}{k} \int\limits_{|u_x| > 1} |a(u_x)|^2 \, e^{-2kz\sqrt{u_x^2 - 1}} \, du_x \ . \tag{3.2.6b}$$

There are only two terms in Eq. (3.2.5) because the integrated interference term can be shown to be identically zero. Furthermore, as is evident from Eq. (3.2.6a), the total homogeneous intensity is conserved on propagation in the sense that $I_{\text{tot}}^{(h)}(z)$ is independent of $z$. Hence we may drop the z-argument in $I_{\text{tot}}^{(h)}(z)$ and we will do so from now on. The conservation of total homogeneous intensity on propagation and related conservation laws are discussed in Refs. 17-20.

By substituting the angular spectrum amplitude from Eq. (3.1.4) into Eqs. (3.2.6a) and (3.2.6b), we can rewrite the total intensities $I_{\text{tot}}^{(h)}$ and $I_{\text{tot}}^{(i)}(z)$ in terms of the field distribution $U(x,0)$ in the plane $z = 0$ as

$$I_{\text{tot}}^{(h)} = \frac{1}{\pi} \int\limits_{-\infty}^{\infty} dx' \int\limits_{-\infty}^{\infty} dx'' \, \frac{\sin[k(x'-x'')]}{(x'-x'')} \, U^*(x',0) \, U(x'',0) \tag{3.2.7a}$$

and

$$I_{\text{tot}}^{(i)}(z) = \int\limits_{-\infty}^{\infty} dx' \int\limits_{-\infty}^{\infty} dx'' \, \mathrm{H}_i(x'-x'',2z) \, U^*(x',0) \, U(x'',0) \ , \tag{3.2.7b}$$

where the kernel $\mathrm{H}_i(x,z)$ is given by Eq. (3.1.6b). Alternatively, using the fact that $I_{\text{tot}}^{(h)}$ is independent of z, we can also express $I_{\text{tot}}^{(h)}$ and $I_{\text{tot}}^{(i)}(z)$ in terms of the field distribution $U(x,z)$ in an arbitrary plane $z = \text{constant} \geq 0$ as follows:

$$I_{\text{tot}}^{(h)} = \frac{1}{\pi} \int\limits_{-\infty}^{\infty} dx' \int\limits_{-\infty}^{\infty} dx'' \, \frac{\sin[k(x'-x'')]}{(x'-x'')} \, U^*(x',z) \, U(x'',z) \tag{3.2.8a}$$

and

$$I_{\text{tot}}^{(i)}(z) = \int\limits_{-\infty}^{\infty} \left|U(x,z)\right|^2 \, dx \; - \; \frac{1}{\pi} \int\limits_{-\infty}^{\infty} dx' \int\limits_{-\infty}^{\infty} dx'' \; \frac{\sin\left[k(x'-x'')\right]}{(x'-x'')} \; U^*(x',z) \, U(x'',z) \qquad .$$

<div align="right">(3.2.8b)</div>

Evidently the total intensities $I_{\text{tot}}(z)$, $I_{\text{tot}}^{(h)}$ and $I_{\text{tot}}^{(i)}(z)$ are useful only if they are finite quantities. In this connection, it should be pointed out that there are physically realistic field distributions that have infinite total intensity but finite total energy flux.[21] Since the field distribution $U(x,0)$ that we are considering is square-integrable [see remarks below Eq. (3.1.4)], $I_{\text{tot}}(0)$ is finite and, consequently, so are $I_{\text{tot}}(z)$, $I_{\text{tot}}^{(h)}$ and $I_{\text{tot}}^{(i)}(z)$ because, as can easily be shown, $I_{\text{tot}}(z) \leq I_{\text{tot}}(0)$, $I_{\text{tot}}^{(h)} < I_{\text{tot}}(0)$ and $I_{\text{tot}}^{(i)}(z) < I_{\text{tot}}(0)$.

Instead of using the total intensities $I_{\text{tot}}^{(h)}$ and $I_{\text{tot}}^{(i)}(z)$ as measures of the relative importance of $U_h(x,z)$ and $U_i(x,z)$, one might consider using the total energy flux $F_{\text{tot}}$ and the total reactive energy $W_{\text{tot}}(z)$ for this purpose. The total energy flux was already discussed in Section 2.1.6 for the 3-D case. The total reactive energy can be introduced in a similar manner from the complex time-averaged energy flux vector $\tilde{\mathbf{F}}$.

For a 2-D scalar field, $\tilde{\mathbf{F}}(x,z)$ can be defined by the expression

$$\tilde{\mathbf{F}}(x,z) \equiv 2i\omega\alpha \, U(x,z)\nabla U^*(x,z) \; ,$$

<div align="right">(3.2.9)</div>

where $\alpha$ is a real, positive constant and $\nabla = \hat{\mathbf{x}}\partial/\partial x + \hat{\mathbf{z}}\partial/\partial z$. Because the real part of $\tilde{\mathbf{F}}$ is just the real energy flux vector $\mathbf{F}$ which we encountered in Sections 2.1.6 and 2.1.7, the total energy flux $F_{\text{tot}}$ across any plane $z = \text{constant} \geq 0$ may evidently be expressed as

$$F_{\text{tot}} = \int\limits_{-\infty}^{\infty} \text{Re}\{\tilde{\mathbf{F}}(x,z) \bullet \hat{\mathbf{z}}\}\, dx \ . \qquad (3.2.10a)$$

Here, since $F_{\text{tot}}$ is independent of the propagation distance $z$, we have omitted its $z$-argument. The total reactive energy $W_{\text{tot}}(z)$ across any plane $z = \text{constant} \geq 0$ may be introduced by the analogous expression[*]

$$W_{\text{tot}}(z) = \int\limits_{-\infty}^{\infty} \text{Im}\{\tilde{\mathbf{F}}(x,z) \bullet \hat{\mathbf{z}}\}\, dx \ . \qquad (3.2.10b)$$

If we use Eqs. (3.1.2)-(3.1.3b) and (3.2.9)-(3.2.10b), we readily find that the total energy flux and the total reactive energy may be written in terms of the angular spectrum amplitude $a(u_x)$ as

$$F_{\text{tot}} = 4\pi\omega\alpha \int\limits_{|u_x| \leq 1} \sqrt{1 - u_x^2}\, \left|a(u_x)\right|^2\, du_x \qquad (3.2.11a)$$

and

$$W_{\text{tot}}(z) = 4\pi\omega\alpha \int\limits_{|u_x| > 1} \sqrt{u_x^2 - 1}\, \left|a(u_x)\right|^2\, e^{-2kz\sqrt{u_x^2 - 1}}\, du_x \ . \qquad (3.2.11b)$$

These equations should be compared with their counterparts for the total intensities, Eqs. (3.2.6a) and (3.2.6b). It is evident that the total energy flux $F_{\text{tot}}$ depends only on the homogeneous contribution, whereas the total reactive energy $W_{\text{tot}}(z)$ depends only on the evanescent one. Because of the mutiplicative factor $\sqrt{u_x^2 - 1}$ in Eq. (3.2.11b), the total reactive energy can diverge in cases when the total evanescent intensity $I_{tot}^{(i)}(z)$ is finite. In fact, for the example considered in Section 3.5, one can show that the total reactive energy does indeed diverge in the plane $z = 0$. For this reason, we chose to use the total intensities $I_{tot}^{(h)}$ and $I_{tot}^{(i)}(z)$ rather than $F_{tot}$ and $W_{tot}(z)$ in our near-field analysis.

---

[*] The electromagnetic analog of this quantity is discussed in Ref. 22.

## 3.3  THE KERNELS $\mathrm{H}_h(x,z)$ AND $\mathrm{H}_i(x,z)$

As was already mentioned in Section 3.1, the kernels $\mathrm{H}_h(x,z)$ and $\mathrm{H}_i(x,z)$ that appear in Eqs. (3.1.5a) and (3.1.5b), and in Eq. (3.2.7b), and that are defined by Eqs. (3.1.6a) and (3.1.6b) cannot be expressed in a closed form. For analytical and numerical work, it is desirable to obtain alternate forms for $\mathrm{H}_h(x,z)$ and $\mathrm{H}_i(x,z)$ that do not contain integrals. In Sections 3.3.1 and 3.3.2, we show how to convert Eqs. (3.1.6a) and (3.1.6b) into two different kinds of expressions that contain infinite series of Bessel functions rather than integrals. It should be pointed out that the approach used in Section 3.3.2 is somewhat similar to that used in Ref. 7 for the 3-D case, but the final results are considerably different.

### 3.3.1  Expressions for $\mathrm{H}_h(x,z)$ and $\mathrm{H}_i(x,z)$ involving Series of Bessel Functions of Integer Order

Let us make the change of variables $u_x = \cos\psi$ in Eq. (3.1.6a). The kernel $\mathrm{H}_h(x,z)$ may then be written in the form

$$\mathrm{H}_h(x,z) \;=\; \frac{1}{2\pi i}\,\frac{\partial T(x,z)}{\partial z}\;,\tag{3.3.1}$$

where the function $T(x,z)$ is given by the expression

$$T(x, z) \;=\; \int\limits_{0}^{\pi} e^{ikx \cos \psi} \, e^{ikz \sin \psi} \; d\psi \; . \qquad (3.3.2)$$

As shown in Appendix B, by expanding the two exponentials in the integrand in terms of Bessel functions, one can express $T(x, z)$ in the form

$$T(x, z) \;=\; \pi J_0\!\left( k \sqrt{x^2 + z^2} \right) \;+\; \pi i \, J_0(kx) \, \mathbf{H}_0(kz)$$

$$+ \; 8i \sum_{m=0}^{\infty} \sum_{n=1}^{\infty} (-1)^n \frac{2m+1}{(2m+1)^2 - 4n^2} \; J_{2n}(kx) \, J_{2m+1}(kz) \, ,$$

$$(3.3.3)$$

where $J_m$ is a Bessel function of the first kind and $m$th order and $\mathbf{H}_m$ is a Struve function of $m$th order.

If we now substitute from Eq. (3.3.3) into Eq. (3.3.1) and make use of the relations[23]

$$\frac{\partial J_0(\beta)}{\partial \beta} \;=\; - \, J_1(\beta) \, , \qquad (3.3.4)$$

$$\frac{\partial J_\nu(\beta)}{\partial \beta} \;=\; J_{\nu-1}(\beta) \;-\; \frac{\nu}{z} J_\nu(\beta) \, , \qquad (3.3.5)$$

and

$$\frac{\partial \, \mathbf{H}_0(\beta)}{\partial \beta} \;=\; \mathbf{H}_{-1}(\beta) \, , \qquad (3.3.6)$$

we obtain the desired formula for the kernel $\mathsf{H}_h(x, z)$:

$$\mathbf{H}_h(x,z) = \frac{ikz}{2\sqrt{x^2+z^2}}\, J_1\left(k\sqrt{x^2+z^2}\right) \;+\; \frac{k}{2}\, J_0(kx)\,\mathbf{H}_{-1}(kz)$$

$$+\; \frac{4k}{\pi} \sum_{m=0}^{\infty} \sum_{n=1}^{\infty} (-1)^n \frac{2m+1}{(2m+1)^2 - 4n^2}\, J_{2n}(kx)$$

$$\times \left[ J_{2m}(kz) \;-\; \frac{2m+1}{kz}\, J_{2m+1}(kz) \right].$$

$$(3.3.7a)$$

In addition, since the kernel $\mathbf{H}_i(x,z)$ is related to $\mathbf{H}_h(x,z)$ by Eq. (3.1.7), we have

$$\mathbf{H}_i(x,z) = -\frac{kz}{2\sqrt{x^2+z^2}}\, Y_1\left(k\sqrt{x^2+z^2}\right) \;-\; \frac{k}{2}\, J_0(kx)\,\mathbf{H}_{-1}(kz)$$

$$-\; \frac{4k}{\pi} \sum_{m=0}^{\infty} \sum_{n=1}^{\infty} (-1)^n \frac{2m+1}{(2m+1)^2 - 4n^2}\, J_{2n}(kx)$$

$$\times \left[ J_{2m}(kz) \;-\; \frac{2m+1}{kz}\, J_{2m+1}(kz) \right],$$

$$(3.3.7b)$$

where $Y_1$ is a Bessel Function of the second kind and first order.

In order to verify that these expressions for $\mathbf{H}_h(x,z)$ and $\mathbf{H}_i(x,z)$ do reduce to Eqs. (3.1.8a) and (3.1.8b), we set $z = 0$ in Eq. (3.3.7a) and make use of the limiting forms [see Ref. 23, pg. 360, Eq. (9.1.7) and pg. 496, Fig. 12.2]

$$J_\nu(\beta) \;\approx\; \frac{(\beta/2)^\nu}{\Gamma(\nu+1)}\,, \qquad\qquad \text{as } \beta \to 0\,, \qquad (3.3.8)$$

$$\mathbf{H}_{-1}(\beta) \;\approx\; \frac{2}{\pi}\,, \qquad\qquad \text{as } \beta \to 0\,. \qquad (3.3.9)$$

Equation (3.3.7a) then becomes

$$\mathsf{H}_h(x,0) \;=\; \frac{k}{\pi} J_0(kx) \;-\; \frac{2k}{\pi} \sum_{n=1}^{\infty} \frac{(-1)^n}{4n^2-1} J_{2n}(kx) \;. \qquad (3.3.10)$$

If we compare this equation with Eq. (3.1.8a), we see that the following identity must hold for the two to be equivalent:

$$\frac{\sin \beta}{\beta} \;=\; J_0(\beta) \;-\; 2 \sum_{n=1}^{\infty} \frac{(-1)^n}{4n^2-1} J_{2n}(\beta) \;. \qquad (3.3.11)$$

This identity does not appear in any of the standard references on Bessel functions. However, we were able to check its validity numerically; thus confirming that Eq. (3.3.7a) for the kernel $\mathsf{H}_h(x,z)$ does yield the correct limiting form, Eq. (3.1.8a), in the plane $z = 0$. Since the kernels $\mathsf{H}_h(x,0)$ and $\mathsf{H}_i(x,0)$ are related by $\mathsf{H}_h(x,0) + \mathsf{H}_i(x,0) = \delta(x)$, Eq. (3.3.7b) evidently also reduces to the correct form, Eq. (3.1.8b).

### 3.3.2 Expressions for $\mathsf{H}_h(x,z)$ and $\mathsf{H}_i(x,z)$ involving Series of Spherical Bessel Functions

We now write the kernel $\mathsf{H}_h(x,z)$ in the form [see Eq. (3.1.6a)]

$$\mathsf{H}_h(x,z) \;=\; \frac{k}{2\pi} \int_{-1}^{1} e^{iku_x x} \, e^{ikz\sqrt{1-u_x^2}} \, du_x \qquad (3.3.12)$$

and use the series expansion

$$e^{ikz\sqrt{1-u_x^2}} \;=\; \sum_{n=0}^{\infty} \frac{i^n}{n!} (kz)^n (1-u_x^2)^{n/2} \;. \qquad (3.3.13)$$

As shown in Appendix C, after substituting from Eq. (3.3.13) into Eq. (3.3.12) and performing the integration with respect to $u_x$, one may express $\mathrm{H}_h(x,z)$ as

$$\mathrm{H}_h(x,z) = \frac{ikz}{2\sqrt{x^2+z^2}}\, J_1\!\left(k\sqrt{x^2+z^2}\right)\; +\; \frac{k}{\pi}\sum_{m=0}^{\infty}(-1)^m\,\frac{2^m\,m!}{(2m)!}\,(kz)^m\,\frac{j_m(kx)}{(kx)^m}\;.$$

$$(3.3.14a)$$

Furthermore, according to Eq. (3.1.7), the similar expression for the kernel $\mathrm{H}_i(x,z)$ is

$$\mathrm{H}_i(x,z) = -\frac{kz}{2\sqrt{x^2+z^2}}\, Y_1\!\left(k\sqrt{x^2+z^2}\right)\; -\; \frac{k}{\pi}\sum_{m=0}^{\infty}(-1)^m\,\frac{2^m\,m!}{(2m)!}\,(kz)^m\,\frac{j_m(kx)}{(kx)^m}\;.$$

$$(3.3.14b)$$

It is apparent that in the plane $z=0$ these equations for $\mathrm{H}_h(x,z)$ and $\mathrm{H}_i(x,z)$ yield the correct limiting forms, Eqs. (3.1.8a) and (3.1.8b), because for $z=0$ the only non-zero term in the summations is the $m=0$ term.

## 3.4  APPROXIMATE RELATIONS FOR THE NEAR FIELD

So far our discussion has been general in the sense that our formulas apply to any propagation distance $z$.  Using the expansions developed in Section 3.3.2 for the kernels $\mathrm{H}_h(x,z)$ and $\mathrm{H}_i(x,z)$, Eqs. (3.3.14a) and (3.3.14b), we will now derive approximate relations for the homogeneous and the evanescent contributions that are valid for propagation distances much smaller than the wavelength, $0 \le kz \ll 1$.  For such distances, we would expect changes in the field $U(x,z)$ from its initial value $U(x,0)$ to be dominated by the decay of the evanescent contribution $U_i(x,z)$, with only slight modifications of the homogeneous contribution $U_h(x,z)$.

When $0 \leq kz << 1$, we can approximate Eq. (3.3.14a) for the kernel $\mathrm{H}_h(x,z)$ by setting $\sqrt{x^2 + z^2} \approx |x|$ and retaining only the first two terms in the sum over $m$:

$$\mathrm{H}_h(x,z) \approx \frac{k}{\pi} \frac{\sin(kx)}{kx} + kz \left[ \frac{ik}{2} \frac{J_1(kx)}{kx} - \frac{k}{\pi} \frac{j_1(kx)}{kx} \right] . \qquad (3.4.1a)$$

For the kernel $\mathrm{H}_i(x,z)$, because there is a singularity at $x = z = 0$ contained in the $Y_1$ term of Eq. (3.3.14b), we do not make any approximations to that term. However, because the $Y_1$ term also dominates the $z$-behavior of $\mathrm{H}_i(x,z)$ in the range $0 \leq kz << 1$, we only need to keep the first term in the sum over $m$. We then obtain the approximate formula

$$\mathrm{H}_i(x,z) \approx -\frac{kz}{2\sqrt{x^2 + z^2}} Y_1\left( k\sqrt{x^2 + z^2} \right) - \frac{k}{\pi} \frac{\sin(kx)}{kx} . \qquad (3.4.1b)$$

It can readily be shown that Eqs. (3.4.1a) and (3.4.1b) reduce to the exact expressions (3.1.8a) and (3.1.8b) in the plane $z = 0$.

By substituting from Eqs. (3.4.1a) and (3.4.1b) into Eqs. (3.1.5a) and (3.1.5b), respectively, and making use of Eq. (3.1.9a), we find that, for $0 \leq kz << 1$, the homogeneous and the evanescent contributions can be approximated by the formulas

$$U_h(x,z) \approx U_h(x,0) + kz \int_{-\infty}^{\infty} \left\{ \frac{ik}{2} \frac{J_1[k(x-x')]}{k(x-x')} - \frac{k}{\pi} \frac{j_1[k(x-x')]}{k(x-x')} \right\} U(x',0) \, dx' ,$$

$$(3.4.2a)$$

$$U_i(x,z) \approx -\frac{kz}{2} \int\limits_{-\infty}^{\infty} \frac{1}{\sqrt{(x-x')^2 + z^2}} \, Y_1\left( k\sqrt{(x-x')^2 + z^2} \right) U(x',0) \, dx' \, - \, U_h(x,0) \, .$$

$$(3.4.2b)$$

From the form of these two equations, we see that for propagation distances much smaller than a wavelength the homogeneous contribution is modified from its value in the plane $z = 0$ by the addition of a small term that is linear in $kz$, whereas the evanescent contribution changes rapidly from its $z = 0$ value. These effects are evident in the example considered in the next section.

## 3.5  DIFFRACTION BY A SLIT

We now consider, as an example, the near-field diffraction of a plane wave incident normally on a slit of width $d$ in an planar opaque screen (see Fig. 3-1), using approximate boundary conditions. We assume that directly behind the slit, in the plane $z = 0^+$, the field may be approximated by the boundary conditions of the Rayleigh-Sommerfeld theory of the first kind, which are given by Eqs. (2.1.31a) and (2.1.31b):

$$U(x,0^+) \; = \; \begin{cases} U^{(\text{inc})}(x,0) \, , & \text{for} \;\; -\dfrac{d}{2} \, \leq \, x \, \leq \, \dfrac{d}{2} \, , \\[2mm] 0 \, , & \text{otherwise} \, . \end{cases} \qquad (3.5.1)$$

For the case of a normally incident plane wave, $U^{(\text{inc})}(x,0) = K$, where $K$ is a constant. It should be pointed out that, for the small slit widths of interest here, one would expect the actual field distribution in the plane $z = 0^+$, i.e., the solution to the rigorous boundary value problem, to be rather different from the field distribution given by the above boundary conditions. (A comparison of exact and approximate boundary conditions will be given in Chapter 4.) Nevertheless, some understanding of propagation in the near field can still be gained by employing these approximate boundary conditions.

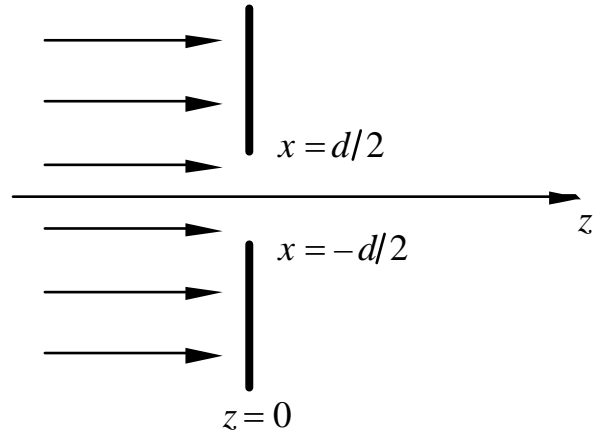

**Figure 3-1** A slit of width d in an opaque screen, illuminated by a normally incident plane wave.

Using Eq. (3.1.4), we find that the angular spectrum amplitude associated with boundary conditions (3.5.1) is

$$a(u_x) = \frac{K}{\pi} \frac{\sin(u_x k d / 2)}{u_x} .$$ (3.5.2)

Consequently, according to Eqs. (3.1.3a) and (3.1.3b), the homogeneous and the evanescent contributions behind the screen ($z \geq 0$) are

$$U_h(x,z) = \frac{K}{\pi} \int\limits_{|u_x| \leq 1} \frac{\sin(u_x kd/2)}{u_x} e^{iku_x x} e^{ikz\sqrt{1-u_x^2}} \, du_x \qquad (3.5.3a)$$

and

$$U_i(x,z) = \frac{K}{\pi} \int\limits_{|u_x| > 1} \frac{\sin(u_x kd/2)}{u_x} e^{iku_x x} e^{-kz\sqrt{u_x^2-1}} \, du_x \; . \qquad (3.5.3b)$$

The corresponding total intensities are [see Eqs. (3.2.6a) and (3.2.6b)]

$$I_{tot}^{(h)} = \frac{4I_{tot}(0)}{\pi kd} \int\limits_0^1 \frac{\sin^2(u_x kd/2)}{u_x^2} \, du_x \qquad (3.5.4a)$$

and

$$I_{tot}^{(i)}(z) = \frac{4I_{tot}(0)}{\pi kd} \int\limits_1^\infty \frac{\sin^2(u_x kd/2)}{u_x^2} e^{-2kz\sqrt{u_x^2-1}} \, du_x \; , \qquad (3.5.4b)$$

where $I_{tot}(0) = |K|^2 d$ is the total intensity in the plane $z = 0$. Equation (3.5.4a) for the total homogeneous intensity can be rewritten in a more compact form as

$$I_{tot}^{(h)} = I_{tot}(0) \left[ \frac{2}{\pi} \, \text{Si}(kd) - \frac{4}{\pi kd} \, \sin^2(kd/2) \right], \qquad (3.5.5)$$

where $\text{Si}(\beta)$ is the sine integral

$$\text{Si}(\beta) \equiv \int\limits_0^\beta \frac{\sin t}{t} \, dt \; . \qquad (3.5.6)$$

Furthermore, in the plane $z = 0$, expressions (3.5.3a) and (3.5.3b) for the homogeneous and the evanescent contributions reduce to

$$U_h(x,0) = \frac{K}{\pi} \left\{ \text{Si}\big[k(x+d/2)\big] - \text{Si}\big[k(x-d/2)\big] \right\}, \qquad (3.5.7a)$$

$$U_i(x,0) \; = \; \begin{cases} \dfrac{K}{\pi} \; \Big\{ \pi \; - \; Si\big[k(x+d/2)\big] \; + \; Si\big[k(x-d/2)\big] \Big\}, & \text{for} \quad -\dfrac{d}{2} \; \leq \; x \; \leq \; \dfrac{d}{2} \,, \\[4mm] -\dfrac{K}{\pi} \; \Big\{ Si\big[k(x+d/2)\big] \; - \; Si\big[k(x-d/2)\big] \Big\}, & \text{otherwise} \,, \end{cases}$$

$$(3.5.7b)$$

and expression (4.4b) for the total evanescent intensity reduces to

$$I_{\text{tot}}^{(i)}(0) \; = \; I_{\text{tot}}(0) \Big[ 1 \; - \; \frac{2}{\pi} \; \text{Si}(kd) \; + \; \frac{4}{\pi kd} \; \sin^2\big(kd/2\big) \Big]. \qquad (3.5.8)$$

We can now examine the field $U(x,z)$, the homogeneous contribution $U_h(x,z)$, the evanescent contribution $U_i(x,z)$, the total intensity $I_{\text{tot}}(z)$, the total homogenous intensity $I_{\text{tot}}^{(h)}$ and the total evanescent intensity $I_{\text{tot}}^{(i)}(z)$ in the near field for specific values of the slit width $d$ and of the propagation distance $z$.

Figures 3-2 through 3-4 depict $|U(x,z)|$, $|U_h(x,z)|$ and $U_i(x,z)$ for $z = 0$, $z = 0.02\lambda$, $z = 0.1\lambda$ and $z = 0.5\lambda$. These figures were computed from Eqs. (3.5.3a) and (3.5.3b) with the use of Fast Fourier Transforms. Figure 3-2, 3-3 and 3-4 pertain to slits of width $d = 0.2\lambda$, $d = 1\lambda$ and $d = 5\lambda$, respectively. As expected, we see that for $z << \lambda$ changes in the field $U(x,z)$ from its initial value $U(x,0)$ are due mostly the decay of the evanescent contribution $U_i(x,z)$. This decay of $U_i(x,z)$ is obviously most important for the case $d = 0.2\lambda$ and, consequently, for this slit width there is a substantial broadening and decrease in the amplitude of the field distribution $U(x,z)$ on propagation from the plane $z = 0$ to $z = 0.5\lambda$.

Figures 3-5 through 3-7 show $I_{\text{tot}}^{(h)}$ and $I_{\text{tot}}^{(i)}(0)$ as functions of the slit width $d$ and $I_{\text{tot}}(z)$ and $I_{\text{tot}}^{(i)}(z)$ as functions of the propagation distance $z$. From Fig. 3-5, we see that for slit widths smaller than about half a wavelength the total evanescent intensity $I_{\text{tot}}^{(i)}(0)$ in the plane $z = 0$ becomes quite appreciable compared with the total

homogeneous intensity $I_{\text{tot}}^{(h)}$. However, as is evident from Fig. 3-7, $I_{\text{tot}}^{(i)}(z)$ decreases very rapidly with $z$ in all cases, although this decay is much more rapid for the cases $d = 1\lambda$ and $d = 5\lambda$ than for $d = 0.2\lambda$, $d = 0.5\lambda$ and $d = 2.5\lambda$.

In order to understand the decay of the total evanescent intensity $I_{\text{tot}}^{(i)}(z)$ with $z$, we can perform an asymptotic expansion of Eq. (3.5.4b) for large $kz$. The first three terms in the asymptotic series for $I_{\text{tot}}^{(i)}(z)$ are then

$$I_{\text{tot}}^{(i)}(z) \sim \frac{4I_{\text{tot}}(0)}{\pi kd} \left\{ \frac{1!}{(2kz)^2} \sin^2(kd/2) + \frac{3!}{(2kz)^4} \left[ \frac{kd}{4} \sin(kd) - \frac{3}{2} \sin^2(kd/2) \right] \right.$$
$$\left. + \frac{5!}{(2kz)^6} \left[ \left( \frac{kd}{4} \right)^2 \cos(kd) - \frac{3kd}{8} \sin(kd) + \frac{15}{8} \sin^2(kd/2) \right] \right\},$$

$$(3.5.9)$$

as we demonstrate in Appendix D. Thus when the slit width $d$ is an integer number of wavelengths, i.e., when $kd = 2m\pi$ (m being an integer), the first two terms in the asymptotic series, which decay as $(kz)^{-2}$ and $(kz)^{-4}$, vanish and the term

$$\frac{4I_{\text{tot}}(0)}{\pi kd} \frac{5!}{(2kz)^6} \left( \frac{kd}{4} \right)^2,$$

which decays as $(kz)^{-6}$, dominates the asymptotic character of the total evanescent intensity. This type of behavior is clearly seen in Fig. 3-7.

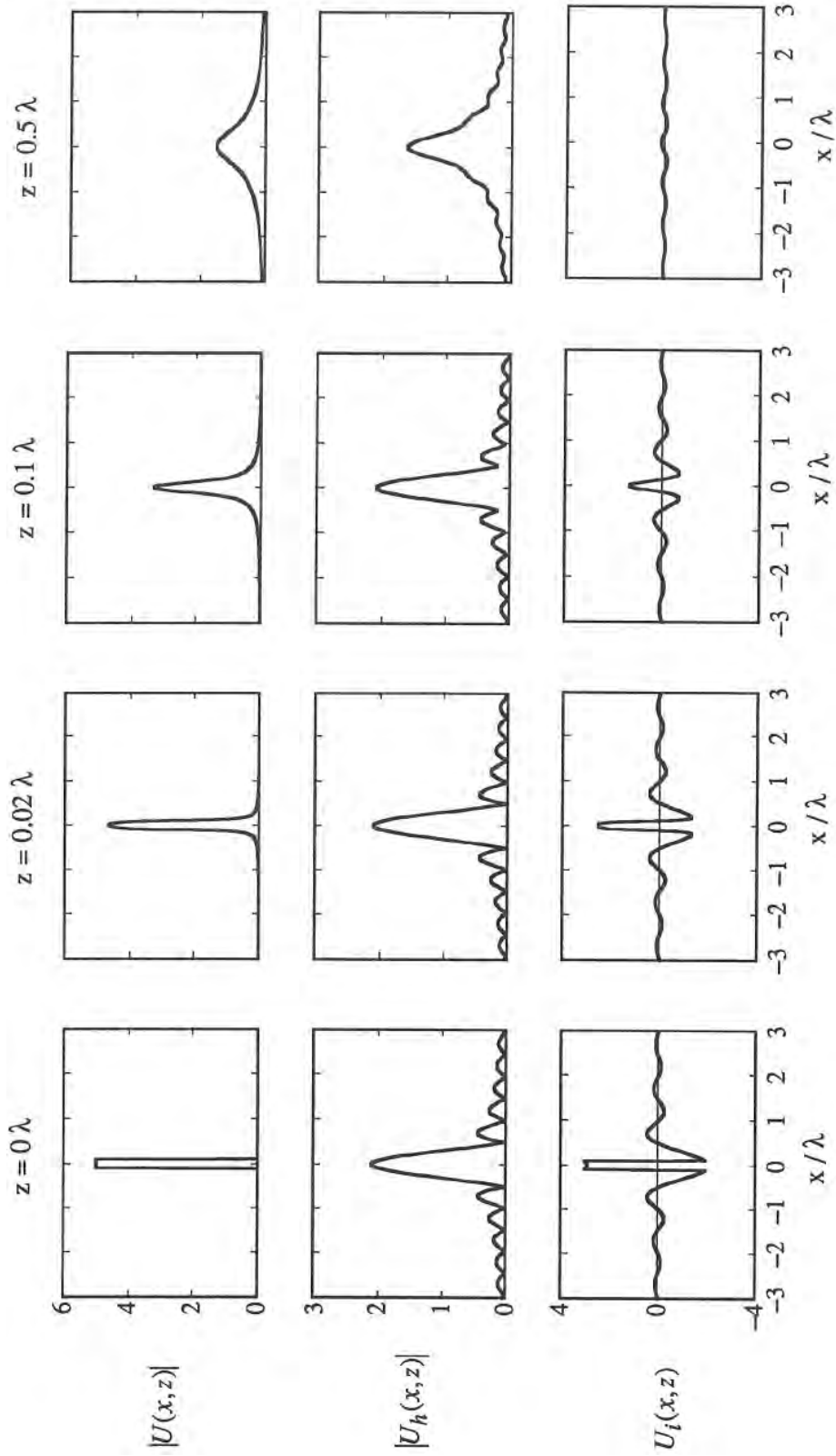

**Figure 3-2** Near-field diffraction patterns showing the field $|U(x,z)|$, the homogeneous contribution $|U_h(x,z)|$ and the evanescent contribution $U_i(x,z)$ for a slit of width $d = 0.2\lambda$, with the choice $K = \lambda/d = 5$.

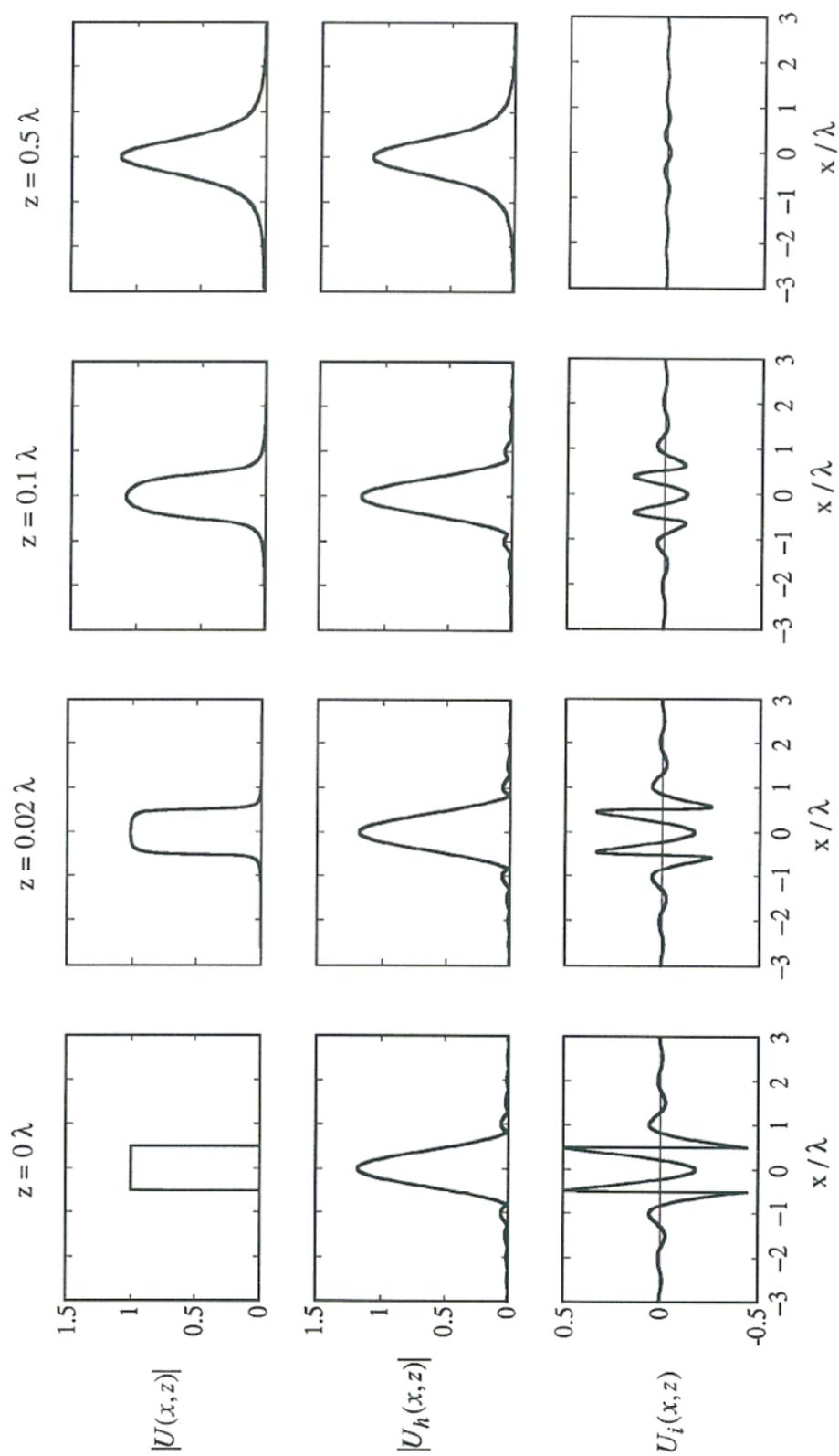

**Figure 3-3** Near-field diffraction patterns showing the field $|U(x,z)|$, the homogeneous contribution $|U_h(x,z)|$ and the evanescent contribution $U_i(x,z)$ for a slit of width $d = 1\lambda$, with the choice $K = \lambda/d = 1$.

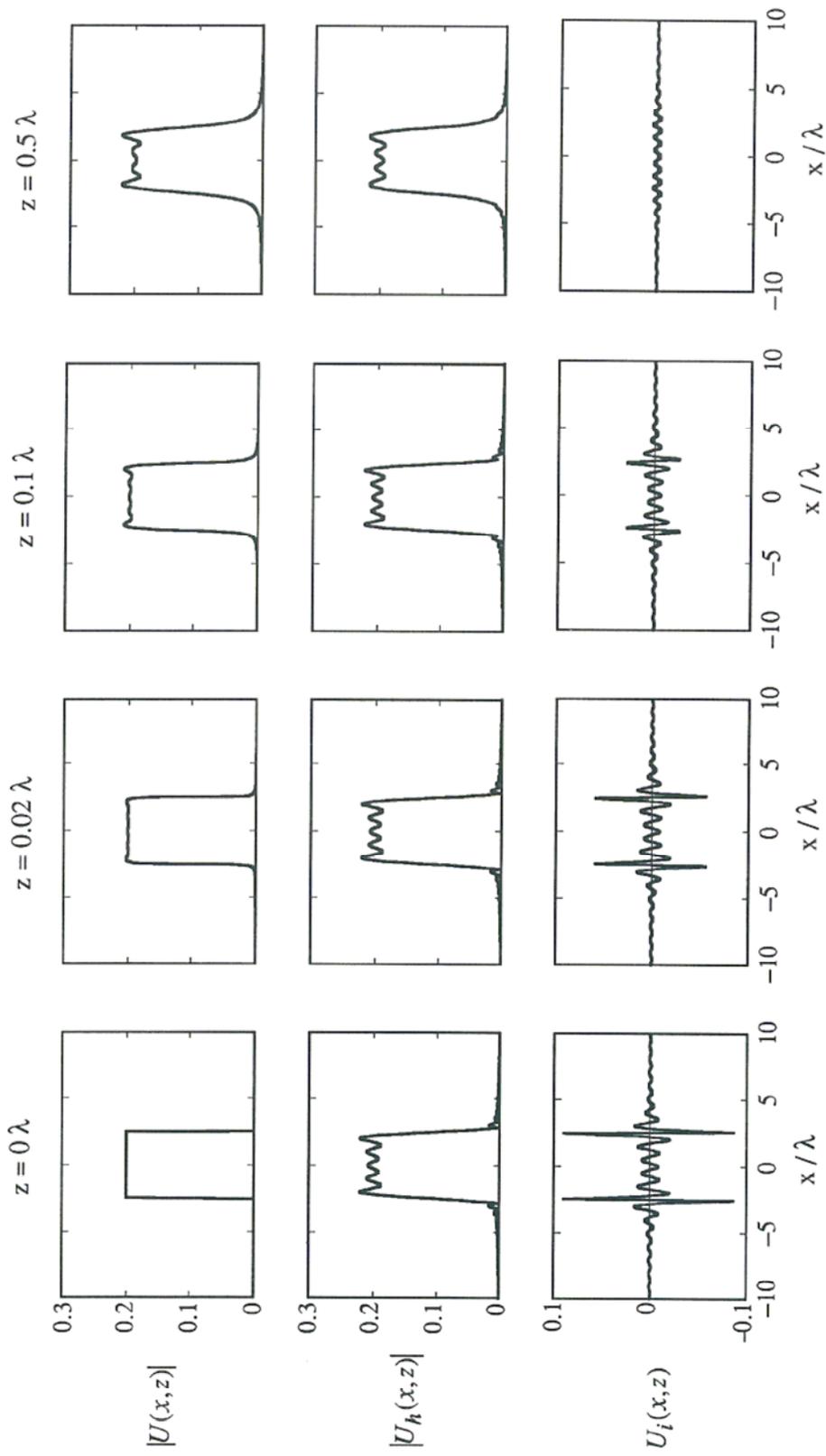

**Figure 3-4** Near-field diffraction patterns showing the field $|U(x,z)|$, the homogeneous contribution $|U_h(x,z)|$ and the evanescent contribution $U_i(x,z)$ for a slit of width $d = 5\lambda$, with the choice $K = \lambda/d = 0.2$.

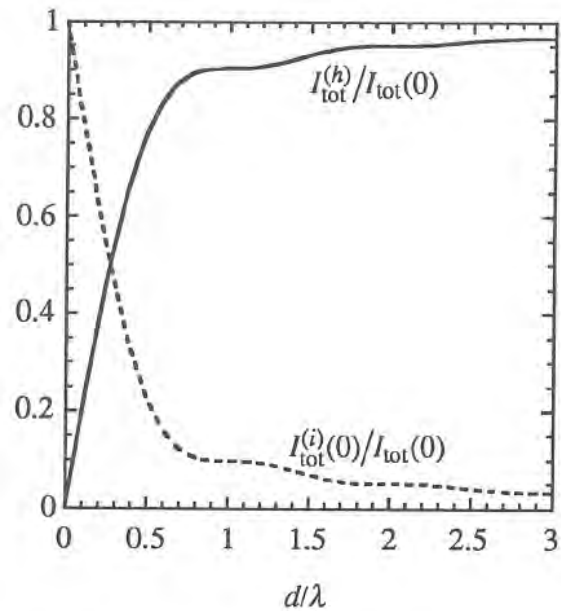

**Figure 3-5** Total homogeneous intensity $I_{\text{tot}}^{(h)}$ (which is independent of $z$) and total evanescent intensity $I_{\text{tot}}^{(i)}(0)$ in the plane $z = 0$ as functions of the slit width $d$. These curves were computed from Eqs. (3.5.5) and (3.5.8).

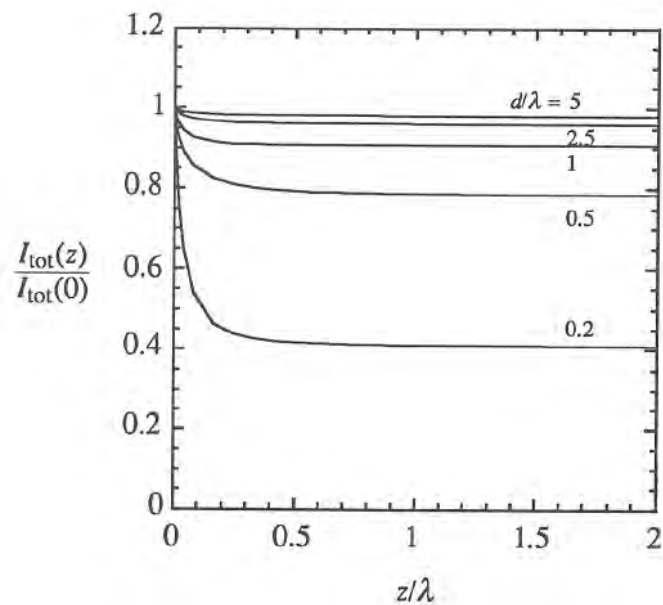

**Figure 3-6** Total intensity $I_{\text{tot}}(z) = I_{\text{tot}}^{(h)} + I_{\text{tot}}^{(i)}(z)$ computed from Eqs. (3.5.4b) and (3.5.5) as a function of the distance $z$ for various slit widths.

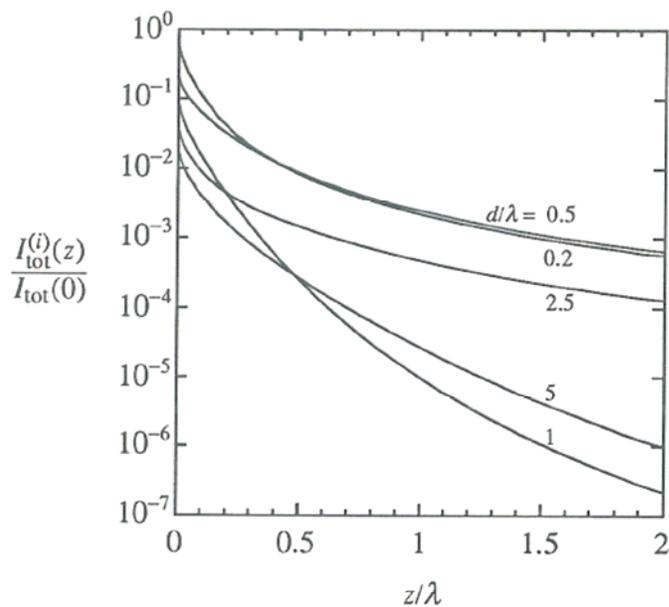

**Figure 3-7** Total evanescent intensity $I_{\text{tot}}^{(i)}(z)$ computed from Eq. (3.5.4b) as a function of the distance $z$ for various slit widths.

## 3.6  CONCLUDING REMARKS

It is clear from our analysis of the homogeneous and the evanescent contributions, and especially from the example presented in Section 3.5, that changes in the field for propagation distances smaller than the wavelength are dominated by the decay of the evanescent contribution.  Although we have considered only two-dimensional scalar fields in this Chapter, most of the analysis can be extended to three-dimensional scalar fields and to vector fields.

CHAPTER 4

# NEAR-FIELD DIFFRACTION OF
# ELECTROMAGNETIC WAVES BY A SLIT

So far we have not imposed rigorous boundary value conditions; in the example presented in Section 3.5, we simply used the approximate boundary conditions of the Rayleigh-Sommerfeld theory of the first kind. In order to obtain a better understanding of the structure of the near-field and to develop more accurate approximate theories for describing near-field diffraction, we now reexamine a problem that has been studied extensively in the literature and that requires the determination of precise boundary values: the diffraction of a plane wave by a slit in a perfectly conducting plane.[1-3] Rather than summing the exact series solution, which is not an efficient procedure for obtaining the field at many points in space, we use the method of finite differences in the time domain to numerically determine the near field. We present numerical results for the field in the vicinity of the slit in Section 4.1 and compare these results with the predictions of the Rayleigh-Sommerfeld theories in Section 4.2.

## 4.1 NUMERICAL RESULTS

We choose a coordinate system in which the conducting screen lies in the plane $z = 0$ and the slit (of width $d$) is parallel to and centered about the $y$-axis (see

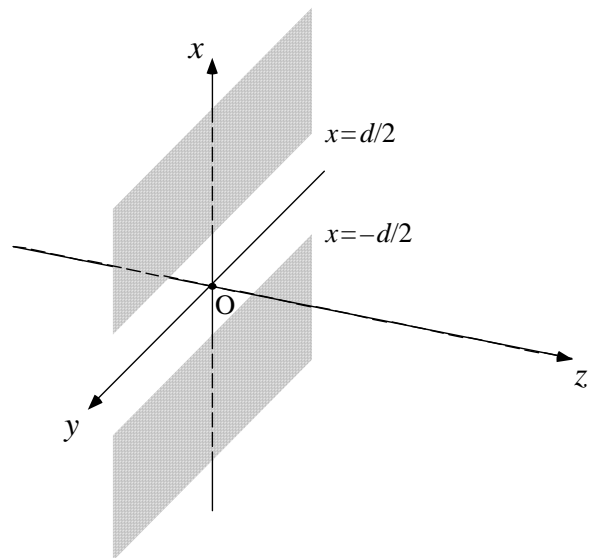

**Figure 4-1** Slit geometry.

Fig. 4-1). We assume that the incident field, which illuminates the slit from the half-space $z < 0$, is monochromatic and two-dimensional, being a function of only the $x$ and $z$-coordinates. By decomposing the incident field into E-polarized and H-polarized contributions, this diffraction problem may be analyzed by two independent scalar theories (see Section 2.2.4). For the E-polarized contribution, the electric field $\mathbf{E}_{\mathrm{EP}}(x, z) = \hat{\mathbf{y}} \, E_y(x, z)$ is parallel to the slit, whereas, for the H-polarized one, the magnetic field $\mathbf{H}_{\mathrm{HP}}(x, z) = \hat{\mathbf{y}} \, H_y(x, z)$ is parallel to it.

To determine the near field accurately, without having to sum a slowly converging series or resorting to approximate theories, we used the finite-difference time-domain (FD-TD) method described in Appendix E. In this approach, both polarizations are treated by applying finite differences to the free-space wave equation. The difference between the two polarizations is in the boundary conditions that are applied on the surface of the perfect conductor.

We first examined the diffraction of a plane wave by a perfectly conducting half-plane in order to verify that our FD-TD program was working properly. We present some of those results here because it is instructive to contrast half-plane diffraction with slit diffraction. Figure 4-2 is a color image of the amplitude and the phase of the total electric field $E_y(x, z)$ in the vicinity of the half-plane when the illuminating field is an E-polarized, normally incident, unit amplitude, plane wave. Figure 4-3 is the corresponding image for H-polarized illumination and shows the amplitude and the phase of the total magnetic field $H_y(x, z)$. To facilitate a comparison between the FD-TD results and the exact analytical solution depicted in Fig. 11.12 of Ref. 4, the amplitude of the magnetic field is shown again as a contour plot in Fig. 4-4. (Note that the orientations of Figs. 4-3 and 4-4 are different.) Some of the numerical errors due to the discretization of the FD-TD method are evident in these contours, especially in those of unit amplitude. However, by comparing Fig. 4-4 with the exact solution, we see that this error is small ($< 2\%$). In fact, it is not visible in the color images.

The numerical results for slit diffraction are shown in Figs. 4-5 through 4-17. By varying the parameters used in the calculations (see Appendix E), we were able to estimate that the error in the amplitude is less than 4 % .

If we examine the field distribution in the half-space $z < 0$, we can readily identify the remnants of the standing wave pattern one would expect to see in front of a perfectly conducting plane. Clearly, with increasing slit size, there is less similarity between the field distribution in the half-space $z < 0$ and such a standing wave pattern. Several edge dislocations,[5,6] where the amplitude is zero and the phase is therefore not defined, are also visible in this region. We have labeled these singular points only in Fig. 4-8, but they are present in all of the color images. It should be pointed out that near dislocations[*] the Poynting vector exhibits some rather unusual behavior.[7,8]

Next, if we consider the plane $z = 0^+$ immediately behind the slit, we see that for E-polarization (see Figs. 4-13 and 4-14) the electric field $E_y(x, 0^+)$ has at least one maximum in the slit and drops to zero at the edge of the slit. Furthermore, the number of maxima increases in a predictable manner as the width of the slit increases. For H-polarization (see Figs. 4-15 and 4-16), the magnetic field $H_y(x, 0^+)$ is very nearly a constant in the slit and decreases on the surface of the perfect conductor with increasing $|x|$. The barely visible oscillations of $H_y(x, 0^+)$ in the slit can be attributed to numerical errors, because the magnetic field in the slit must be equal to the incident magnetic field (see Section 2.2.1). Although $H_y(x, 0^+)$ does not have zero value in the shadow region, i.e., for $|x| > d/2$, there is obviously no energy flow through the perfect conductor since the tangential electric field is identically zero on the surface.

In the plane $z = 0^+$, the E-polarized field $E_y(x, 0^+)$ is obviously very different from the H-polarized field $H_y(x, 0^+)$. However, after propagating a distance of just a few wavelengths along the $z$-direction, $E_y(x, z)$ and $H_y(x, z)$ become similar to each other, as is evident from Fig. 4-17. This phenomenon may be explained by examining the spatial Fourier transforms of $E_y(x, 0^+)$ and $H_y(x, 0^+)$. One would then notice that the main differences between $E_y(x, 0^+)$ and $H_y(x, 0^+)$ are in the higher spatial frequencies[†]. Since, in the angular spectrum representation (see Section 2.1.3), these higher spatial frequencies give rise either to homogeneous

---

[*] We examine another type of dislocation, the vortex or screw dislocation, in Chapter 6.
[†] This type of analysis is used in Ref. 9 to show that the main differences between the Rayleigh-Sommerfeld theory of the first kind, the Rayleigh-Sommerfeld theory of the second kind and the Kirchhoff theory are also due to high spatial frequencies

plane wave that propagate at a large angle with respect to the $z$-axis or to evanescent plane waves that propagate along the $x$-axis and decay exponentially along the $z$-axis, $E_y(x, z)$ and $H_y(x, z)$ become more similar to each other with increasing $z$, for fixed $x$.

It is interesting to locate, in the color images, the points where the amplitude has a non-zero extremum and to examine the behavior of the phase in the vicinity of these points. As expected from the two theorems of Section 2.1.7, the surfaces of constant phase are relatively "flat" there. Furthermore, near maxima of the amplitude these surfaces are spaced further apart than the corresponding surfaces for a plane wave, whereas near minima they are spaced closer together than those for a plane wave. This behavior of the phase is particularly noticeable along the $z$-axis. As an example, in Fig. 4-8 we have labeled all the points where the amplitude has an extremum ($\times$ = maximum, $\circ$ = non-zero minimum, $\bullet$ = zero).

## 4.2 COMPARISON OF NUMERICAL RESULTS WITH PREDICTIONS OF APPROXIMATE THEORIES

We can now compare the FD-TD numerical results for the slit with the predictions of the Rayleigh-Sommerfeld theories of the first and second kinds (see Section 2.1.5). We will restrict this comparison to the plane $z = 0^+$ immediately behind the slit because that is where we would expect the greatest differences between the numerical results and the approximate theories.

*text continued after Fig. 4-17*

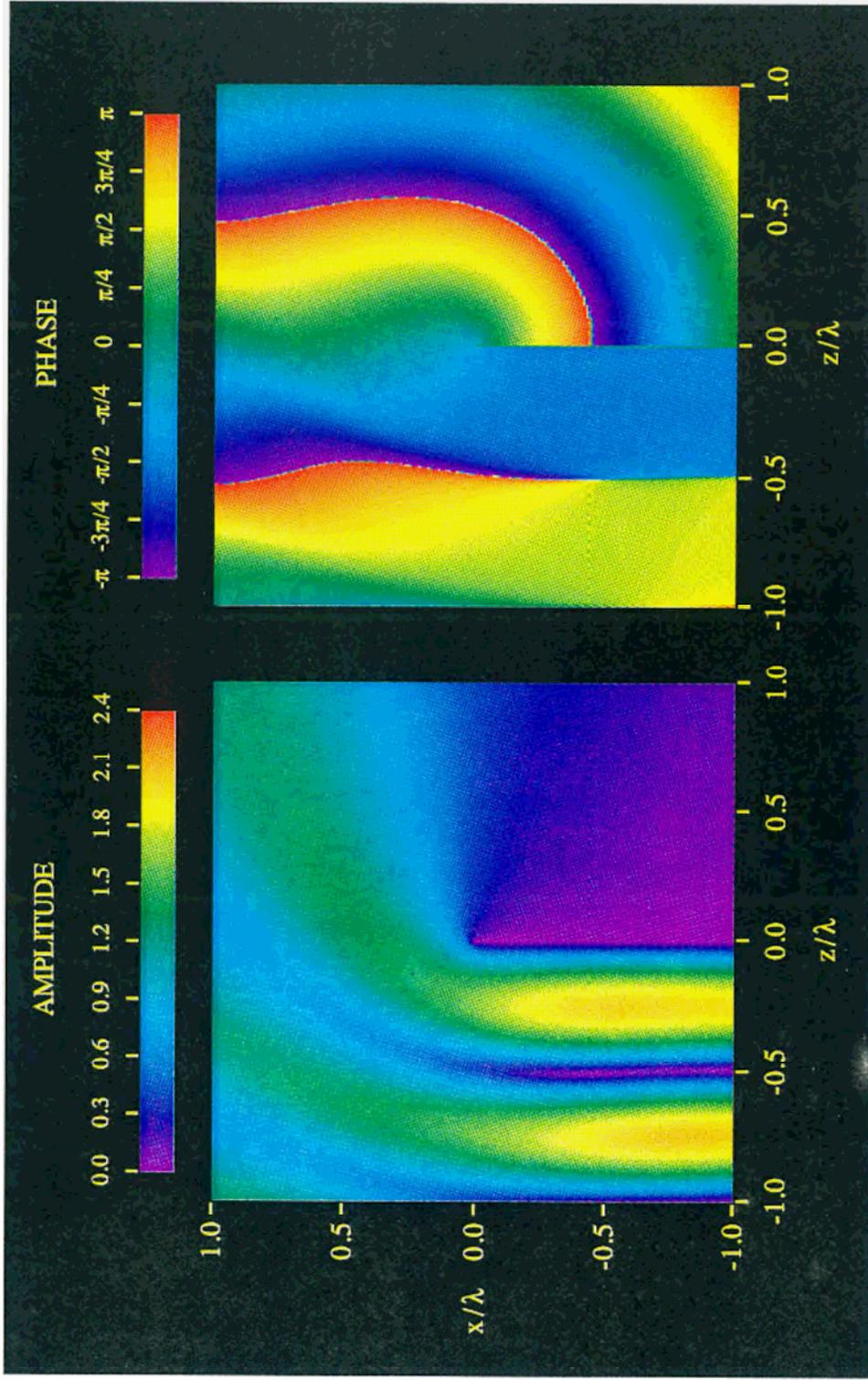

**Figure 4-2** Diffraction of an E-polarized, unit amplitude, normally incident, plane wave by a perfectly conducting half-plane: Images showing the amplitude and the phase of the electric field $E_y(x,z)$. The half-plane begins at the origin and continues along the negative $x$-axis. The illuminating plane wave is incident from the half-space $z < 0$.

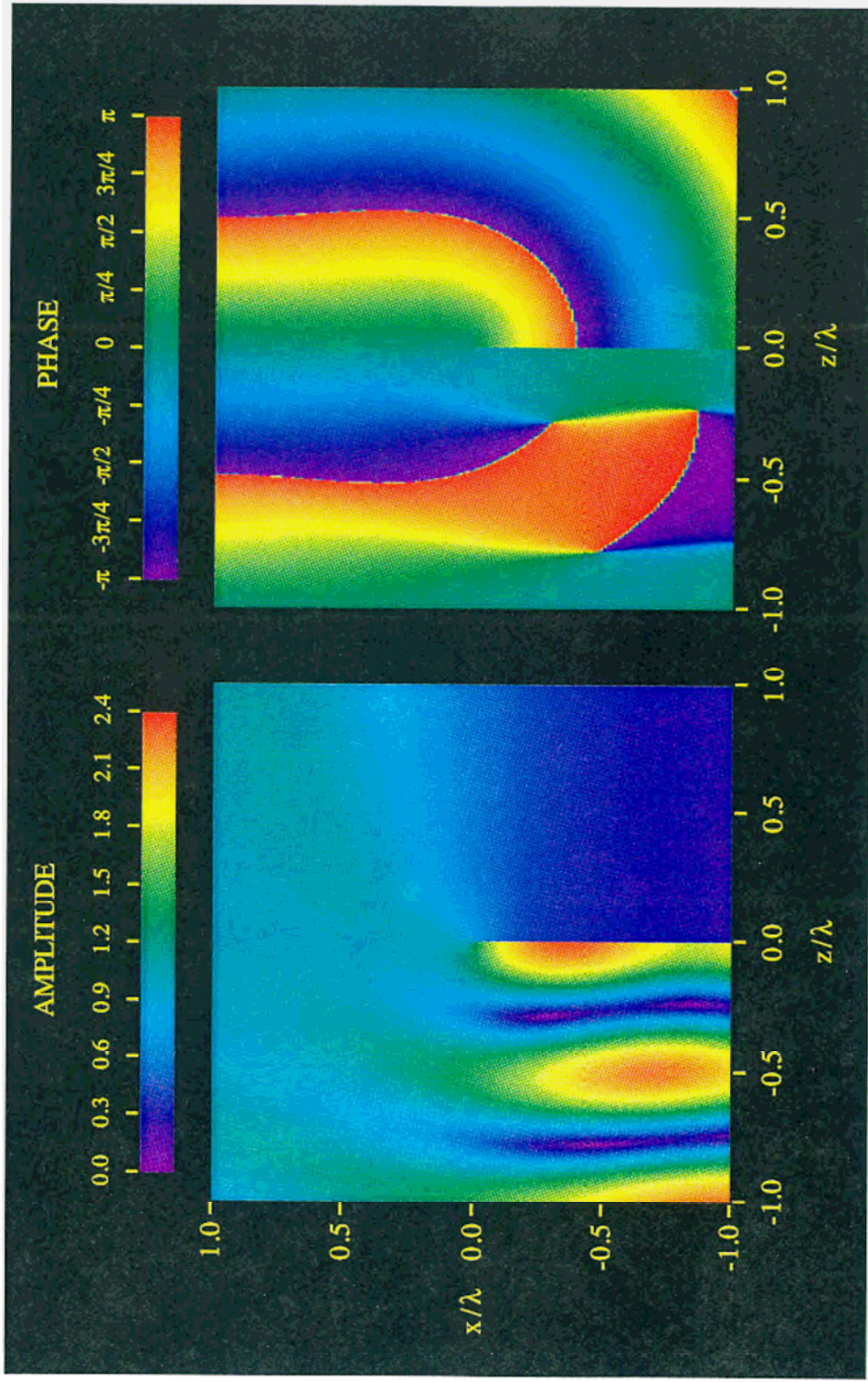

**Figure 4-3** Diffraction of an H-polarized, unit amplitude, normally incident, plane wave by a perfectly conducting half-plane: Images showing the amplitude and the phase of the magnetic field $H_y(x,z)$. The half-plane begins at the origin and continues along the negative $x$-axis. The illuminating plane wave is incident from the half-space $z < 0$.

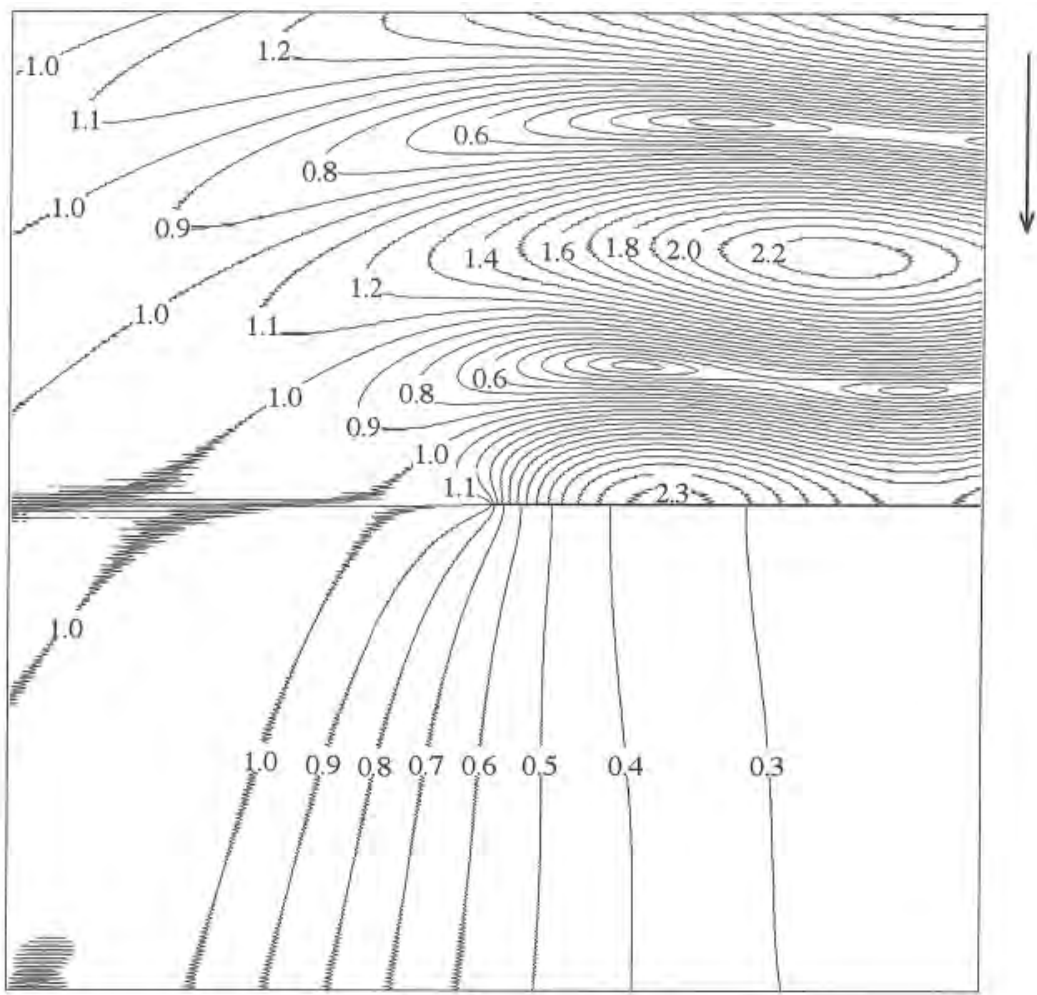

**Figure 4-4** Diffraction of an H-polarized, unit amplitude, normally incident, plane wave by a perfectly conducting half-plane: Same numerical results as in Fig. 4-3, but with contours of constant amplitude of $H_y(x,z)$ shown. To facilitate comparison with published results, here the half-plane begins at the center of the plot and continues to the right, and the plane wave is normally incident from the top, as indicated by the arrow. The size of the region is the same as in Fig. 4-3, namely two wavelengths per side.

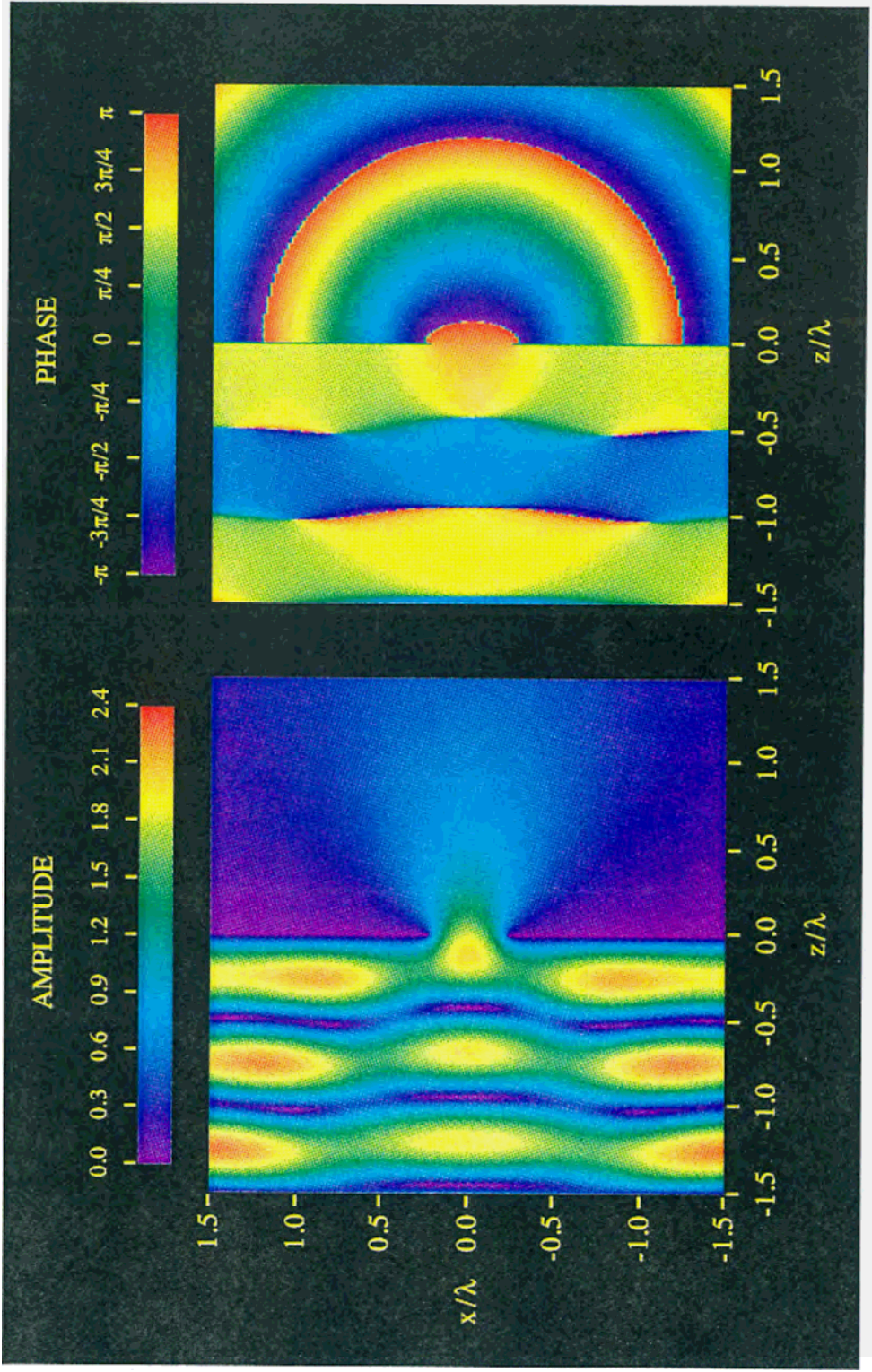

**Figure 4-5** Diffraction of an E-polarized, unit amplitude, normally incident, plane wave by a slit of width $d = 0.5\lambda$ in a perfectly conducting plane: Images showing the amplitude and the phase of the electric field $E_y(x,z)$.

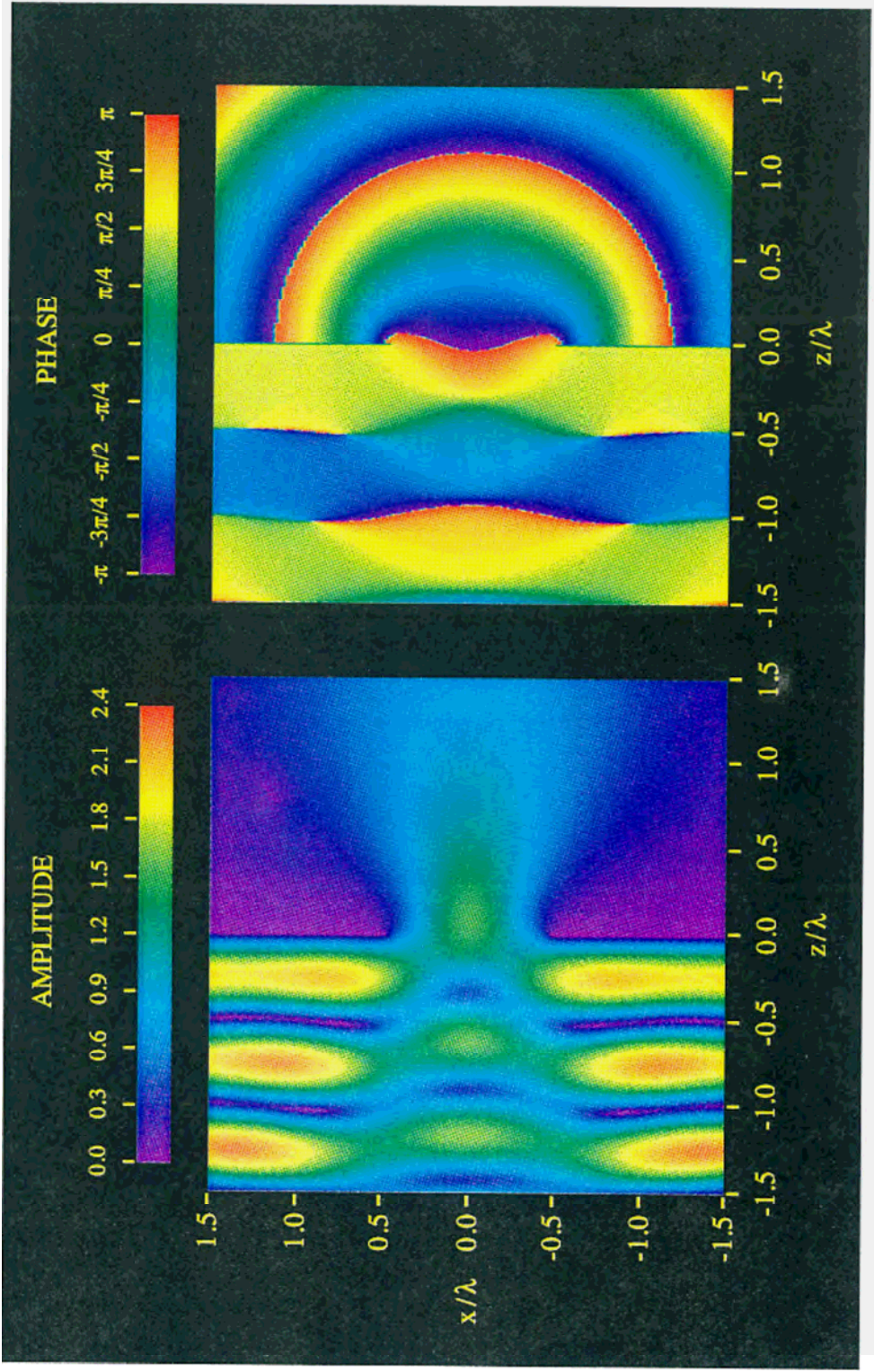

**Figure 4-6** Diffraction of an E-polarized, unit amplitude, normally incident, plane wave by a slit of width $d = 1\lambda$ in a perfectly conducting plane: Images showing the amplitude and the phase of the electric field $E_y(x,z)$.

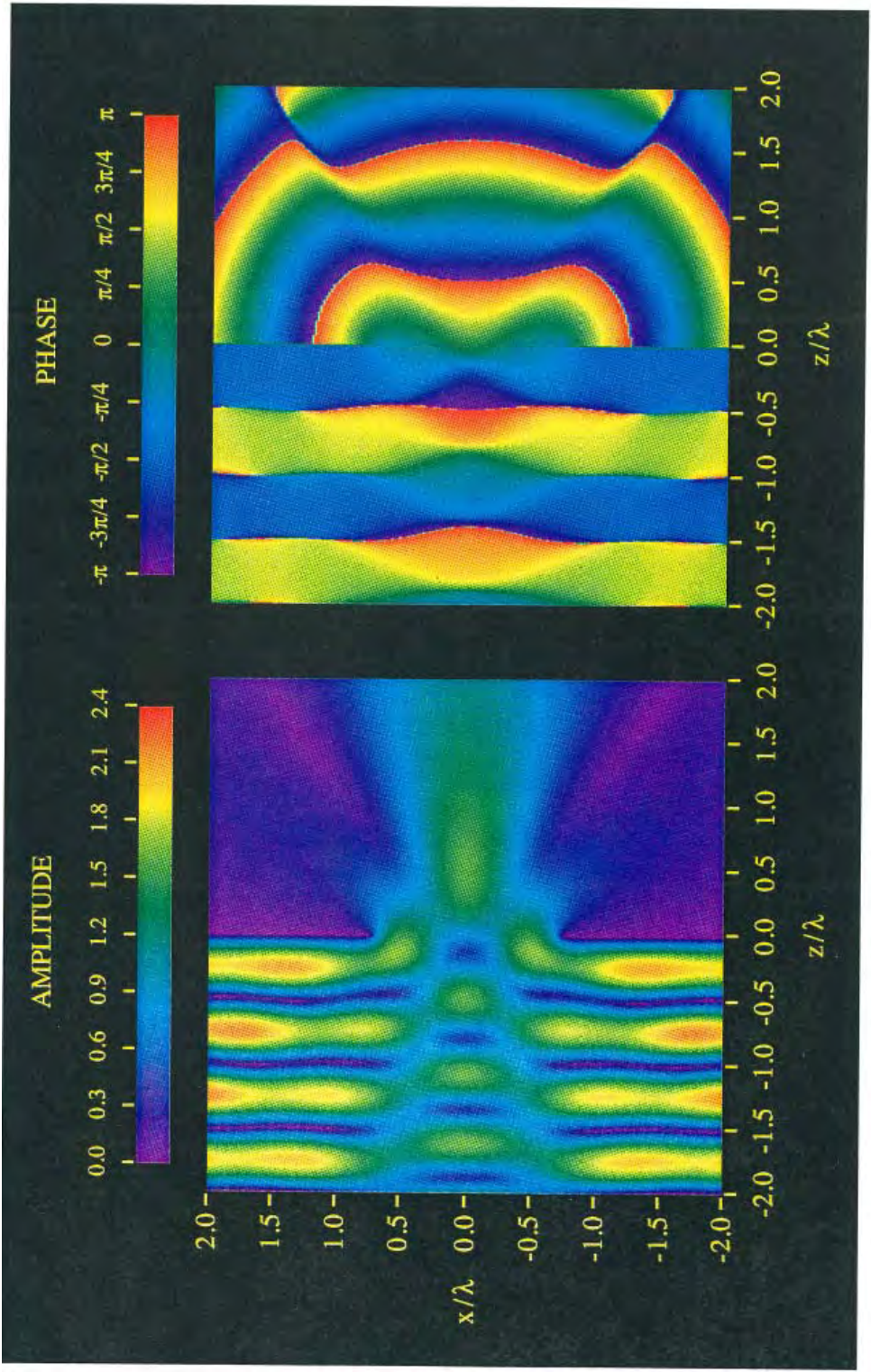

**Figure 4-7** Diffraction of an E-polarized, unit amplitude, normally incident, plane wave by a slit of width $d = 1.5\lambda$ in a perfectly conducting plane: Images showing the amplitude and the phase of the electric field $E_y(x,z)$.

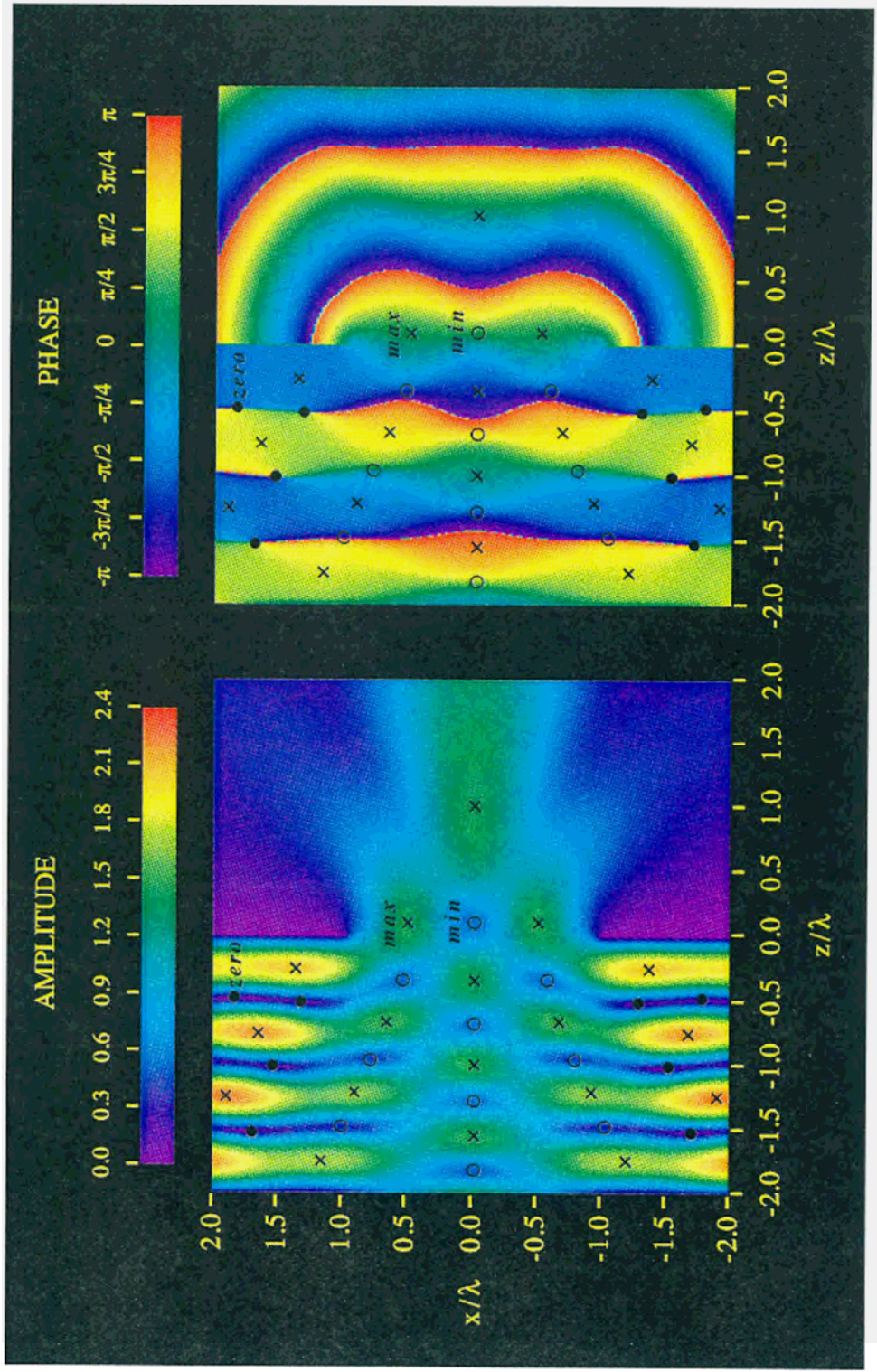

**Figure 4-8** Diffraction of an E-polarized, unit amplitude, normally incident, plane wave by a slit of width $d = 2\lambda$ in a perfectly conducting plane: Images showing the amplitude and the phase of the electric field $E_y(x,z)$.

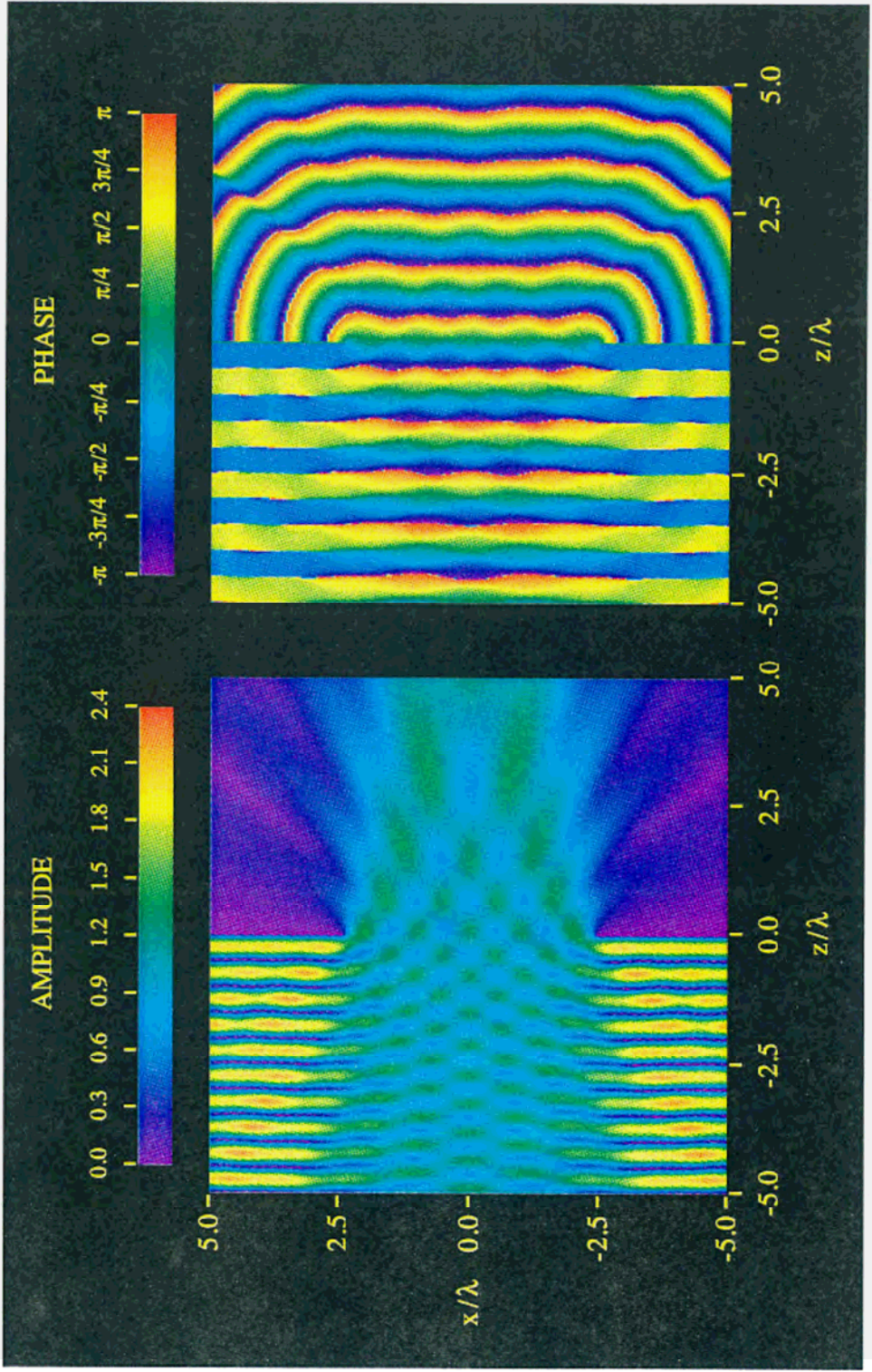

**Figure 4-9** Diffraction of an E-polarized, unit amplitude, normally incident, plane wave by a slit of width $d = 5\lambda$ in a perfectly conducting plane: Images showing the amplitude and the phase of the electric field $E_y(x,z)$.

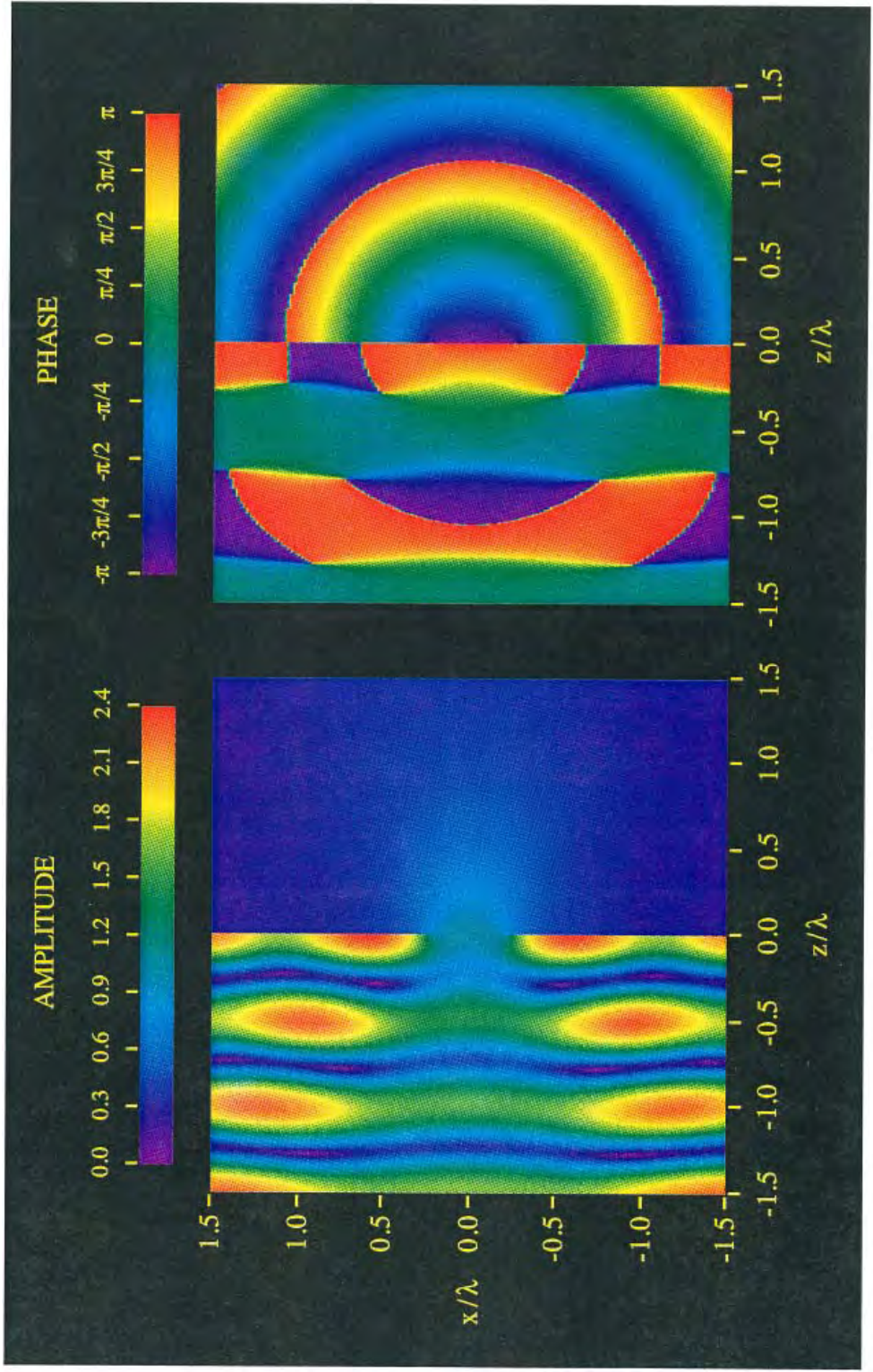

**Figure 4-10** Diffraction of an H-polarized, unit amplitude, normally incident, plane wave by a slit of width $d = 0.5\lambda$ in a perfectly conducting plane: Images showing the amplitude and the phase of the magnetic field $H_y(x,z)$.

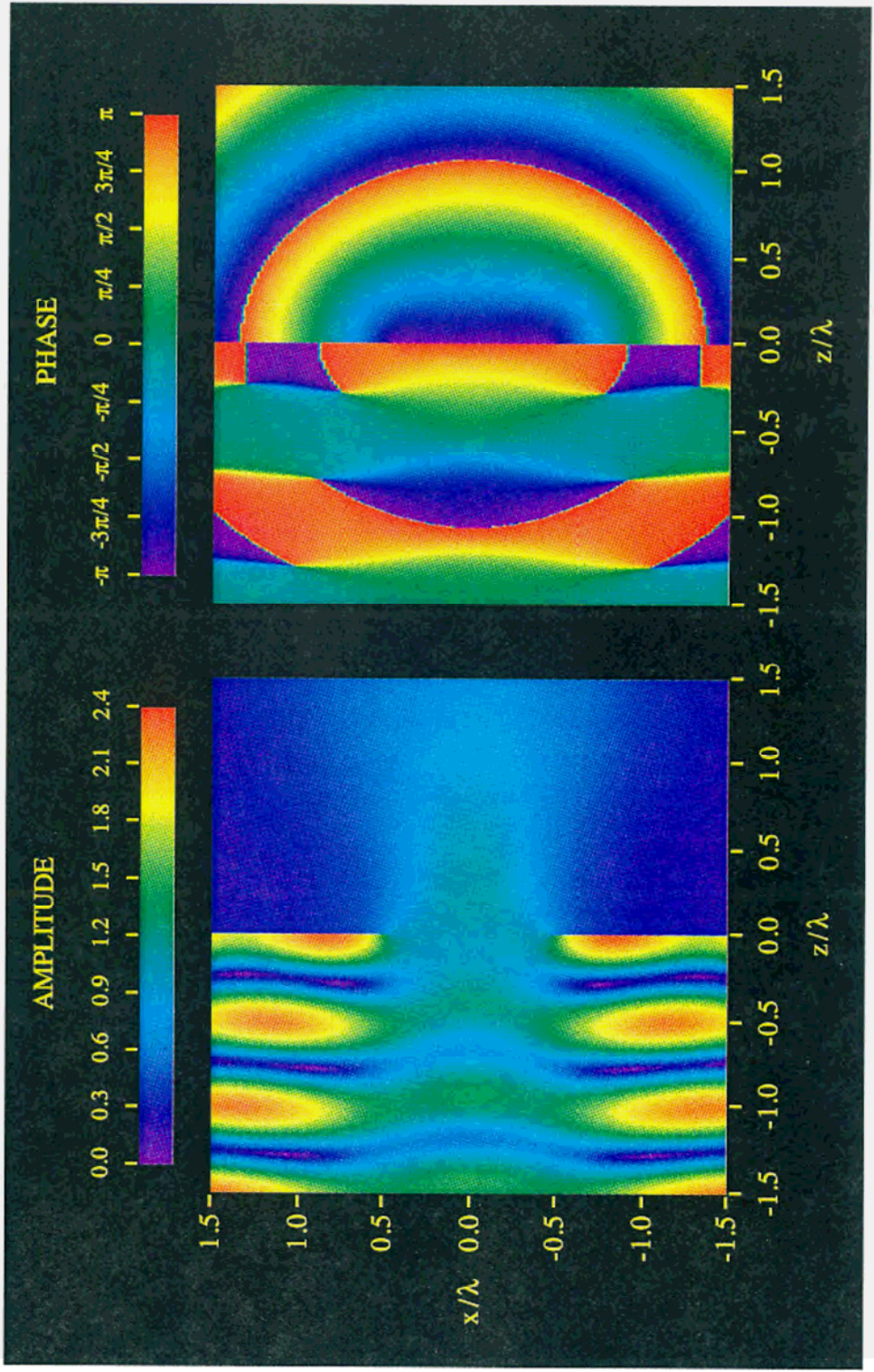

**Figure 4-11** Diffraction of an H-polarized, unit amplitude, normally incident, plane wave by a slit of width $d = 1\lambda$, in a perfectly conducting plane: Images showing the amplitude and the phase of the magnetic field $H_y(x,z)$.

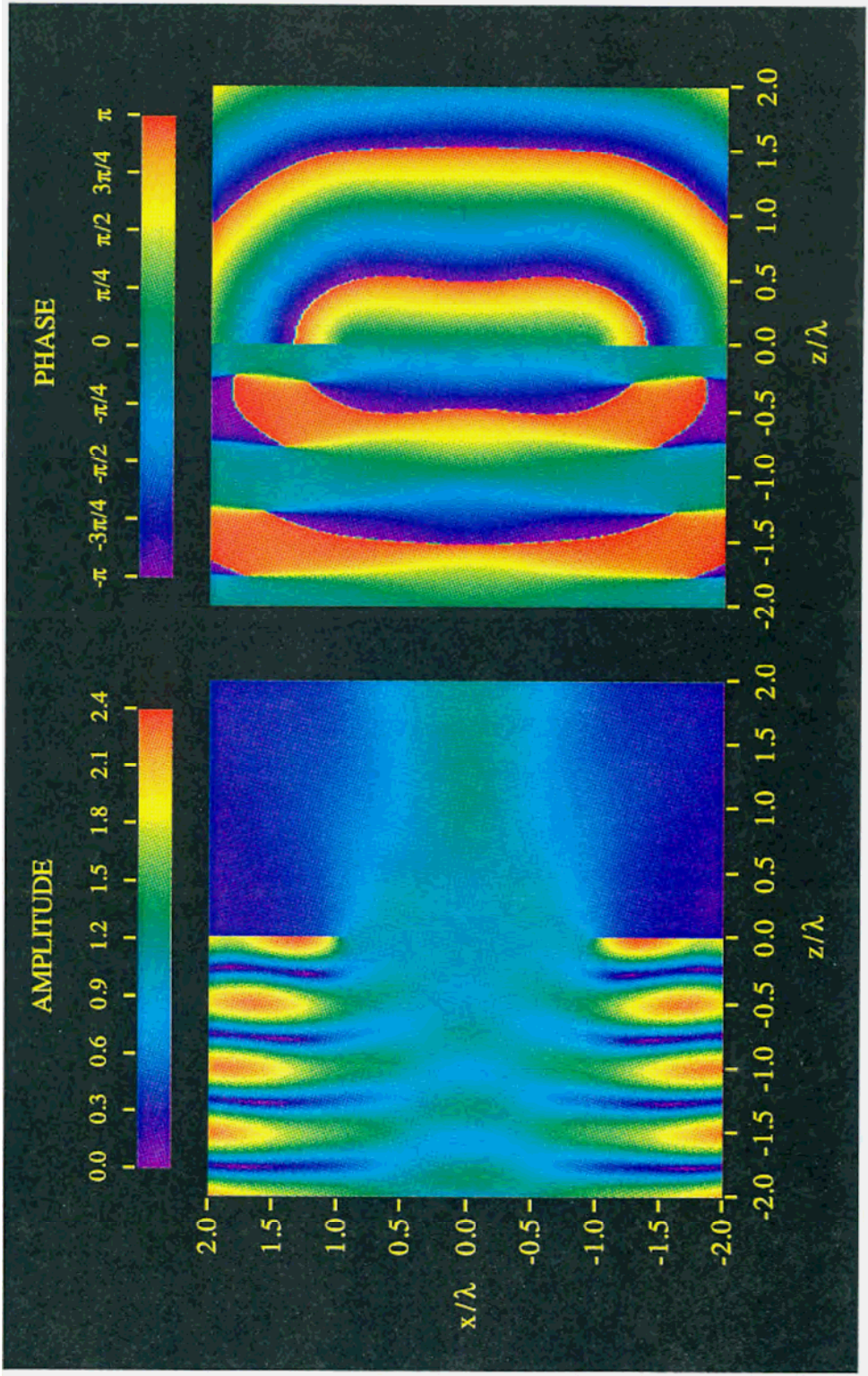

**Figure 4-12** Diffraction of an H-polarized, unit amplitude, normally incident, plane wave by a slit of width $d = 2\lambda$ in a perfectly conducting plane: Images showing the amplitude and the phase of the magnetic field $H_y(x,z)$.

**E-POLARIZATION**

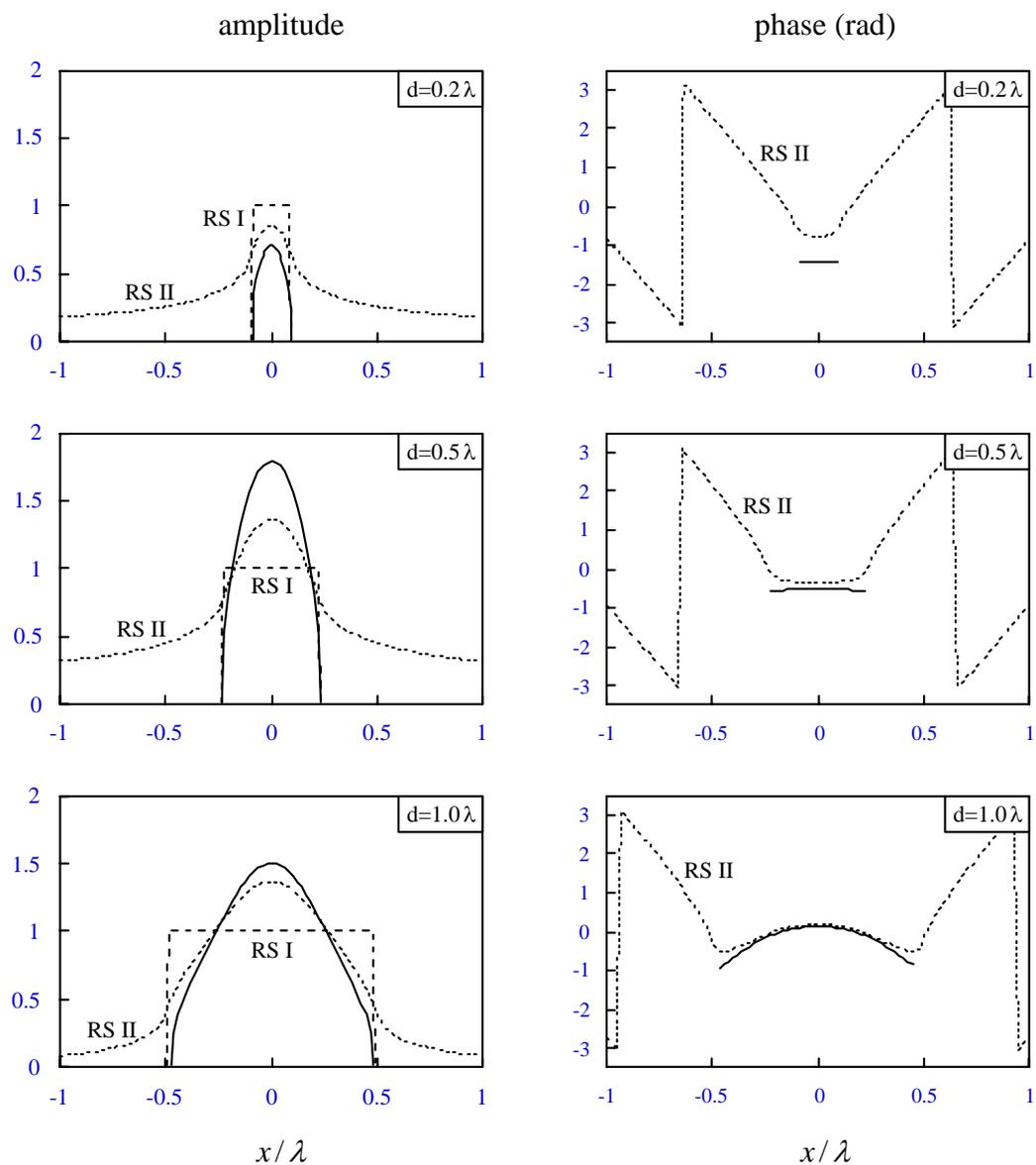

**Figure 4-13** Field distribution in the plane $z = 0^+$ for slits of width $d = 0.2\lambda$, $d = 0.5\lambda$ and $d = 1\lambda$. Comparison of numerical results (solid lines) for an E-polarized plane wave, showing the amplitude and phase of the electric field $E_y(x, 0^+)$, with the predictions of the Rayleigh-Sommerfeld theories of the first (RS I: dashed lines) and second (RS II: dotted lines) kinds.

**E-POLARIZATION**

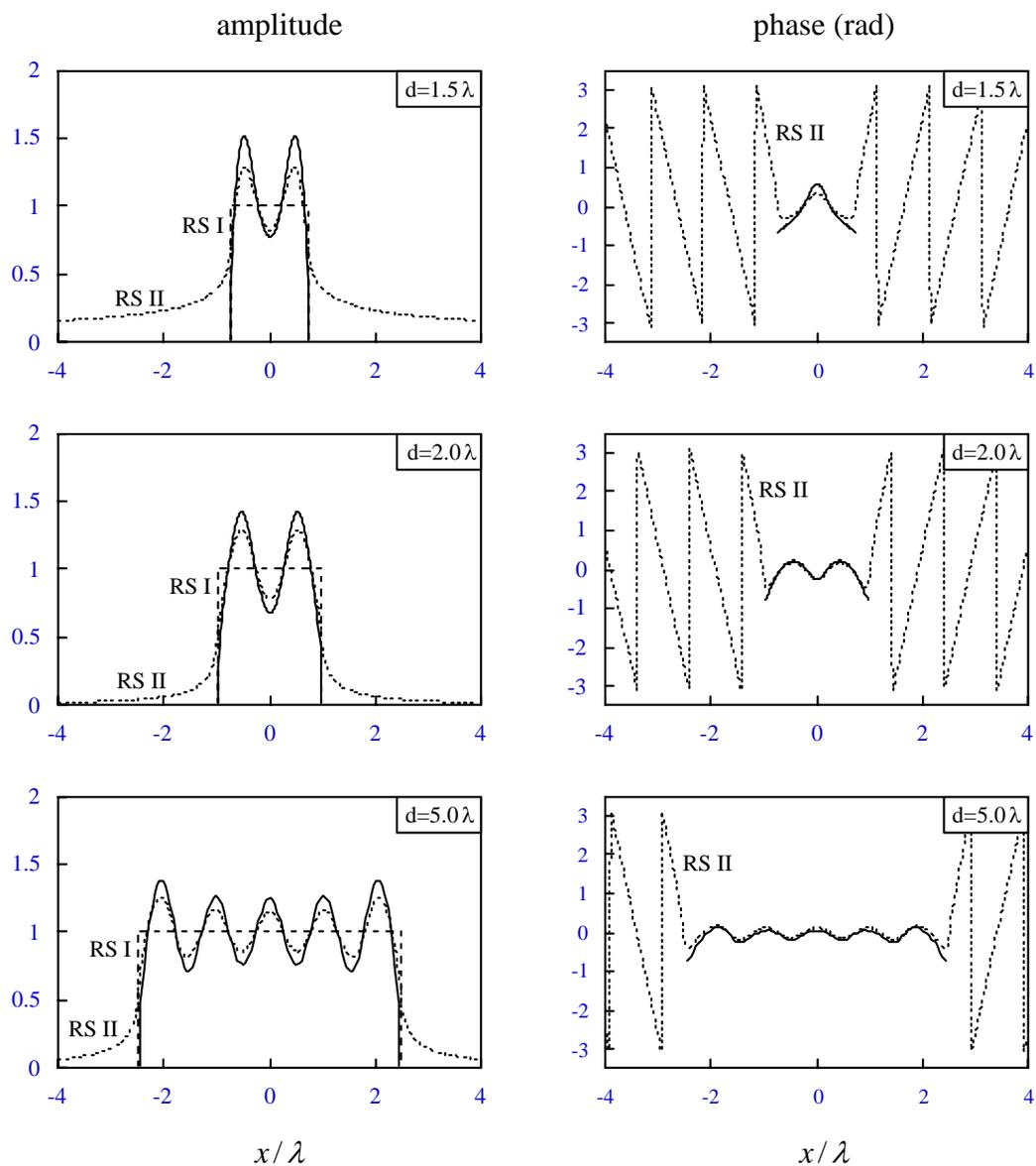

**Figure 4-14** Field distribution in the plane $z = 0^+$ for slits of width $d = 1.5\lambda$, $d = 2\lambda$ and $d = 5\lambda$. Comparison of numerical results (solid lines) for an E-polarized plane wave, showing the amplitude and phase of the electric field $E_y(x, 0^+)$, with the predictions of the Rayleigh-Sommerfeld theories of the first (RS I: dashed lines) and second (RS II: dotted lines) kinds.

**H-POLARIZATION**

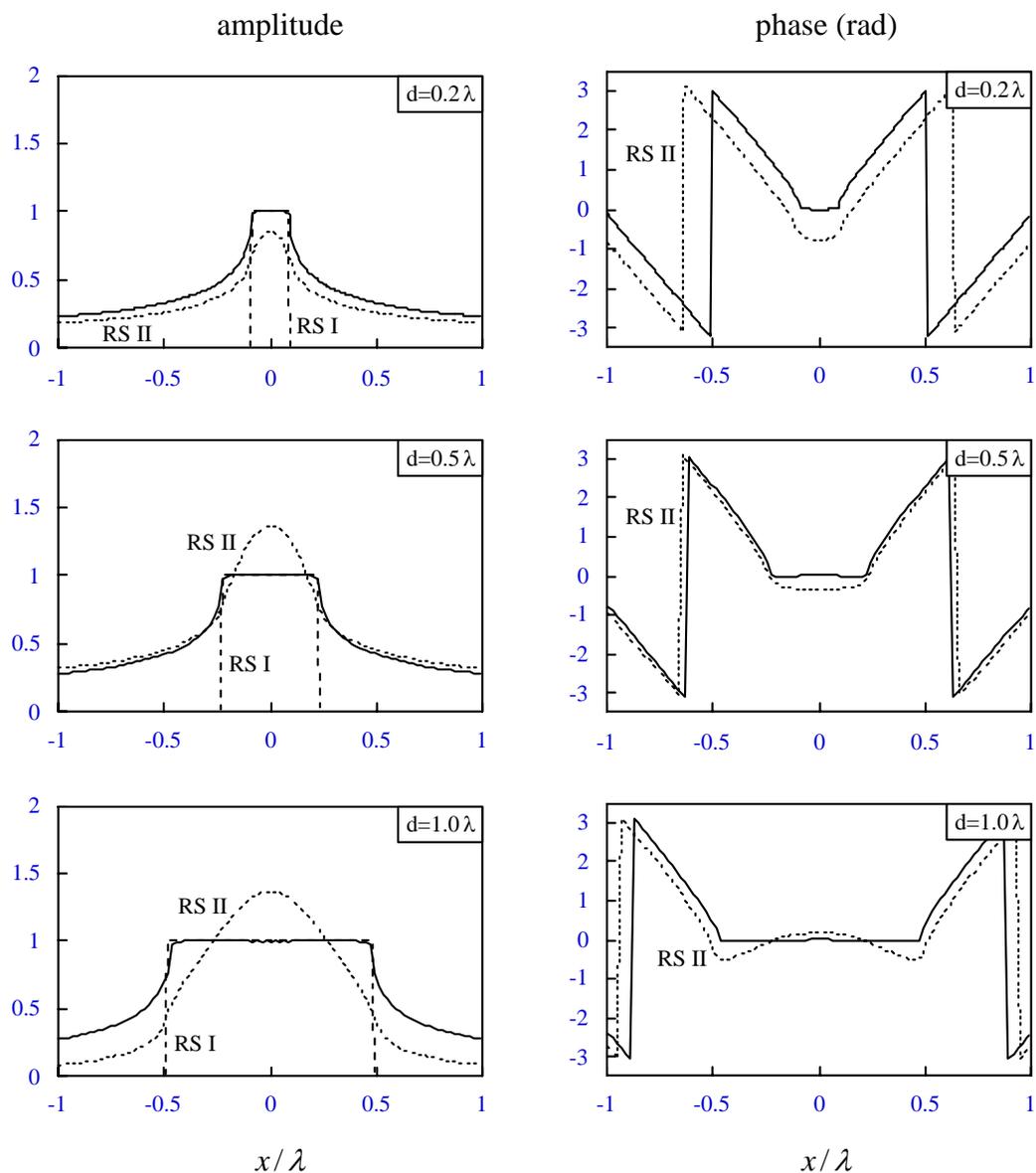

amplitude             phase (rad)

**Figure 4-15** Field distribution in the plane $z = 0^+$ for slits of width $d = 0.2\lambda$, $d = 0.5\lambda$ and $d = 1\lambda$. Comparison of numerical results (solid lines) for an H-polarized plane wave, showing the amplitude and phase of the magnetic field $H_y(x, 0^+)$, with the predictions of the Rayleigh-Sommerfeld theories of the first (RS I: dashed lines) and second (RS II: dotted lines) kinds.

**H-POLARIZATION**

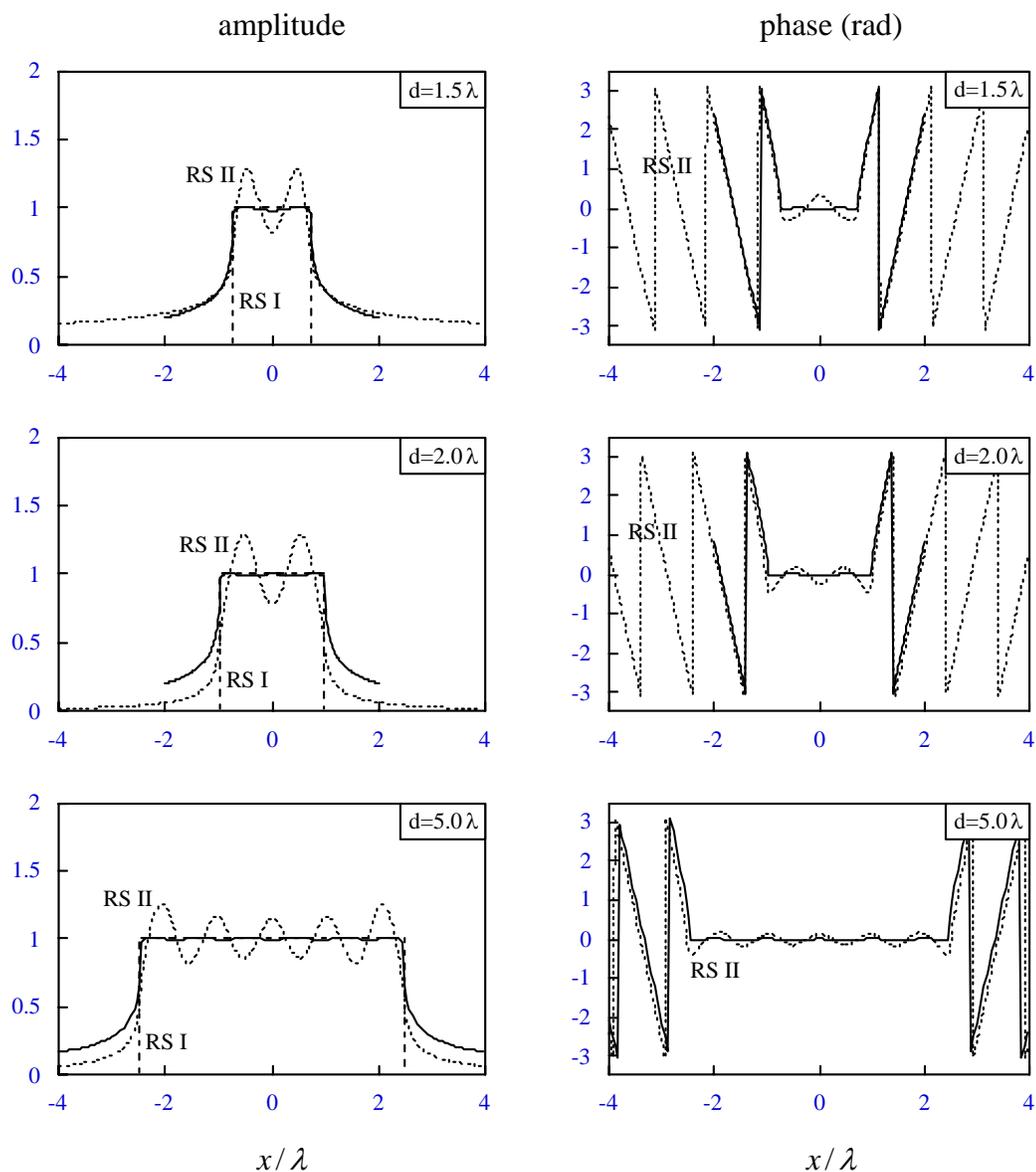

**Figure 4-16** Field distribution in the plane $z = 0^+$ for slits of width $d = 1.5\lambda$, $d = 2\lambda$ and $d = 5\lambda$. Comparison of numerical results (solid lines) for an H-polarized plane wave, showing the amplitude and phase of the magnetic field $H_y(x, 0^+)$, with the predictions of the Rayleigh-Sommerfeld theories of the first (RS I: dashed lines) and second (RS II: dotted lines) kinds.

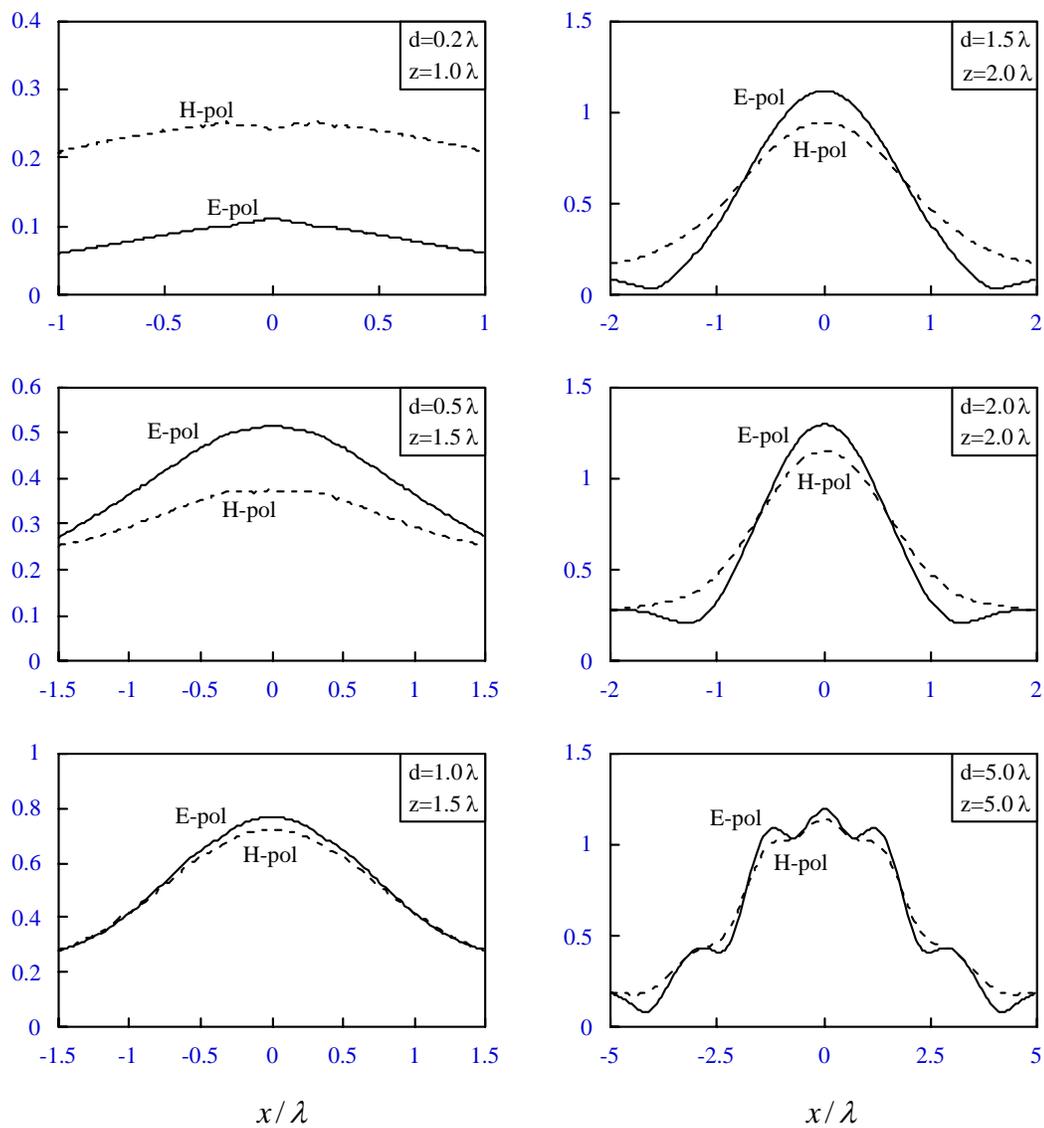

**Figure 4-17** Comparison of numerical results for E and H-polarizations in a plane $z$ = constant > 0 behind the slit for various slit widths $d$. The amplitude of the electric field $E_y(x,z)$ is shown for E-polarization and that of the magnetic field $H_y(x,z)$ for H-polarization.

### 4.2.1  Predictions of Approximate Theories

We need to determine the field distribution in the plane $z = 0^+$ predicted by the approximate theories when the illuminating field is a unit amplitude, normally incident, plane wave, i.e., when

$$U^{(\text{inc})}(x,z) = e^{ikz} . \qquad (4.2.1)$$

For the Rayleigh-Sommerfeld theory of the first kind, the boundary value $U^{(I)}(x,0^+)$ of the field is equal to the incident field in the illuminated region and is zero in the shadow region. Therefore,

$$U^{(I)}(x,0^+) = 0 \qquad\qquad \text{for } |x| > \frac{d}{2} , \qquad (4.2.2a)$$

$$U^{(I)}(x,0^+) = 1 \qquad\qquad \text{for } |x| \leq \frac{d}{2} . \qquad (4.2.2b)$$

The result $U^{(II)}(x,0^+)$ for the Rayleigh-Sommerfeld theory of the second kind is more complicated. To derive it, we require the two-dimensional version of Eq. (2.1.34):

$$U^{(II)}(x,0^+) = -\frac{1}{2\pi} \int\limits_{-d/2}^{d/2} \frac{\partial U^{(\text{inc})}(x',z')}{\partial z'} \bigg|_{z'=0} G(x-x',z)\,dx' . \quad (4.2.3)$$

Here $G(x,z)$ is the two-dimensional Green function,

$$G(x,z) = i\pi H_0^{(1)}\!\left( k\sqrt{x^2 + z^2} \right) , \qquad (4.2.4)$$

$H_n^{(1)}$ being the Hankel function of the first kind and $n$th order. If we let $z = 0^+$ in Eq. (4.2.3) and use Eqs. (4.2.1) and (4.2.4), we find that

$$U^{(II)}(x, 0^+) \;=\; \frac{k}{2} \int_{-d/2}^{d/2} H_0^{(1)}\bigl(k|x - x'|\bigr)\, dx' \;. \qquad (4.2.5)$$

Let us examine the above integral in the three regions $x > d/2$, $|x| \leq d/2$ and $x < -d/2$. After a simple change of variables, we obtain the expressions

$$U^{(II)}(x, 0^+) \;=\; \frac{k}{2} \int_{x-d/2}^{x+d/2} H_0^{(1)}(ks)\, ds \qquad\qquad \text{for } x \,>\, d/2 \;,$$

$$(4.2.6a)$$

$$U^{(II)}(x, 0^+) \;=\; \frac{k}{2} \int_{0}^{x+d/2} H_0^{(1)}(ks)\, ds \;+\; \frac{k}{2} \int_{0}^{-x+d/2} H_0^{(1)}(ks)\, ds \qquad \text{for } |x| \,\leq\, d/2 \;,$$

$$(4.2.6b)$$

$$U^{(II)}(x, 0^+) \;=\; \frac{k}{2} \int_{-x-d/2}^{-x+d/2} H_0^{(1)}(ks)\, ds \qquad\qquad \text{for } x \,<\, -d/2 \;.$$

$$(4.2.6c)$$

If we define the function $F(\beta)$ by the integral

$$F(\beta) \;\equiv\; \int_{0}^{\beta} H_0^{(1)}(\alpha)\, d\alpha \;, \qquad (4.2.7)$$

Eqs. (4.2.6a)-(4.2.6c) may be rewritten in the simple form

$$U^{(II)}(x, 0^+) = \frac{1}{2} \left\{ F\big[k\big(|x| + d/2\big)\big] - F\big[k\big(|x| - d/2\big)\big] \right\} \qquad \text{for } |x| > d/2 \; ,$$

$$(4.2.8a)$$

$$U^{(II)}(x, 0^+) = \frac{1}{2} \left\{ F\big[k(x + d/2)\big] + F\big[-k(x - d/2)\big] \right\} \qquad \text{for } |x| \leq d/2 \; .$$

$$(4.2.8b)$$

Equations (4.2.8a) and (4.2.8b) are the desired expressions for the boundary value of the field given by the Rayleigh-Sommerfeld theory of the second kind. It should be pointed out that the function $F(\beta)$ may also be expressed in closed form as[10]

$$F(\beta) = \beta H_0^{(1)}(\beta) + \frac{\pi \beta}{2} \left\{ \mathbf{H}_0(\beta) \, H_1^{(1)}(\beta) - \mathbf{H}_1(\beta) \, H_0^{(1)}(\beta) \right\}, \quad (4.2.9)$$

where $\mathbf{H}_n$ is the Struve function of $n$th order.

In the shadow region for large $|x|$, more specifically for $k\big(|x| - d/2\big) >> 1$, we can obtain an approximate relation for $U^{(II)}(x, 0^+)$. We substitute from Eq. (4.2.7) into Eq. (4.2.8a) and use the first two terms in Hankel's asymptotic expansion for $H_0^{(1)}(\alpha)$ [see Ref. 10, pg. 364, Eq. (9.2.7)],

$$H_0^{(1)}(\alpha) \sim e^{-i\pi/4} \, e^{i\alpha} \sqrt{\frac{2}{\pi \alpha}} \left( 1 - \frac{i}{8\alpha} \right) . \qquad (4.2.10)$$

Equation (4.2.8a) then becomes

$$U^{(II)}(x, 0^+) \approx \frac{k \, e^{-i\pi/4}}{\sqrt{2\pi}} \int\limits_{|x|-d/2}^{|x|+d/2} e^{iks} \left[ \frac{1}{(ks)^{1/2}} - \frac{i}{8(ks)^{3/2}} \right] ds$$

$$\approx \frac{k \, e^{-i\pi/4}}{\sqrt{2\pi}} \, e^{ik|x|} \int\limits_{-d/2}^{d/2} e^{iks'} \left[ \frac{1}{\big(k|x| + ks'\big)^{1/2}} - \frac{i}{8\big(k|x| + ks'\big)^{3/2}} \right] ds' \; .$$

$$(4.2.11)$$

We can approximate the quantities in the square brackets on the second line of Eq. (4.2.11) by their Taylor series, keeping only terms of order $(k|x|)^{-1/2}$ and $(k|x|)^{-3/2}$:

$$\frac{1}{(k|x| + ks')^{1/2}} \; - \; \frac{i}{8(k|x| + ks')^{3/2}} \; \approx \; \frac{1}{(k|x|)^{1/2}} \; - \; \frac{1}{(k|x|)^{3/2}} \left( \frac{i}{8} \; + \; \frac{s'}{2} \right) .$$

$$(4.2.12)$$

After integration, Eq. (4.2.11) then simplifies to

$$U^{(II)}(x, 0^+) \; \approx \; \frac{e^{-i\pi/4}}{\sqrt{2\pi}} \; \frac{e^{ik|x|}}{\sqrt{k|x|}} \left\{ 2\sin(kd/2) \; + \; \frac{i}{k|x|} \left[ \frac{kd}{2} \cos(kd/2) \; - \; \frac{5}{4}\sin(kd/2) \right] \right\} .$$

$$(4.2.13)$$

Hence, deep into the shadow region where $k(|x| - d/2) >> 1$, $U^{(II)}(x, 0^+)$ decays as $(k|x|)^{-1/2}$, except when the width $d$ of the slit is an integer multiple of the wavelength, in which case it decays as $(k|x|)^{-3/2}$.

### 4.2.2   Comparison of Results

We can now proceed with our comparison. The FD-TD numerical results for the electric field $E_y(x, 0^+)$ for the case of E-polarized illumination are shown in Figs. 4-13 and 4-14 together with the predictions of the two Rayleigh-Sommerfeld theories, $U^{(I)}(x, 0^+)$ and $U^{(II)}(x, 0^+)$. Figures 4-15 and 4-16 contain the corresponding plots for H-polarized illumination. Because the phase of $U^{(I)}(x, 0^+)$ in the slit is simply equal to zero, it is not shown in Figs. 4-13 through 4-16.

From the plots, we see that neither $U^{(I)}(x, 0^+)$ nor $U^{(II)}(x, 0^+)$ agree well with the numerical results for all values of $x$. Nevertheless, there is a reasonable agreement over a certain range of $x$-values. For E-polarization, $E_y(x, 0^+)$ resembles $U^{(II)}(x, 0^+)$ in the slit ($|x| \leq d/2$) and $E_y(x, 0^+) = U^{(I)}(x, 0^+) = 0$ in the shadow region ($|x| > d/2$). For H-polarization, $H_y(x, 0^+) = U^{(I)}(x, 0^+) = 1$ in the slit and $H_y(x,0^+)$ resembles $U^{(II)}(x,0^+)$ in the shadow region. Although the predictions of the usual Kirchhoff theory (see Section 2.1.5) are not shown in these plots, it is evident that, since for that theory $U^{(K)}(x,0^+) \ = \ \frac{1}{2}\Big[U^{(I)}(x,0^+) + U^{(II)}(x,0^+)\Big]$, the agreement between $U^{(K)}(x,0^+)$ and the numerical results is rather poor for all $x$-values.

As was explained in Section 4.2.1, when the width of the slit is equal to an integer number of wavelengths, i.e., when $d = m\lambda$ ($m$ being a positive integer), the field distribution $U^{(II)}(x,0^+)$ predicted by the Rayleigh-Sommerfeld theory of the second kind decays more rapidly into the shadow region. From the figures for H-polarization, Figs. 4-15 and 4-16, we see that this more rapid decay causes a substantial discrepancy between $\big|U^{(II)}(x,0^+)\big|$ and $\big|H_y(x,0^+)\big|$ for $|x| > d/2$ when $d = m\lambda$. However, since the agreement between $\big|U^{(II)}(x,0^+)\big|$ and $\big|H_y(x,0^+)\big|$ in the shadow region is excellent for $d = 0.2\lambda$, $0.5\lambda$ and $1.5\lambda$, one would expect a good agreement in that region in general, provided that $d \neq m\lambda$.

From the preceeding discussion, it is appears that, by modifying the Rayleigh-Sommerfeld boundary values, we can obtain a much better approximation to the exact results than is provided by the Rayleigh-Sommerfeld theories themselves. Specifically, we can introduce two sets of modified boundary values (denoted by superscripts $M1$ and $M2$):

$$
\begin{aligned}
U^{(M1)}(x,0^+) \ &= \ U^{(I)}(x,0^+) \\
&= \ 0 \qquad\qquad\qquad\qquad \text{for } |x| \ > \ \frac{d}{2} \ ,
\end{aligned}
$$

$$\text{(4.2.14a)}$$

$$U^{(M1)}(x,0^+) = U^{(II)}(x,0^+)$$

$$= \frac{1}{2}\left\{F\big[k(x + d/2)\big] + F\big[-k(x - d/2)\big]\right\} \quad \text{for } |x| \leq \frac{d}{2} ,$$

$$(4.2.14b)$$

and

$$U^{(M2)}(x,0^+) = U^{(II)}(x,0^+)$$

$$= \frac{1}{2}\left\{F\big[k(|x| + d/2)\big] - F\big[k(|x| - d/2)\big]\right\} \quad \text{for } |x| > \frac{d}{2} ,$$

$$(4.2.15a)$$

$$U^{(M2)}(x,0^+) = U^{(I)}(x,0^+)$$

$$= 1 \quad \text{for } |x| \leq \frac{d}{2} .$$

$$(4.2.15b)$$

The first set, Eqs. (4.2.14a) and (4.2.14b), should yield a reasonable approximation to the electric field $E_y(x,0^+)$ for E-polarization, whereas the second, Eqs. (4.2.15a) and (4.2.15b), should yield a reasonable approximation to the magnetic field $H_y(x,0^+)$ for H-polarization. However, as is evident from Figs. 4-13 through 4-16, the error between these modified boundary values and the exact results depends on the width of the slit and, for certain widths, it can be significant.

One can obtain new approximate theories of diffraction by propagating the boundary values $U^{(M1)}(x,0^+)$ and $U^{(M2)}(x,0^+)$ into the half-space $z \geq 0$. Such new theories, and their generalizations for three-dimensional electromagnetic fields, will be discussed in Section 5.1.

# CHAPTER 5

# NEW METHODS IN APERTURE DIFFRACTION

In this Chapter, we propose some new approaches for treating diffraction by an aperture in a thin screen. Except for a few preliminary results, these new methods are still relatively untested.

## 5.1 IMPROVED DIFFRACTION THEORIES FOR THE NEAR FIELD

We first consider two new theories of diffraction that are modifications of the scalar Rayleigh-Sommerfeld theories. These new theories should provide better approximations to near fields of apertures in Dirichlet-type and in Neumann-type screens[*]. We then discuss the analogous modifications of the $e$ and the $m$ electromagnetic theories of diffraction.

### 5.1.1 New Scalar Theories

We can generalize the modified boundary values $U^{(M1)}$ and $U^{(M2)}$ that were introduced in Section 4.2.2 so that they apply to any field incident from the half-space $z < 0$ and to an aperture of arbitrary shape (see Fig. 2-1 for notation). In the plane $z = 0^+$ immediately behind the screen, we then have, instead of Eqs. (4.2.14a) and (4.2.14b),

---

[*] As was already mentioned in Section 2.1.5, the Rayleigh-Sommerfeld theories are often used as approximations for fields diffracted by apertures in Dirichlet-type or Neumann-type screens, even though these theories were not originally intended for such screens.

$$U^{(M1)}(\boldsymbol{\rho}, 0^+) \; = \; 0 \qquad\qquad \text{on S} \; , \qquad\qquad (5.1.1a)$$

$$U^{(M1)}(\boldsymbol{\rho}, 0^+) \; = \; U^{(II)}(\boldsymbol{\rho}, 0^+) \qquad \text{in A} \qquad\qquad (5.1.1b)$$

and, instead of Eqs. (4.2.15a) and (4.2.15b),

$$U^{(M2)}(\boldsymbol{\rho}, 0^+) \; = \; U^{(II)}(\boldsymbol{\rho}, 0^+) \qquad \text{on S} \; , \qquad\qquad (5.1.2a)$$

$$U^{(M2)}(\boldsymbol{\rho}, 0^+) \; = \; U^{(\text{inc})}(\boldsymbol{\rho}, 0) \qquad \text{in A} \; . \qquad\qquad (5.1.2b)$$

Here $U^{(II)}(\boldsymbol{\rho}, 0^+)$ is simply the boundary value of the field given by the Rayleigh-Sommerfeld theory of the second kind [see Eq. (2.1.34)],

$$U^{(II)}(\boldsymbol{\rho}, 0^+) \; = \; -\frac{1}{2\pi} \int_A \frac{\partial U^{(\text{inc})}(\boldsymbol{\rho}', z')}{\partial z'} \bigg|_{z'=0} G(\boldsymbol{\rho} - \boldsymbol{\rho}', 0) \; d^2\rho' \; , \quad (5.1.3)$$

$$G(\boldsymbol{\rho}, z) \; = \; \frac{e^{ik\sqrt{\rho^2 + z^2}}}{\sqrt{\rho^2 + z^2}} \; . \qquad\qquad (5.1.4)$$

If we use Rayleigh's first diffraction formula (2.1.10) to propagate the modified boundary values $U^{(M1)}(\boldsymbol{\rho}, 0^+)$ and $U^{(M2)}(\boldsymbol{\rho}, 0^+)$ into the half-space $z \geq 0$, we obtain the expressions

$$U^{(M1)}(\boldsymbol{\rho}, z) \; = \; -\frac{1}{2\pi} \int_A U^{(II)}(\boldsymbol{\rho}, 0^+) \; \frac{\partial G(\boldsymbol{\rho} - \boldsymbol{\rho}', z)}{\partial z} \; d^2\rho' \qquad (5.1.5)$$

and

$$U^{(M2)}(\boldsymbol{\rho},z) \;=\; -\frac{1}{2\pi}\int_{\mathrm{A}} U^{(\mathrm{inc})}(\boldsymbol{\rho},0)\;\frac{\partial\,G(\boldsymbol{\rho}-\boldsymbol{\rho}',z)}{\partial z}\;d^2\rho'$$

$$-\frac{1}{2\pi}\int_{\mathrm{S}} U^{(II)}(\boldsymbol{\rho},0^+)\;\frac{\partial\,G(\boldsymbol{\rho}-\boldsymbol{\rho}',z)}{\partial z}\;d^2\rho'\;. \tag{5.1.6}$$

These are the two new approximate diffraction formulas. It should be noted that their average is equal to the field given by the Kirchhoff theory, i.e.,

$$\frac{1}{2}\left[U^{(M1)}(\boldsymbol{\rho},z)\;+\;U^{(M2)}(\boldsymbol{\rho},z)\right]=\frac{1}{2}\left[U^{(I)}(\boldsymbol{\rho},z)\;+\;U^{(II)}(\boldsymbol{\rho},z)\right]$$

$$=\;U^{(K)}(\boldsymbol{\rho},z)\;. \tag{5.1.7}$$

It is beyond the scope of this work to examine the full ramifications of these new approximate scalar theories or to determine how well, in general, their predictions agree with exact results. Nevertheless, from our discussion in Section 4.2.2, we would expect the first (second) modified theory $U^{(M1)}(\boldsymbol{\rho},z)$ $\left[U^{(M2)}(\boldsymbol{\rho},z)\right]$ to provide a reasonable approximation to the field behind a Dirichlet-type (Neumann-type) screen. Furthermore, the new theories agree with the exact boundary conditions over a portion of the aperture plane: $U^{(II)}(\boldsymbol{\rho},0^+)\;=\;U_D(\boldsymbol{\rho},0)=0$ on S and $U^{(M2)}(\boldsymbol{\rho},0^+)\;=U_N(\boldsymbol{\rho},0)=U^{(\mathrm{inc})}(\boldsymbol{\rho},0)$ in A.

### 5.1.2   New Electromagnetic Theories

We can use an analogous approach to obtain two new electromagnetic theories of diffraction from the *m* and *e*-theories discussed in Section 2.2.3. These new electromagnetic theories, which we denote by the superscripts *M*1 and *M*2, should yield a better approximation to the diffracted fields behind an aperture in a perfectly conducting screen.

One of the major inadequacies of the *e*-theory is that it gives a non-zero value for the tangential electric field on the surface of the perfect conductor. We can remedy this deficiency by introducing the modified boundary values

$$\hat{\mathbf{z}} \times \mathbf{E}^{(M1)}(\boldsymbol{\rho}, 0^+) \; = \; 0 \qquad\qquad \text{on S} \qquad (5.1.8a)$$

$$\hat{\mathbf{z}} \times \mathbf{E}^{(M1)}(\boldsymbol{\rho}, 0^+) \; = \; \hat{\mathbf{z}} \times \mathbf{E}^{(e)}(\boldsymbol{\rho}, 0^+) \qquad \text{in A} \; , \qquad (5.1.8b)$$

where $\mathbf{E}^{(e)}(\boldsymbol{\rho}, 0^+)$ is the boundary value of the electric field in the *e*-theory,

$$\mathbf{E}^{(e)}(\boldsymbol{\rho}, 0^+) \; = \; \lim_{z \to 0^+} \frac{i}{2\pi k} \nabla \times \nabla \times \int_A \Big[ \hat{\mathbf{z}} \times \mathbf{H}^{(\text{inc})}(\boldsymbol{\rho}', 0) \Big] G(\boldsymbol{\rho} - \boldsymbol{\rho}', z) \, d^2 \rho' \; .$$

$$(5.1.9)$$

We can now apply Eqs. (2.2.11a) and (2.2.11b) to determine the electric and magnetic fields predicted by the first modified theory for the half-space $z \geq 0$. They are given by the formulas

$$\mathbf{E}^{(M1)}(\boldsymbol{\rho}, z) \; = \; \frac{1}{2\pi} \nabla \times \int_A \Big[ \hat{\mathbf{z}} \times \mathbf{E}^{(e)}(\boldsymbol{\rho}', 0^+) \Big] G(\boldsymbol{\rho} - \boldsymbol{\rho}', z) \, d^2 \rho' \; , \quad (5.1.10a)$$

$$\mathbf{H}^{(M1)}(\boldsymbol{\rho}, z) \; = \; -\frac{i}{2\pi k} \nabla \times \nabla \times \int_A \Big[ \hat{\mathbf{z}} \times \mathbf{E}^{(e)}(\boldsymbol{\rho}', 0^+) \Big] G(\boldsymbol{\rho} - \boldsymbol{\rho}', z) \, d^2 \rho' \; .$$

$$(5.1.10b)$$

It is possible to improve the *m*-theory in a similar manner by setting the *x* and *y*-components of the magnetic field in the aperture equal to their correct values, namely, equal to those of the incident magnetic field (see Section 2.2.1),

$$\hat{\mathbf{z}} \times \mathbf{H}^{(M2)}(\boldsymbol{\rho}, 0^+) \; = \; \hat{\mathbf{z}} \times \mathbf{H}^{(\text{inc})}(\boldsymbol{\rho}, 0) \qquad \text{in A} \; , \qquad (5.1.11a)$$

$$\hat{\mathbf{z}} \times \mathbf{H}^{(M2)}(\boldsymbol{\rho},0^+) \; = \; \hat{\mathbf{z}} \times \mathbf{H}^{(m)}(\boldsymbol{\rho},0^+) \qquad \text{on S .} \qquad (5.1.11b)$$

In these formulas, $\mathbf{H}^{(M2)}(\boldsymbol{\rho},0^+)$ is the boundary value of the magnetic field given by the *m*-theory,

$$\mathbf{H}^{(m)}(\boldsymbol{\rho},0^+) \; = \; \lim_{z \to 0^+} \; -\frac{i}{2\pi k} \nabla \times \nabla \times \int_A \Big[ \hat{\mathbf{z}} \times \mathbf{E}^{(\text{inc})}(\boldsymbol{\rho}',0) \Big] G(\boldsymbol{\rho}-\boldsymbol{\rho}',z) \, d^2\rho' \; .$$

$$(5.1.12)$$

By again making use of Eqs. (2.2.11a) and (2.2.11b), we find that the electric and magnetic fields predicted by this second modified theory for the half-space $z \geq 0$ are

$$\mathbf{H}^{(M2)}(\boldsymbol{\rho},z) \; = \; \frac{1}{2\pi} \nabla \times \int_A \Big[ \hat{\mathbf{z}} \times \mathbf{H}^{(\text{inc})}(\boldsymbol{\rho}',0) \Big] G(\boldsymbol{\rho}-\boldsymbol{\rho}',z) \, d^2\rho'$$

$$+ \; \frac{1}{2\pi} \nabla \times \int_S \Big[ \hat{\mathbf{z}} \times \mathbf{H}^{(m)}(\boldsymbol{\rho}',0^+) \Big] G(\boldsymbol{\rho}-\boldsymbol{\rho}',z) \, d^2\rho' \; ,$$

$$(5.1.13a)$$

$$\mathbf{E}^{(M2)}(\boldsymbol{\rho},z) \; = \; \frac{i}{2\pi k} \nabla \times \nabla \times \int_A \Big[ \hat{\mathbf{z}} \times \mathbf{H}^{(\text{inc})}(\boldsymbol{\rho}',0) \Big] G(\boldsymbol{\rho}-\boldsymbol{\rho}',z) \, d^2\rho'$$

$$+ \; \frac{i}{2\pi k} \nabla \times \nabla \times \int_S \Big[ \hat{\mathbf{z}} \times \mathbf{H}^{(m)}(\boldsymbol{\rho}',0^+) \Big] G(\boldsymbol{\rho}-\boldsymbol{\rho}',z) \, d^2\rho' \; .$$

$$(5.1.13b)$$

Even though both of the new electromagnetic diffraction theories satisfy Maxwell's equation, they obviously give very different predictions for the fields transmitted through an aperture in a perfectly conducting screen. It is still neccessary to determine which one of the two theories is better and under what conditions.

It should be pointed out that for both scalar and electromagnetic fields the new theories are more difficult to employ than the usual approximate theories because they require additional integrations. A comparison with exact results for several different cases is therefore still needed in order to decide whether the additional computation efforts are warranted.

## 5.2 ITERATIVE FOURIER-BASED ALGORITHM FOR APERTURE DIFFRACTION

We will now describe a new algorithm for solving rigorous diffraction problems that involve apertures in infinitely thin screens*. For these types of problems, this algorithm requires far less computer time and memory than the finite-difference time-domain method (for references see Appendix E), and also has some advantages over the method of moments[2,3] because it does not require matrix inversion. It can be implemented very effectively with fast Fourier transforms (FFT's).

When the angular spectrum representation is used to analyze rigorous aperture diffraction, the mixed boundary value problem is posed in terms of dual integral equations for the angular spectrum amplitude of the transmitted field. The new algorithm is an iterative solution of such integral equations which bears some resemblance to the iterative algorithms used in phase retrieval.[4-7] Here, we will discuss its implementation for scalar fields diffracted by apertures in either Dirichlet-type or Neumann-type screens, and for electromagnetic fields diffracted by apertures

---

in perfectly conducting screens. It may be possible to adapt the algorithm to other situations as well.

### 5.2.1 Dual Scalar Integral Equations

We begin by deriving the dual integral equations[8-13] for scalar fields. It will become apparent later that, in fact, these integral equations can be written down by inspection with the help of Eqs. (2.1.8) and (2.1.9). Nevertheless, the derivation is instructive because it shows how to compute the reflected field once the transmitted field has been determined, and because it provides a proof of Eqs. (2.1.8) and (2.1.9).

Let us use the angular spectrum representation (see Section 2.1.3) for the incident and the reflected fields in the half-space $z \leq 0$,

$$U^{(\text{inc})}(\boldsymbol{\rho},z) \; = \; \int a^{(\text{inc})}(\mathbf{u}_\perp) \, e^{ik\mathbf{u}_\perp \cdot \boldsymbol{\rho}} \, e^{iku_z z} \, d^2 u_\perp \; , \qquad (5.2.1\text{a})$$

$$U^{(\text{rfl})}(\boldsymbol{\rho},z) \; = \; \int a^{(\text{rfl})}(\mathbf{u}_\perp) \, e^{ik\mathbf{u}_\perp \cdot \boldsymbol{\rho}} \, e^{-iku_z z} \, d^2 u_\perp \qquad (5.2.1\text{b})$$

and for the transmitted field in the half-space $z \geq 0$,

$$U^{(\text{trn})}(\boldsymbol{\rho},z) \; = \; \int a^{(\text{trn})}(\mathbf{u}_\perp) \, e^{ik\mathbf{u}_\perp \cdot \boldsymbol{\rho}} \, e^{iku_z z} \, d^2 u_\perp \; , \qquad (5.2.1\text{c})$$

$$u_z \; = \; \sqrt{1 - u_\perp^2} \qquad \text{for } u_\perp \leq 1 \; , \qquad (5.2.2\text{a})$$

$$u_z \; = \; i\sqrt{u_\perp^2 - 1} \qquad \text{for } u_\perp > 1 \; . \qquad (5.2.2\text{b})$$

The continuity conditions (2.1.4) and (2.1.5) require that

$$U^{(\text{inc})}(\boldsymbol{\rho},0^-) \; + \; U^{(\text{rfl})}(\boldsymbol{\rho},0^-) \; = \; U^{(\text{trn})}(\boldsymbol{\rho},0^+) \qquad\qquad \text{in A}$$

$$(5.2.3)$$

and

$$\frac{\partial U^{(\mathrm{inc})}(\boldsymbol{\rho},z)}{\partial z}\bigg|_{z=0^-} + \frac{\partial U^{(\mathrm{rfl})}(\boldsymbol{\rho},z)}{\partial z}\bigg|_{z=0^-} = \frac{\partial U^{(\mathrm{trn})}(\boldsymbol{\rho},z)}{\partial z}\bigg|_{z=0^+} \quad \text{in A .}$$

$$(5.2.4)$$

Therefore, in the aperture, the following integral relations must be fulfilled by the angular spectrum amplitudes $a^{(\mathrm{inc})}(\mathbf{u}_\perp)$, $a^{(\mathrm{rfl})}(\mathbf{u}_\perp)$ and $a^{(\mathrm{trn})}(\mathbf{u}_\perp)$:

$$\int \left[ a^{(\mathrm{inc})}(\mathbf{u}_\perp) + a^{(\mathrm{rfl})}(\mathbf{u}_\perp) \right] e^{ik\mathbf{u}_\perp \cdot \boldsymbol{\rho}}\, d^2u_\perp = \int a^{(\mathrm{trn})}(\mathbf{u}_\perp)\, e^{ik\mathbf{u}_\perp \cdot \boldsymbol{\rho}}\, d^2u_\perp \qquad \text{in A ,}$$

$$(5.2.5)$$

$$\int u_z \left[ a^{(\mathrm{inc})}(\mathbf{u}_\perp) - a^{(\mathrm{rfl})}(\mathbf{u}_\perp) \right] e^{ik\mathbf{u}_\perp \cdot \boldsymbol{\rho}}\, d^2u_\perp = \int u_z a^{(\mathrm{trn})}(\mathbf{u}_\perp)\, e^{ik\mathbf{u}_\perp \cdot \boldsymbol{\rho}}\, d^2u_\perp \quad \text{in A .}$$

$$(5.2.6)$$

*(a) Dirichlet-Type Screen*

On the surface of a Dirichlet-type screen, since the field is equal to zero, the angular spectrum amplitudes must also satisfy the relations

$$\int \left[ a^{(\mathrm{inc})}(\mathbf{u}_\perp) + a^{(\mathrm{rfl})}(\mathbf{u}_\perp) \right] e^{ik\mathbf{u}_\perp \cdot \boldsymbol{\rho}}\, d^2u_\perp = 0 \qquad \text{on S ,} \quad (5.2.7)$$

$$\int a^{(\mathrm{trn})}(\mathbf{u}_\perp)\, e^{ik\mathbf{u}_\perp \cdot \boldsymbol{\rho}}\, d^2u_\perp = 0 \qquad\qquad \text{on S .} \quad (5.2.8)$$

If we add Eq. (5.2.7) to the left-hand side of Eq. (5.2.5), add Eq. (5.2.8) to the right-hand side and take the Fourier transform of the resulting equation, we find that the angular spectrum amplitude of the reflected field can be obtained directly from the angular spectrum amplitudes of the incident and the transmitted fields using the formula

$$a^{(\text{rfl})}(\mathbf{u}_\perp) = a^{(\text{trn})}(\mathbf{u}_\perp) - a^{(\text{inc})}(\mathbf{u}_\perp) \ . \qquad (5.2.9)$$

Then, according to Eqs. (5.2.6) and (5.2.8), for a Dirichlet-type screen the angular spectrum amplitude of the transmitted field obeys the dual integral equations

$$\int a^{(\text{trn})}(\mathbf{u}_\perp) \, e^{ik\mathbf{u}_\perp \cdot \boldsymbol{\rho}} \, d^2 u_\perp = 0 \qquad\qquad \text{on } \mathrm{S}\,,$$

$$\int u_z \, a^{(\text{trn})}(\mathbf{u}_\perp) \, e^{ik\mathbf{u}_\perp \cdot \boldsymbol{\rho}} \, d^2 u_\perp = -\frac{i}{k} \left. \frac{\partial U^{(\text{inc})}(\boldsymbol{\rho}, z)}{\partial z} \right|_{z=0} \qquad \text{in } \mathrm{A}\,. \qquad (5.2.10)$$

The first of these equations ensures that the transmitted field is zero on the screen and the second that the $z$-derivative of the transmitted field in the aperture is equal to the $z$-derivative of the incident field. The second equation is therefore equivalent to Eq. (2.1.8).

*(b) Neumann-Type Screen*

On a Neumann-type screen, since the normal derivative of the field is identically zero, in lieu of Eqs. (5.2.7) and (5.2.8) we have

$$\int u_z \Big[a^{(\text{inc})}(\mathbf{u}_\perp) - a^{(\text{rfl})}(\mathbf{u}_\perp)\Big] e^{ik\mathbf{u}_\perp \cdot \boldsymbol{\rho}} \, d^2 u_\perp = 0 \qquad \text{on } \mathrm{S}\,, \qquad (5.2.11)$$

$$\int u_z \, a^{(\text{trn})}(\mathbf{u}_\perp) \, e^{ik\mathbf{u}_\perp \cdot \boldsymbol{\rho}} \, d^2 u_\perp = 0 \qquad\qquad \text{on } \mathrm{S}\,. \qquad (5.2.12)$$

From these relations and Eq. (5.2.6), we find that $a^{(\text{rfl})}(\mathbf{u}_\perp)$ is now related to $a^{(\text{inc})}(\mathbf{u}_\perp)$ and $a^{(\text{trn})}(\mathbf{u}_\perp)$ by the expression

$$a^{(\text{rfl})}(\mathbf{u}_\perp) = a^{(\text{inc})}(\mathbf{u}_\perp) - a^{(\text{trn})}(\mathbf{u}_\perp) \qquad\qquad (5.2.13)$$

The dual integral equations for a Neumann-type screen are then

$$\left.\begin{array}{ll} \int u_z\, a^{(\mathrm{trn})}(\mathbf{u}_\perp)\, e^{ik\mathbf{u}_\perp\bullet\boldsymbol{\rho}}\, d^2u_\perp \;=\; 0 & \text{on S}\,, \\[2ex] \int a^{(\mathrm{trn})}(\mathbf{u}_\perp)\, e^{ik\mathbf{u}_\perp\bullet\boldsymbol{\rho}}\, d^2u_\perp \;=\; U^{(\mathrm{inc})}(\boldsymbol{\rho},0) & \text{in A}. \end{array}\right\} \qquad (5.2.14)$$

In this case, the first equation ensures that the normal derivative of the transmitted field is zero on the screen and the second that the transmitted field in the aperture is equal to the incident field, as required by Eq. (2.1.9).

### 5.2.2   Iterative Fourier-Based Algorithm for Scalar Fields

An iterative approach for solving both Eq. (5.2.10) and Eq. (5.2.14) becomes evident from the Fourier transform relations between the field in the plane $z = 0^+$ and the angular spectrum amplitude $a^{(\mathrm{trn})}(\mathbf{u}_\perp)$ (see Section 2.1.3),

$$U^{(0)}(\boldsymbol{\rho}) \;=\; \int a^{(\mathrm{trn})}(\mathbf{u}_\perp)\, e^{ik\mathbf{u}_\perp\bullet\boldsymbol{\rho}}\, d^2u_\perp , \qquad (5.2.15a)$$

$$a^{(\mathrm{trn})}(\mathbf{u}_\perp) \;=\; \left(\frac{k}{2\pi}\right)^2 \int U^{(0)}(\boldsymbol{\rho}')\, e^{-ik\mathbf{u}_\perp\bullet\boldsymbol{\rho}'}\, d^2\rho' , \qquad (5.2.15b)$$

and the analogous relations between the $z$-derivative of the field in that plane and $a^{(\mathrm{trn})}(\mathbf{u}_\perp)$,

$$U_z^{(0)}(\boldsymbol{\rho}) \;=\; ik\, u_z \int a^{(\mathrm{trn})}(\mathbf{u}_\perp)\, e^{ik\mathbf{u}_\perp\bullet\boldsymbol{\rho}}\, d^2u_\perp , \qquad (5.2.16a)$$

$$a^{(\mathrm{trn})}(\mathbf{u}_\perp) \;=\; -\frac{i}{ku_z}\left(\frac{k}{2\pi}\right)^2 \int U_z^{(0)}(\boldsymbol{\rho})\, e^{-ik\mathbf{u}_\perp\bullet\boldsymbol{\rho}'}\, d^2\rho' . \qquad (5.2.16b)$$

Here we have introduced the compact notation

$$U^{(0)}(\boldsymbol{\rho}) \equiv U^{(\mathrm{trn})}(\boldsymbol{\rho},0^+) \; , \; U_z^{(0)}(\boldsymbol{\rho}) \equiv \frac{\partial U^{(\mathrm{trn})}(\boldsymbol{\rho},z)}{\partial z}\bigg|_{z=0^+} . \quad (5.2.17)$$

The iterative algorithm then consists of the following steps:

0) Assume initially that $U_z^{(0)}(\boldsymbol{\rho})$ is given by the boundary values of the Rayleigh-Sommerfeld theory of the second kind (RS II).

1) Use Eq. (5.2.16b) to calculate the angular spectrum amplitude $a^{(\mathrm{trn})}(\mathbf{u}_\perp)$.

2) Use Eq. (5.2.15a) to calculate $U^{(0)}(\boldsymbol{\rho})$. Over the portion of the aperture plane where the value of $U^{(0)}(\boldsymbol{\rho})$ is prescribed, set it equal to that value.

3) Use Eq. (5.2.15b) to calculate $a^{(\mathrm{trn})}(\mathbf{u}_\perp)$.

4) Use Eq. (5.2.16a) to calculate $U_z^{(0)}(\boldsymbol{\rho})$. Over the portion of the aperture plane where the value of $U_z^{(0)}(\boldsymbol{\rho})$ is prescribed, set it equal to that value. Go back to step #1.

The necessary calculations for Dirichlet-type screens are shown explicitly in Fig. 5-1 and those for Neumann-type screens in Fig. 5-2. All the integrations can be performed with FFT's. The algorithm can also be started at step #3 with $U^{(0)}(\boldsymbol{\rho})$ given by the boundary values of the Rayleigh-Sommerfeld theory of the first kind (RS I). Other initial values of $U_z^{(0)}(\boldsymbol{\rho})$ or $U^{(0)}(\boldsymbol{\rho})$ can also be used. The ones just mentioned are the simplest.

The preceding steps do not guarantee the correct result since, without edge conditions, the rigorous diffraction problem does not necessarily have a unique solution (see Section 2.1.1). However, it is relatively simple to introduce a small "loss" into the algorithm so that the solution that obeys edge conditions is chosen.

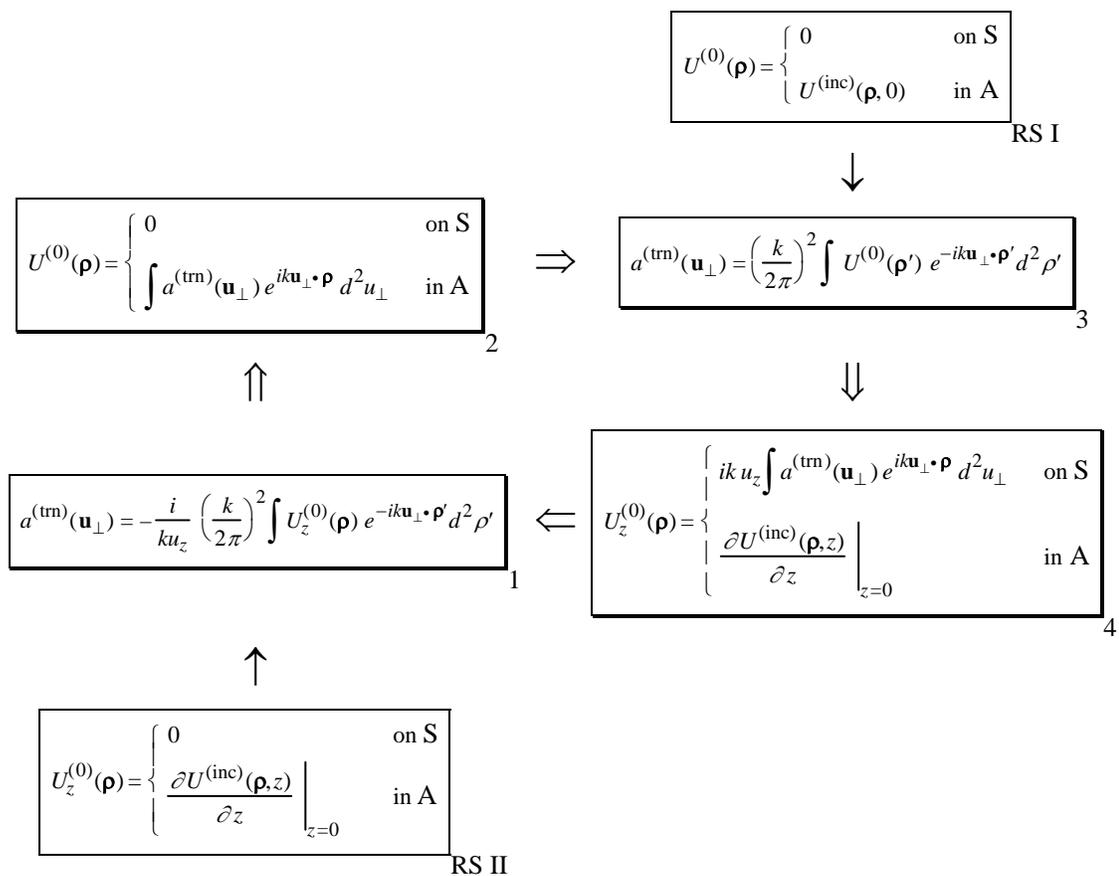

**Figure 5-1** Block diagram of the iterative Fourier-based algorithm for a Dirichlet-type screen.

This loss can be, for example, a weakly attenuating filter applied to the angular spectrum amplitude $a^{(\text{trn})}(\mathbf{u}_\perp)$ for large values of $u_\perp$. The filtering favors the solution with the smallest values of $\left| a^{(\text{trn})}(\mathbf{u}_\perp) \right|$ for large $u_\perp$. Hence, in the space domain, the algorithm favors a form for $U^{(0)}(\boldsymbol{\rho})$ that is free of singularities, which is precisely the one that satisfies edge conditions. Even with this type of filtering, there no assurance that, as the number of iterations is increased, the results generated by the algorithm will converge to the correct solution. The algorithm could simply oscillate between several different results without ever converging. Nevertheless, the numerical results presented in the next section suggest that, in fact, the algorithm converges to the correct solution very rapidly.

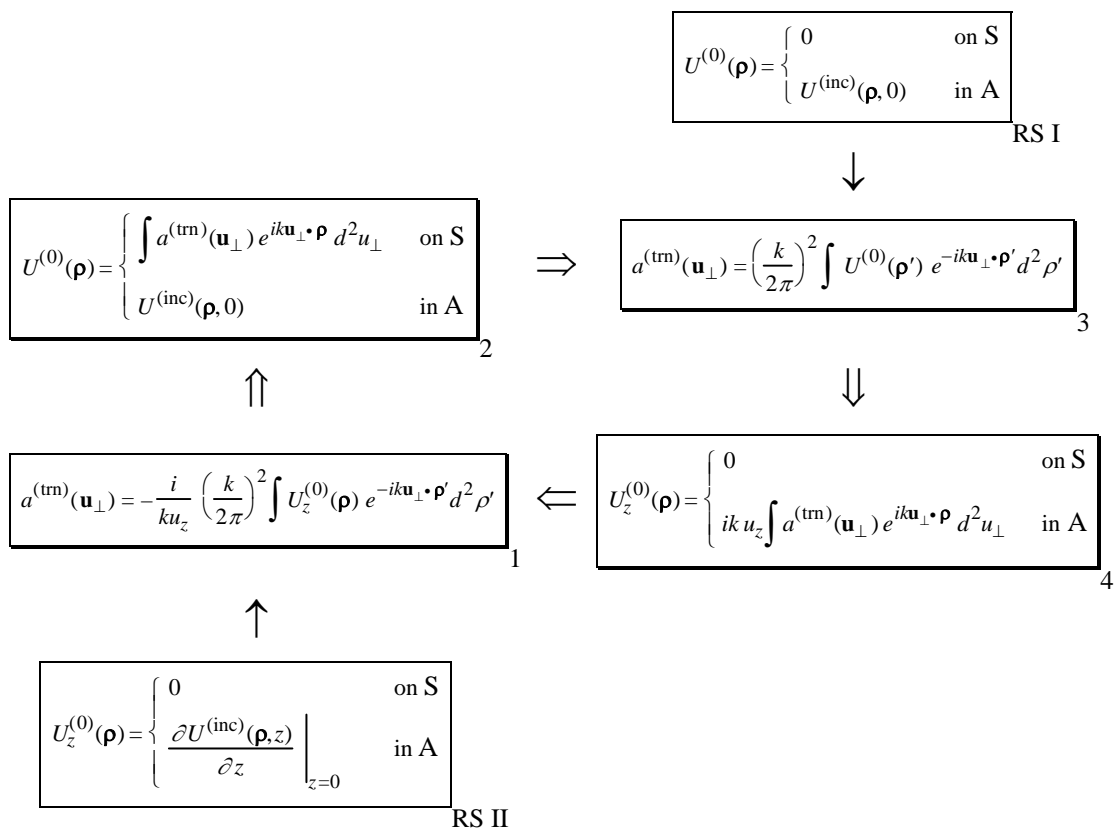

**Figure 5-2** Block diagram of the iterative Fourier-based algorithm for a Neumann-type screen.

It should be pointed out that step #1 of the algorithm could create difficulties because there is a singularity present on the circle $u_\perp = 1$ ($u_z = 0$). If the algorithm is implemented with FFT's, the singularity can be avoided by choosing the values of $\mathbf{u}_\perp$ where $a^{(\mathrm{trn})}(\mathbf{u}_\perp)$ is sampled so that they are not too close to $u_\perp = 1$.

Let us examine step #2 of the first iteration through the algorithm when the starting values are the boundary values of the Rayleigh-Sommerfeld theory of the second kind. For a Dirchlet-type screen (see Fig. 5-1), the field $U^{(0)}(\rho)$ at this step is simply equal to the aperture plane field $U^{(M1)}(\rho, 0^+)$ of the first modified theory of Section 5.1.1 and, for a Neumann-type screen (see Fig. 5-2), it is equal to the field $U^{(M2)}(\rho, 0^+)$ of the second modified theory. Therefore, these modified theories can be viewed as the first in a series of iterative improvements to the Rayleigh-Sommerfeld theory of the second kind.

### 5.2.3 Numerical Results

In order to test the new algorithm, we applied it to the simple two-dimensional diffraction problem considered in Chapters 3 and 4: a slit illuminated by a unit amplitude, normally incident, plane wave. The necessary one-dimensional integrations were all performed with FFT's and a weak exponential filter of the form $\exp\{-0.1(u_\perp - 25)\}$ was applied to $a^{(\text{trn})}$ for $u_\perp > 25$. The numerical results for slits of width $d = 0.5\lambda$ and $d = 2\lambda$ in a Dirichlet-type screen are shown in Fig. 5-3. By comparing these plots with the corresponding plots in Figs. 4-13 and 4-14, we see that the algorithm does produce the correct solution. In fact, it converges to this solution very rapidly.

### 5.2.4 Dual Integral Equations and the Iterative Fourier-Based Algorithm for Electromagnetic Fields

To obtain the dual electromagnetic integral equations, we use an approach similar to that for scalar fields. We represent the incident, reflected and transmitted electric fields by angular spectra of plane waves (see Section 2.2.2):

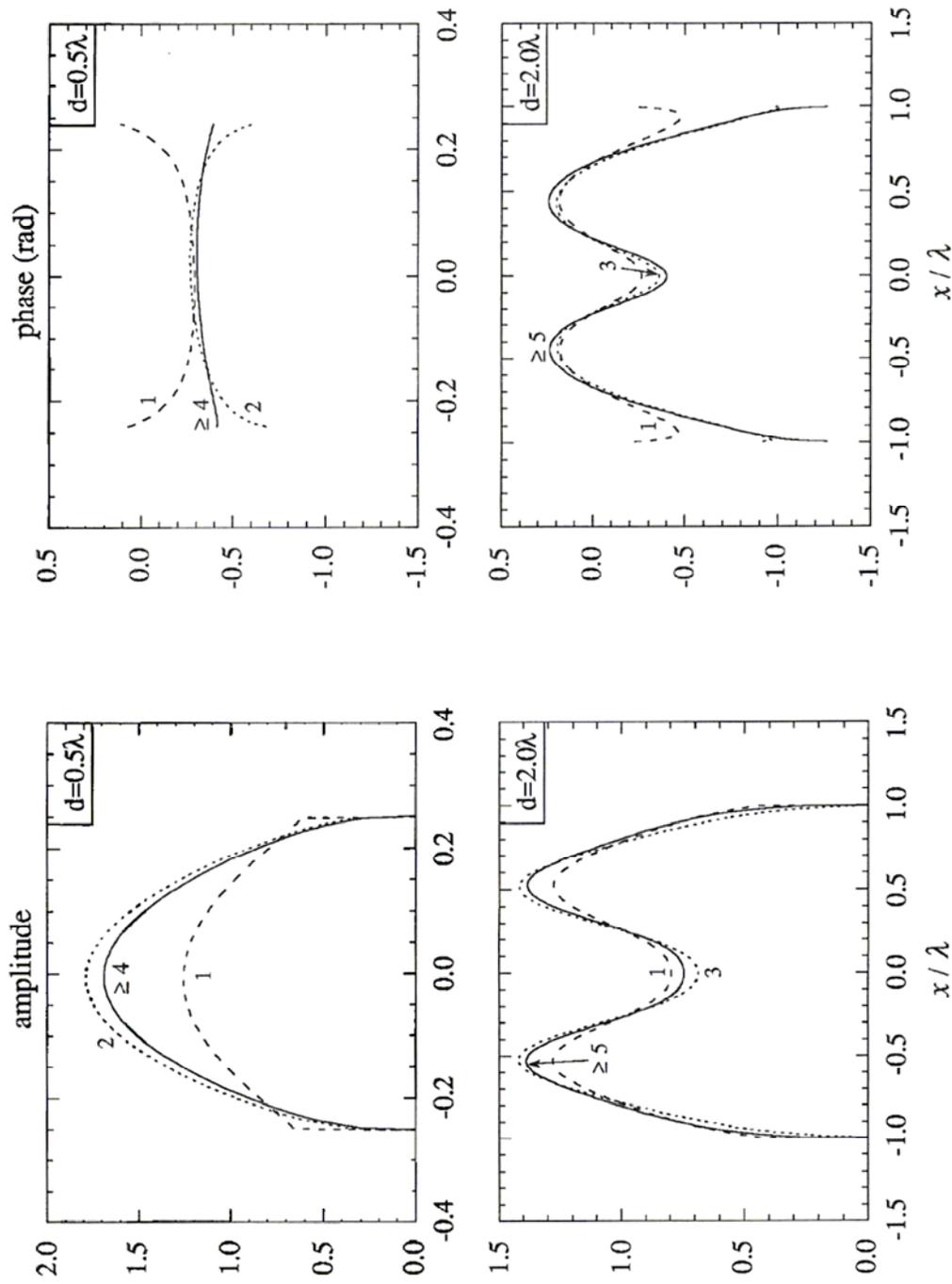

**Figure 5-3** Aperture field obtained from iterative Fourier-based algorithm for the case of a unit amplitude, normally incident, plane wave diffracted by a slit in a Dirichlet-type screen. Number of iterations is shown next to curves.

$$\mathbf{E}^{(\text{inc})}(\boldsymbol{\rho}, z) = \int \left[ \mathbf{e}_\perp^{(\text{inc})}(\mathbf{u}_\perp) - \hat{\mathbf{z}} \frac{\mathbf{u}_\perp \cdot \mathbf{e}_\perp^{(\text{inc})}(\mathbf{u}_\perp)}{u_z} \right] e^{ik\mathbf{u}_\perp \cdot \boldsymbol{\rho}} e^{iku_z z} \, d^2 u_\perp \; ,$$

$$(5.2.18a)$$

$$\mathbf{E}^{(\text{rfl})}(\boldsymbol{\rho}, z) = \int \left[ \mathbf{e}_\perp^{(\text{rfl})}(\mathbf{u}_\perp) + \hat{\mathbf{z}} \frac{\mathbf{u}_\perp \cdot \mathbf{e}_\perp^{(\text{rfl})}(\mathbf{u}_\perp)}{u_z} \right] e^{ik\mathbf{u}_\perp \cdot \boldsymbol{\rho}} e^{-iku_z z} \, d^2 u_\perp \; ,$$

$$(5.2.18b)$$

$$\mathbf{E}^{(\text{trn})}(\boldsymbol{\rho}, z) = \int \left[ \mathbf{e}_\perp^{(\text{trn})}(\mathbf{u}_\perp) - \hat{\mathbf{z}} \frac{\mathbf{u}_\perp \cdot \mathbf{e}_\perp^{(\text{trn})}(\mathbf{u}_\perp)}{u_z} \right] e^{ik\mathbf{u}_\perp \cdot \boldsymbol{\rho}} e^{iku_z z} \, d^2 u_\perp \; .$$

$$(5.2.18c)$$

The transverse vectorial angular spectrum amplitude $\mathbf{e}_\perp^{(\text{inc})}(\mathbf{u}_\perp)$ is related to the transverse electric field $\mathbf{E}_\perp^{(\text{inc})} = \mathbf{E}^{(\text{inc})} - \hat{\mathbf{z}} \cdot \mathbf{E}^{(\text{inc})}$ in the plane $z = 0^+$ by the Fourier transform

$$\mathbf{e}_\perp^{(\text{inc})}(\mathbf{u}_\perp) = \left( \frac{k}{2\pi} \right)^2 \int \mathbf{E}_\perp^{(\text{inc})}(\boldsymbol{\rho}', 0^+) \, e^{-ik\mathbf{u}_\perp \cdot \boldsymbol{\rho}'} d^2 \rho' \; , \qquad (5.2.19)$$

and analogous expressions relate $\mathbf{e}_\perp^{(\text{rfl})}(\mathbf{u}_\perp)$ and $\mathbf{e}_\perp^{(\text{trn})}(\mathbf{u}_\perp)$ to $\mathbf{E}_\perp^{(\text{rfl})}(\boldsymbol{\rho}, 0^+)$ and $\mathbf{E}_\perp^{(\text{trn})}(\boldsymbol{\rho}, 0^+)$. Furthermore, the magnetic fields $\mathbf{H}^{(\text{inc})}$, $\mathbf{H}^{(\text{rfl})}$ and $\mathbf{H}^{(\text{trn})}$ are given by equations identical to (5.2.18a)-(5.2.18c) with the substitutions

$$\mathbf{E}^{(\text{inc})} \Rightarrow \mathbf{H}^{(\text{inc})} \; , \; \; \mathbf{E}^{(\text{rfl})} \Rightarrow \mathbf{H}^{(\text{rfl})} \; , \; \; \mathbf{E}^{(\text{trn})} \Rightarrow \mathbf{H}^{(\text{trn})} \qquad (5.2.20)$$

and

$$\mathbf{e}^{(\text{inc})} \Rightarrow \mathbf{h}_\perp^{(\text{inc})} \; , \; \; \mathbf{e}_\perp^{(\text{rfl})} \Rightarrow \mathbf{h}_\perp^{(\text{rfl})} \; , \; \; \mathbf{e}_\perp^{(\text{trn})} \Rightarrow \mathbf{h}_\perp^{(\text{trn})} \; . \qquad (5.2.21)$$

To ensure the continuity of the electric and magnetic fields in the aperture, the transverse vectorial angular spectrum amplitudes must obey the two conditions

$$\int \left\{ \mathbf{e}_\perp^{(inc)}(\mathbf{u}_\perp) + \mathbf{e}_\perp^{(rfl)}(\mathbf{u}_\perp) - \frac{\hat{\mathbf{z}}}{u_z} \mathbf{u}_\perp \cdot \left[ \mathbf{e}_\perp^{(inc)}(\mathbf{u}_\perp) - \mathbf{e}_\perp^{(rfl)}(\mathbf{u}_\perp) \right] \right\} e^{ik\mathbf{u}_\perp \cdot \boldsymbol{\rho}} \, d^2 u_\perp$$

$$= \int \left[ \mathbf{e}_\perp^{(trn)}(\mathbf{u}_\perp) - \hat{\mathbf{z}} \frac{\mathbf{u}_\perp \cdot \mathbf{e}_\perp^{(trn)}(\mathbf{u}_\perp)}{u_z} \right] e^{ik\mathbf{u}_\perp \cdot \boldsymbol{\rho}} \, d^2 u_\perp \qquad \text{in A}$$

$$(5.2.22)$$

and

$$\int \left\{ \mathbf{h}_\perp^{(inc)}(\mathbf{u}_\perp) + \mathbf{h}_\perp^{(rfl)}(\mathbf{u}_\perp) - \frac{\hat{\mathbf{z}}}{u_z} \mathbf{u}_\perp \cdot \left[ \mathbf{h}_\perp^{(inc)}(\mathbf{u}_\perp) - \mathbf{h}_\perp^{(rfl)}(\mathbf{u}_\perp) \right] \right\} e^{ik\mathbf{u}_\perp \cdot \boldsymbol{\rho}} \, d^2 u_\perp$$

$$= \int \left[ \mathbf{h}_\perp^{(trn)}(\mathbf{u}_\perp) - \hat{\mathbf{z}} \frac{\mathbf{u}_\perp \cdot \mathbf{h}_\perp^{(trn)}(\mathbf{u}_\perp)}{u_z} \right] e^{ik\mathbf{u}_\perp \cdot \boldsymbol{\rho}} \, d^2 u_\perp \qquad \text{in A}.$$

$$(5.2.23)$$

Furthermore, since on the surface of the perfect conductor both the tangential electric field and the normal magnetic field must be zero, the angular spectrum amplitudes must also satisfy

$$\int \left[ \mathbf{e}_\perp^{(inc)}(\mathbf{u}_\perp) + \mathbf{e}_\perp^{(rfl)}(\mathbf{u}_\perp) \right] e^{ik\mathbf{u}_\perp \cdot \boldsymbol{\rho}} \, d^2 u_\perp = 0 \qquad \text{on S}, \quad (5.2.24)$$

$$\int \mathbf{e}_\perp^{(trn)}(\mathbf{u}_\perp) \, e^{ik\mathbf{u}_\perp \cdot \boldsymbol{\rho}} \, d^2 u_\perp = 0 \qquad \text{on S}, \quad (5.2.25)$$

$$\int \frac{1}{u_z} \mathbf{u}_\perp \cdot \left[ \mathbf{h}_\perp^{(inc)}(\mathbf{u}_\perp) + \mathbf{h}_\perp^{(rfl)}(\mathbf{u}_\perp) \right] e^{ik\mathbf{u}_\perp \cdot \boldsymbol{\rho}} \, d^2 u_\perp = 0 \qquad \text{on S}, \quad (5.2.26)$$

$$\int \frac{\mathbf{u}_\perp \cdot \mathbf{h}_\perp^{(trn)}(\mathbf{u}_\perp)}{u_z} \, e^{ik\mathbf{u}_\perp \cdot \boldsymbol{\rho}} \, d^2 u_\perp = 0 \qquad \text{on S}. \quad (5.2.27)$$

From Eqs. (5.2.22) through (5.2.27), it can readily be shown that the following relationships hold:

$$\mathbf{e}_\perp^{(\mathrm{rfl})}(\mathbf{u}_\perp) = \mathbf{e}_\perp^{(\mathrm{trn})}(\mathbf{u}_\perp) - \mathbf{e}_\perp^{(\mathrm{inc})}(\mathbf{u}_\perp) \ , \tag{5.2.28}$$

$$\mathbf{h}_\perp^{(\mathrm{rfl})}(\mathbf{u}_\perp) = \mathbf{h}_\perp^{(\mathrm{inc})}(\mathbf{u}_\perp) - \mathbf{h}_\perp^{(\mathrm{trn})}(\mathbf{u}_\perp) \ . \tag{5.2.29}$$

Therefore, according to Eqs. (5.2.23) and (5.2.25), the dual integral equations[*] for the transmitted fields are

$$\left.\begin{array}{ll} \displaystyle\int \mathbf{e}_\perp^{(\mathrm{trn})}(\mathbf{u}_\perp)\, e^{ik\mathbf{u}_\perp \cdot \boldsymbol{\rho}}\, d^2u_\perp = 0 & \text{on S}\,, \\[4mm] \displaystyle\int \mathbf{h}_\perp^{(\mathrm{trn})}(\mathbf{u}_\perp)\, e^{ik\mathbf{u}_\perp \cdot \boldsymbol{\rho}}\, d^2u_\perp = \mathbf{H}_\perp^{(\mathrm{inc})}(\boldsymbol{\rho},0) & \text{in A}. \end{array}\right\} \tag{5.2.30}$$

These equations are obviously coupled since the angular spectrum amplitudes $\mathbf{e}_\perp^{(\mathrm{trn})}(\mathbf{u}_\perp)$ and $\mathbf{h}_\perp^{(\mathrm{trn})}(\mathbf{u}_\perp)$ are related to each other by the expressions

$$\begin{aligned} \mathbf{h}^{(\mathrm{trn})}(\mathbf{u}_\perp) &= \mathbf{u} \times \mathbf{e}^{(\mathrm{trn})}(\mathbf{u}_\perp) \\[2mm] &= \mathbf{u} \times \left[ \mathbf{e}_\perp^{(\mathrm{trn})}(\mathbf{u}_\perp) - \hat{\mathbf{z}}\, \frac{\mathbf{u}_\perp \cdot \mathbf{e}_\perp^{(\mathrm{trn})}(\mathbf{u}_\perp)}{u_z} \right], \end{aligned} \tag{5.2.31a}$$

$$\begin{aligned} \mathbf{e}^{(\mathrm{trn})}(\mathbf{u}_\perp) &= -\mathbf{u} \times \mathbf{h}^{(\mathrm{trn})}(\mathbf{u}_\perp) \\[2mm] &= -\mathbf{u} \times \left[ \mathbf{h}_\perp^{(\mathrm{trn})}(\mathbf{u}_\perp) - \hat{\mathbf{z}}\, \frac{\mathbf{u}_\perp \cdot \mathbf{h}_\perp^{(\mathrm{trn})}(\mathbf{u}_\perp)}{u_z} \right]. \end{aligned} \tag{5.2.31b}$$

---

[*] There are actually four integral equations to solve simultaneously because each vector integral equation corresponds to a pair of scalar equations.

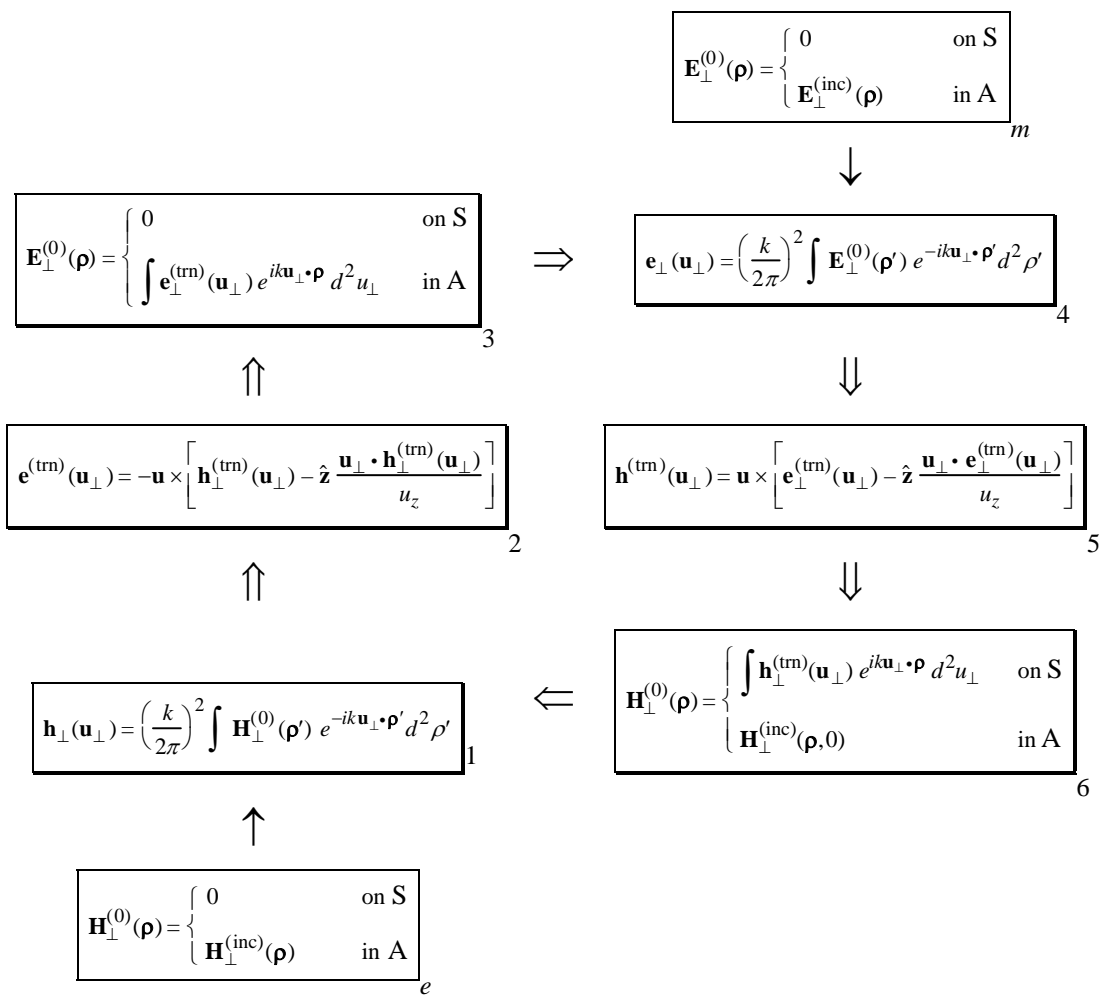

**Figure 5-4** Block diagram of the iterative Fourier-based algorithm for a perfectly conducting screen. Here $\mathbf{E}_\perp^{(0)}(\boldsymbol{\rho}) \equiv \mathbf{E}_\perp^{(\mathrm{trn})}(\boldsymbol{\rho}, 0^+)$ and $\mathbf{H}_\perp^{(0)}(\boldsymbol{\rho}) \equiv \mathbf{H}_\perp^{(\mathrm{trn})}(\boldsymbol{\rho}, 0^+)$.

The iterative Fourier-based algorithm for solving Eq. (5.2.30) is similar to the scalar version, but it is considerably more complicated (see Fig. 5-4). In order to pick out the one solution that satisfies edge conditions, a weakly attenuating filter can again be applied to the angular spectrum amplitudes $\mathbf{e}_\perp^{(\mathrm{trn})}(\mathbf{u}_\perp)$ and $\mathbf{h}_\perp^{(\mathrm{trn})}(\mathbf{u}_\perp)$. However, unlike the scalar case, the correct solution to the electromagnetic aperture diffraction problem does contain singularities.  Fortunately, that solution is the least singular one so it is still favored by the attenuating filter.

As indicated in Fig. 5-4, the algorithm can be initiated with the *e*-theory at step #1 or with the *m*-theory at step #4.  It can readily be shown that, if the *e*-theory is used to start the algorithm, then in the first iteration the transverse electric field $\mathbf{E}_{\perp}^{(0)}(\boldsymbol{\rho})$ at step #3 is equal to the transverse vector component of aperture plane electric field $\mathbf{E}^{(M1)}(\boldsymbol{\rho}, 0^+)$ of the first modified electromagnetic theory.  On the other hand, if the *m*-theory is used, then at step #6 of the first iteration the transverse magnetic field $\mathbf{H}_{\perp}^{(0)}(\boldsymbol{\rho})$ is equal to the transverse component of the magnetic field $\mathbf{H}^{(M2)}(\boldsymbol{\rho}, 0^+)$ of the second modified theory.  Hence the two modified electromagnetic theories are first order improvements to the *e* and the *m*-theories.

CHAPTER 6

# OPTICAL VORTICES AND THEIR
# EFFECTS ON DIFFRACTION

The study of dislocations in an optical field, where the amplitude is zero and, consequently, the phase is not defined, provides an interesting insight into diffraction. Ever since the seminal work of Nye and Berry on these fine structures of wavefields,[1-4] there has been a considerable effort to understand their occurrence under various conditions.[5-16] We have already encountered edge dislocations in the two-dimensional diffraction patterns of Chapter 4. We now turn our attention to optical vortices (screw dislocations). For these types of dislocations, which can only appear in three-dimensional diffraction, the surfaces of constant phase are of the form of spiral staircases. We begin with a brief discussion of the vortices present in Bessel beams and then examine a specific situation where a vortex dramatically affects the diffraction of a field.

## 6.1 VORTICES IN BESSEL BEAMS

In a plane $z$ = constant, the field in the vicinity of an optical vortex that lies along the $z$-axis can be expressed as

$$U(\rho,\varphi,z) \; = \; C \, \rho^{|m|} \, e^{im\varphi} \; , \qquad\qquad (6.1.1)$$

where $\rho = \sqrt{x^2 + y^2}$, $\cos\varphi = x/\rho$, $\sin\varphi = y/\rho$ and $m$ is an integer, sometimes called the strength (or charge) of the vortex. Hence an $m$th-order Bessel beam propagating into the half-space $z > 0$, which is of the form (see Section 2.1.4)

$$B_{u_\perp}^{(m)}(\rho,\varphi,z) \;=\; J_m(ku_\perp\rho)\; e^{im\varphi}\; e^{iku_z z}, \qquad (6.1.2)$$

possesses a vortex of strength $m$ along the $z$-axis, since for small arguments the Bessel function can be approximated by[17]

$$J_\nu(\beta) \;\approx\; \frac{(\beta/2)^\nu}{\Gamma(\nu+1)}\;, \quad \text{as } \beta \to 0\;. \qquad (6.1.3)$$

This applies both to nondiffracting Bessel beams, for which

$$u_z \;=\; \sqrt{1 - u_\perp^2}\;, \qquad u_\perp \le 1\;, \qquad (6.1.4a)$$

and to evanescent ones, for which

$$u_z \;=\; i\sqrt{u_\perp^2 - 1}\;, \qquad u_\perp > 1\;. \qquad (6.1.4a)$$

It is instructive to examine the behavior of the energy flux vector $\mathbf{F}$ associated with these two types of Bessel beams. As discussed in Section 2.1.7, $\mathbf{F}$ may be written in the form

$$\mathbf{F} \;=\; 2\omega\alpha\, A^2 \nabla\phi\;, \qquad (6.1.5)$$

where $A$ is the amplitude of the field and $\phi$ is its phase. It follows that

$$\mathbf{F} \;=\; 2\omega\alpha\left(\hat{\boldsymbol{\varphi}}\,\frac{m}{\rho} \;+\; \hat{\mathbf{z}}\,ku_z\right) J_m^2(ku_\perp\rho) \qquad (6.1.6a)$$

and

$$\mathbf{F} \;=\; 2\omega\alpha\,\hat{\boldsymbol{\varphi}}\,\frac{m}{\rho}\;J_m^2(ku_\perp\rho)\,e^{-2kz\sqrt{u_\perp^2-1}}\;, \qquad (6.1.6b)$$

respectively, for nondiffracting and for evanescent Bessel beams, $\hat{\boldsymbol{\varphi}}$ being a unit vector in the azimuthal direction. From this pair of expressions, we see that in nondiffracting Bessel beams the flow lines of the energy flux vector are spirals about the $z$-axis that are right (left) handed for positive (negative) $m$, whereas in evanescent Bessel beams they are concentric circles that decrease exponentially in number as $z$ increases. These two cases are illustrated in Figs. 6-1 and 6-2.

## 6.2 SPIRAL PHASE PLATE ILLUMINATED BY A GAUSSIAN BEAM

To illustrate the dramatic effect that the presence of a vortex can have on the diffraction of a field, we now examine the field emerging from an $m$th-order spiral phase plate, with transmission function $t_m(\varphi) = e^{im\varphi}$, when the plate is illuminated by a Gaussian beam. This specific situation, and its generalization to illumination by a Laguerre-Gaussian beam, has already been examined in Ref. 18 by means of a decomposition of the output field in the complete set of Laguerre-Gaussian beams. However, explicit results were presented only for the far field. Here we wish to understand how the vortex along the $z$-axis affects the field as it propagates from near zone to far zone. Instead of using a mode decomposition, we will show that there is a relatively simple closed-form expression for the output field. Since the input Gaussian beam is paraxial, the output is also paraxial[*] and, consequently, evanescent waves play no role in this situation.

---

[*] From Eq. (6.2.4) it is apparent that the output beam is paraxial if the condition $kw_o \gg 1$ is satisfied, which is precisely the requirement for the input beam to be paraxial.

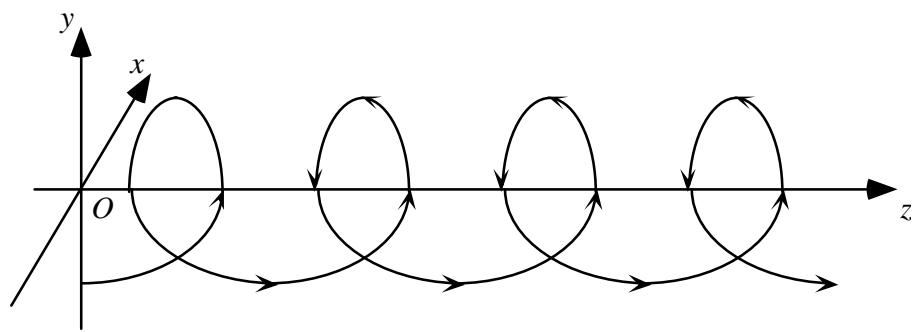

**Figure 6-1**  Typical flow line of the energy flux vector of a nondiffracting Bessel beam with positive *m*.

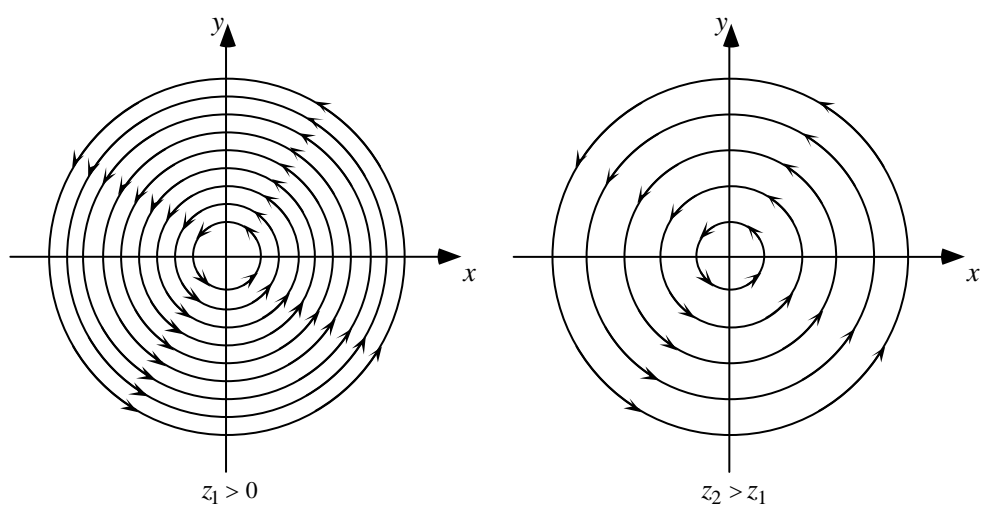

**Figure 6-2**  Flow lines of the energy flux vector of an evanescent Bessel beam with positive *m*.

So that the phase of the field emerging from the spiral phase plate is a continuous function of the angle $\varphi$, and so that there is only a single dislocation in the field, we take *m* to be an integer.  This also ensures that the amplitude of the field is rotationally symmetric.  In practice, the transmission function $t_m(\varphi)$ of the phase plate can be obtained from a transparent plate of uniform refractive index with thickness that varies appropriately with $\varphi$, or from one of constant thickness with the correct refractive index profile. The first approach was used in the experimental work described in Ref. 18.

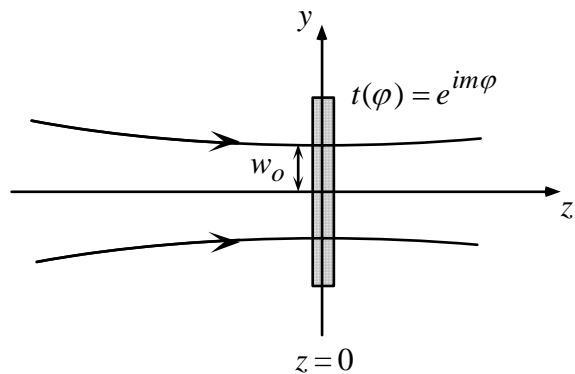

**Figure 6-3** Spiral phase plate illuminated by a Gaussian beam of width $w_o$.

For simplicity, we assume that the waist of the illuminating beam coincides with the plane $z = 0^-$, which is the location of the phase plate (see Fig. 6-3). The field in the plane $z = 0^+$ behind the plate is then given by the expression

$$U_m(\rho, \varphi, 0^+) \;=\; A\, e^{-\rho^2/w_o^2}\, e^{im\varphi} \;, \qquad (6.2.1)$$

$w_o$ being the width of the beam waist.

We can use the Bessel-beam representation, in the paraxial approximation, to propagate the field given by Eq. (6.2.1). The field in the half-space $z > 0$ is then of the form (see Section 2.1.4)

$$U_m(\rho, \varphi, z) \;=\; \sum_{n=-\infty}^{\infty} \int_0^\infty c_n(u_\perp)\, \mathrm{B}_{u_\perp}^{(n)}(\rho, \varphi, z)\, du_\perp \;, \qquad (6.2.2)$$

$$\mathrm{B}_{u_\perp}^{(n)}(\rho,\varphi,z) \;=\; J_n(ku_\perp\rho)\; e^{in\varphi}\; e^{-iku_\perp^2 z}\; e^{ikz}\;, \tag{6.2.3}$$

with coefficients

$$
\begin{aligned}
c_n(u_\perp) \;&=\; \frac{k^2}{2\pi}\, u_\perp \int\limits_0^\infty \int\limits_0^{2\pi} U_m(\rho,\varphi,0^+)\, e^{-in\varphi}\, J_n(ku_\perp\rho)\,\rho\, d\rho\, d\varphi \\
&=\; \delta_{mn}\, Ak^2 u_\perp \int\limits_0^\infty e^{-\rho^2/w_o^2}\, J_m(ku_\perp\rho)\,\rho\, d\rho\;.
\end{aligned}
\tag{6.2.4}
$$

From Eq. (6.2.4), we see that the beam $U_m(\rho,\varphi,z)$ contains only $m$th-order Bessel beams. If we substitute from Eqs. (6.2.3) and (6.2.4) into Eq. (6.2.2) and make the change of variables $s = u_\perp^2$, we find that

$$
\begin{aligned}
U_m(\rho,\varphi,z) \;=\; \frac{Ak^2}{2}\, e^{im\varphi}\, e^{ikz} \int\limits_0^\infty & \left[\int\limits_0^\infty J_{|m|}(k\rho\sqrt{s})\, J_{|m|}(k\rho'\sqrt{s})\, e^{-iksz/2}\, ds\right] \\
& \times\, e^{-\rho'^2/w_o^2}\, \rho'\, d\rho'\;.
\end{aligned}
\tag{6.2.5}
$$

The order of the Bessel functions in this expression depends only on the absolute value of $m$ because $J_{-m} = (-1)^m J_m$.

We can perform the integration over $s$ in Eq. (6.2.5) by making use of the relations[19]

$$\int\limits_0^\infty J_m(a\sqrt{s})\, J_m(b\sqrt{s})\, \sin(cs)\, ds \;=\; \frac{1}{c}\, J_m\!\left(\frac{ab}{2c}\right)\, \cos\!\left(\frac{a^2+b^2}{4c} - \frac{m\pi}{2}\right) \tag{6.2.6a}$$

and

$$\int\limits_0^\infty J_m\left(a\sqrt{s}\,\right) J_m\left(b\sqrt{s}\,\right) \cos(cs)\, ds \;=\; \frac{1}{c}\, J_m\!\left(\frac{ab}{2c}\right)\, \sin\!\left(\frac{a^2+b^2}{4c}\;-\;\frac{m\pi}{2}\right).$$

$$(6.2.6b)$$

We then obtain the expression

$$U_m(\rho,\varphi,z) \;=\; -i^{|m|+1}\, A k^2\, e^{ik\rho^2/2z}\, e^{im\varphi}\, \frac{e^{ikz}}{kz}$$
$$\times \int\limits_0^\infty J_{|m|}\!\left(\frac{k\rho\rho'}{z}\right)\, \exp\!\left[-\rho'^2\!\left(\frac{1}{w_o^2}\;-\;\frac{ik}{2z}\right)\right]\rho'\, d\rho',$$

$$(6.2.7)$$

which can be simplified further with help of the integral [see Ref. 19, pg. 717, Eq. (6.631 7)]

$$\int\limits_0^\infty e^{-as^2} J_m(bs)\, s\, ds \;=\; \frac{\sqrt{\pi}\, b}{8 a^{3/2}}\, e^{-b^2/8a} \left[\, I_{\frac{m}{2}-\frac{1}{2}}\!\left(b^2/8a\right) - I_{\frac{m}{2}+\frac{1}{2}}\!\left(b^2/8a\right)\right],$$

$$(\mathrm{Re}\, a > 0,\; m > -2),$$

$$(6.2.8)$$

$I_\nu$ being a modified Bessel function of the first kind and $\nu$th order.

Finally, after some straightforward algebra, we obtain the following relatively simple result for the field in the half-space $z > 0$:

$$U_m(\rho,\varphi,z) \;=\; -i^{|m|+1}\sqrt{\pi}\; \frac{A}{k}\; \frac{\rho z\, g^{3/2}(z)}{w^3(z)}\, e^{ik\rho^2/2z}\, e^{im\varphi}\, e^{ikz}$$
$$\times\, e^{-\rho^2 g(z)/2w^2(z)} \left[\, I_{\frac{|m|}{2}-\frac{1}{2}}\!\left(\frac{\rho^2 g(z)}{2w^2(z)}\right) \;-\; I_{\frac{|m|}{2}+\frac{1}{2}}\!\left(\frac{\rho^2 g(z)}{2w^2(z)}\right)\right].$$

$$(6.2.9)$$

Here

$$g(z) \;=\; 1 \;+\; i\frac{z_r}{z} \;, \tag{6.2.10}$$

$$w(z) \;=\; w_o \left[ 1 \;+\; \left(\frac{z}{z_r}\right)^2 \right]^{1/2} \tag{6.2.11}$$

and

$$z_r = kw_o^2 \big/ 2 \tag{6.2.12}$$

is the Rayleigh range of the Gaussian beam that illuminates the phase plate. It should be noted that Eq. (6.2.11) is identical to the formula for the width of a Gaussian beam propagating in free space. Furthermore, as expected, the amplitude of $U_m(\rho,\varphi,z)$ is rotationally symmetric about the $z$-axis.

Let us rewrite Eq. (6.2.9) as the product

$$U_m(\rho,\varphi,z) \;=\; U_0(\rho,\varphi,z)\, M_m(\rho,\varphi,z) \;, \tag{6.2.13}$$

where $U_0(\rho,\varphi,z)$ is an ordinary Gaussian beam,

$$U_0(\rho,\varphi,z) \;=\; A\,\frac{w_o}{w(z)}\, e^{-\rho^2/w^2(z)}\, e^{ik\rho^2/2R(z)}\, e^{-i\psi(z)}\, e^{ikz} \;, \tag{6.2.14}$$

with

$$R(z) = z\left[ 1 \;+\; \left(\frac{z_r}{z}\right)^2 \right], \tag{6.2.15}$$

$$\sin\psi(z) \;=\; \frac{z}{\sqrt{z^2 + z_r^2}} \;, \quad \cos\psi(z) \;=\; \frac{z_r}{\sqrt{z^2 + z_r^2}} \;, \tag{6.2.16}$$

and $M_m(\rho,\varphi,z)$ is the modifying function

$$M_m(\rho,\varphi,z) \;=\; i^{|m|}\,\frac{\sqrt{\pi}}{2}\,\frac{\rho\,g^{1/2}(z)}{w(z)}\,e^{im\varphi}\,e^{\rho^2 g(z)\big/2w^2(z)}$$

$$\times\left[I_{\frac{|m|}{2}-\frac{1}{2}}\!\left(\frac{\rho^2 g(z)}{2w^2(z)}\right)\;-\;I_{\frac{|m|}{2}+\frac{1}{2}}\!\left(\frac{\rho^2 g(z)}{2w^2(z)}\right)\right]. \qquad (6.2.17)$$

If $m$ is an even integer, the modified Bessel functions in Eq. (6.2.17) can be expressed explicitly in terms of hyperbolic sines and cosines. When $m = 0$, since there is no phase plate, Eq. (6.2.13) should reduce to the expression for an ordinary Gaussian beam and, therefore, the modifying function $M_0(\rho,\varphi,z)$ should be equal to unity. With the help of the expressions [see Ref. 17, pg. 443, Eqs. (10.2.13) and (10.2.14)]

$$\sqrt{\frac{\pi}{2\gamma}}\,I_{-1/2}(\gamma)\;=\;\frac{\cosh\gamma}{\gamma}\;, \qquad (6.2.18a)$$

$$\sqrt{\frac{\pi}{2\gamma}}\,I_{1/2}(\gamma)\;=\;\frac{\sinh\gamma}{\gamma}\;, \qquad (6.2.18b)$$

it is straightforward to verify that indeed $M_0(\rho,\varphi,z)=1$.

Let us examine the behavior of the beam $U_m(\rho,\varphi,z)$ near the $z$-axis. For this purpose, we make use of the limiting form [see Ref. 17, pg. 375, Eq. (9.6.7)]

$$I_\nu(\beta)\;\approx\;\frac{(\beta/2)^\nu}{\Gamma(\nu+1)}\qquad\text{as }\beta\to 0\;. \qquad (6.2.19)$$

Therefore, as $\rho\to 0$, the limiting forms of the modifying function $M_m(\rho,\varphi,z)$ and of the field $U_m(\rho,\varphi,z)$ are

$$M_m(\rho,\varphi,z)\;\approx\;\frac{i^{|m|}\sqrt{\pi}}{2^{|m|}\,\Gamma\!\left(\frac{|m|}{2}+\frac{1}{2}\right)}\,\frac{g^{|m|/2}(z)}{w^{|m|}(z)}\,\rho^{|m|}\,e^{im\varphi} \qquad (6.2.20)$$

and

$$U_m(\rho,\varphi,z) \;\approx\; A\,\frac{i^{|m|}\sqrt{\pi}}{2^{|m|}\,\Gamma\!\left(\frac{|m|}{2}+\frac{1}{2}\right)}\,\frac{w_o}{w(z)}\left[\frac{g(z)}{w^2(z)}\right]^{\frac{|m|}{2}}e^{-i\psi(z)}\,e^{ikz}\,\rho^{|m|}\,e^{im\varphi}$$

$$\approx\; A\,\frac{i^{3|m|/2}\sqrt{\pi}}{2^{3|m|/2}\,\Gamma\!\left(\frac{|m|}{2}+\frac{1}{2}\right)}\,\frac{w_o}{w(z)}\left[\frac{k\,w_o}{z\,w(z)}\right]^{\frac{|m|}{2}}$$

$$\times\,\exp\!\left[-i\left(1+\frac{|m|}{2}\right)\psi(z)\right]e^{ikz}\,\rho^{|m|}\,e^{im\varphi}\;.$$

$$(6.2.21)$$

We see that an $m$th-order spiral phase plate causes a vortex of strength $m$, $\rho^{|m|}e^{im\varphi}$, to appear in the field along the $z$-axis. It is apparent that, for $m \neq 0$, the rapidity with which the field changes from its zero on axis value depends on the propagation distance. For distances $z$ much smaller than the Rayleigh range, we can approximate Eq. (6.2.21) by the expression

$$U_m(\rho,\varphi,z) \;\approx\; A\,\frac{i^{3|m|/2}\sqrt{\pi}}{2^{|m|}\,\Gamma\!\left(\frac{|m|}{2}+\frac{1}{2}\right)}\left[\frac{k}{2z}\right]^{\frac{|m|}{2}}\rho^{|m|}\,e^{im\varphi}\;,\qquad\left(\rho\to 0,\,z<<z_r\right).$$

$$(6.2.22)$$

Hence, near the $z$-axis, the closer one approaches the plane $z = 0^{+}$, the more rapidly the field increases radially from zero. This is to be expected since, in the limit $z \to 0^{+}$, the field must be non-zero on the $z$-axis [see Eq. (6.2.1)].

Figures 6-4 and 6-5 depict the behavior of the beam $U_m(\rho,\varphi,z)$ and of the modifying function $M_m(\rho,\varphi,z)$ in the half-space $z > 0$ for $m = 1$ and $m = 2$. For comparison, an ordinary Gaussian beam $U_0(\rho,\varphi,z)$ is also shown in these figures. As we would anticipate from our preceding discussion, the width of the hole in the center of the beams $U_1(\rho,\varphi,z)$ and $U_2(\rho,\varphi,z)$ increases with increasing $z$.

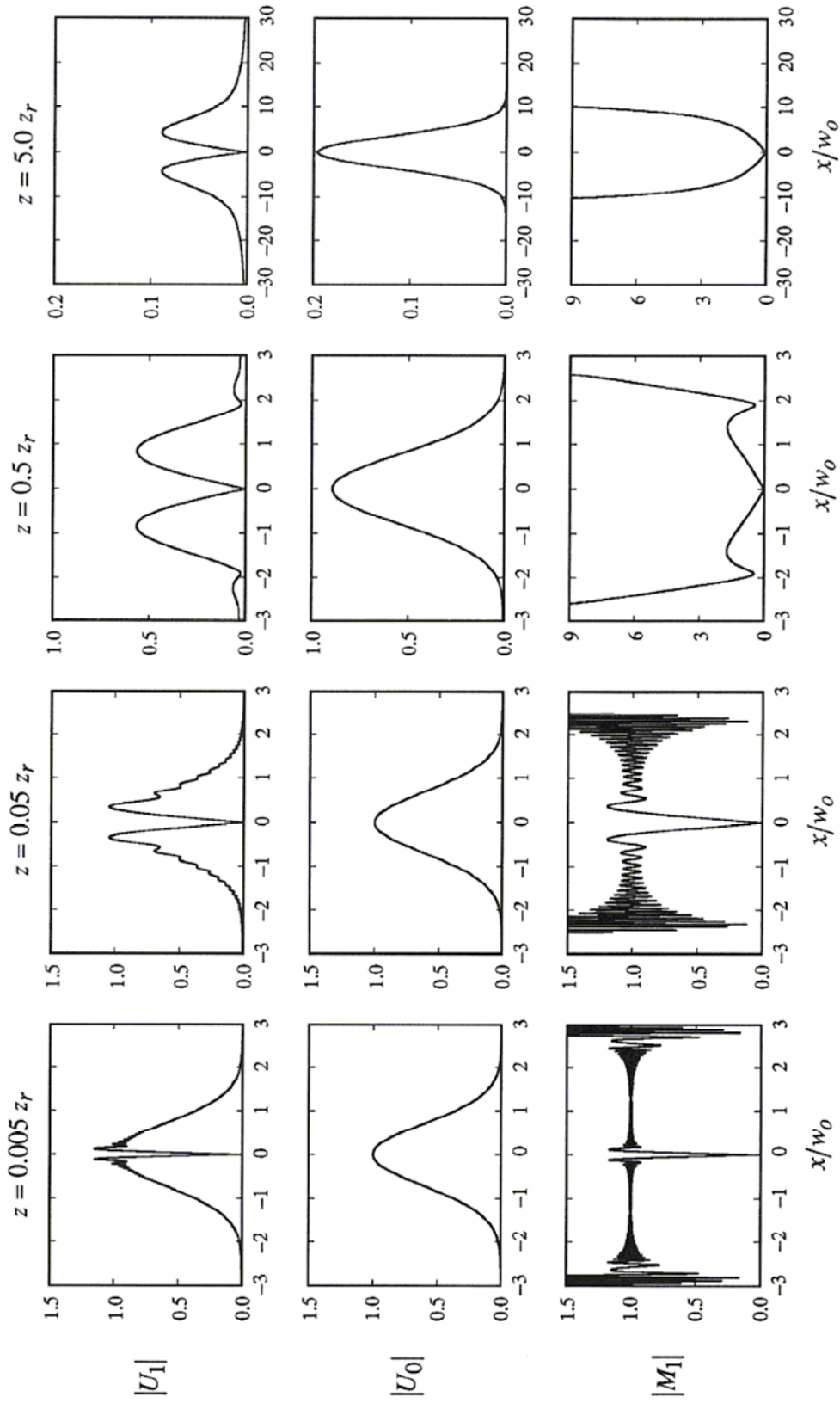

**Figure 6-4** (top) A Gaussian beam that has passed through a spiral phase plate with $m = 1$, (middle) an ordinary Gaussian beam in free space, and (bottom) the modifying function for $m = 1$.

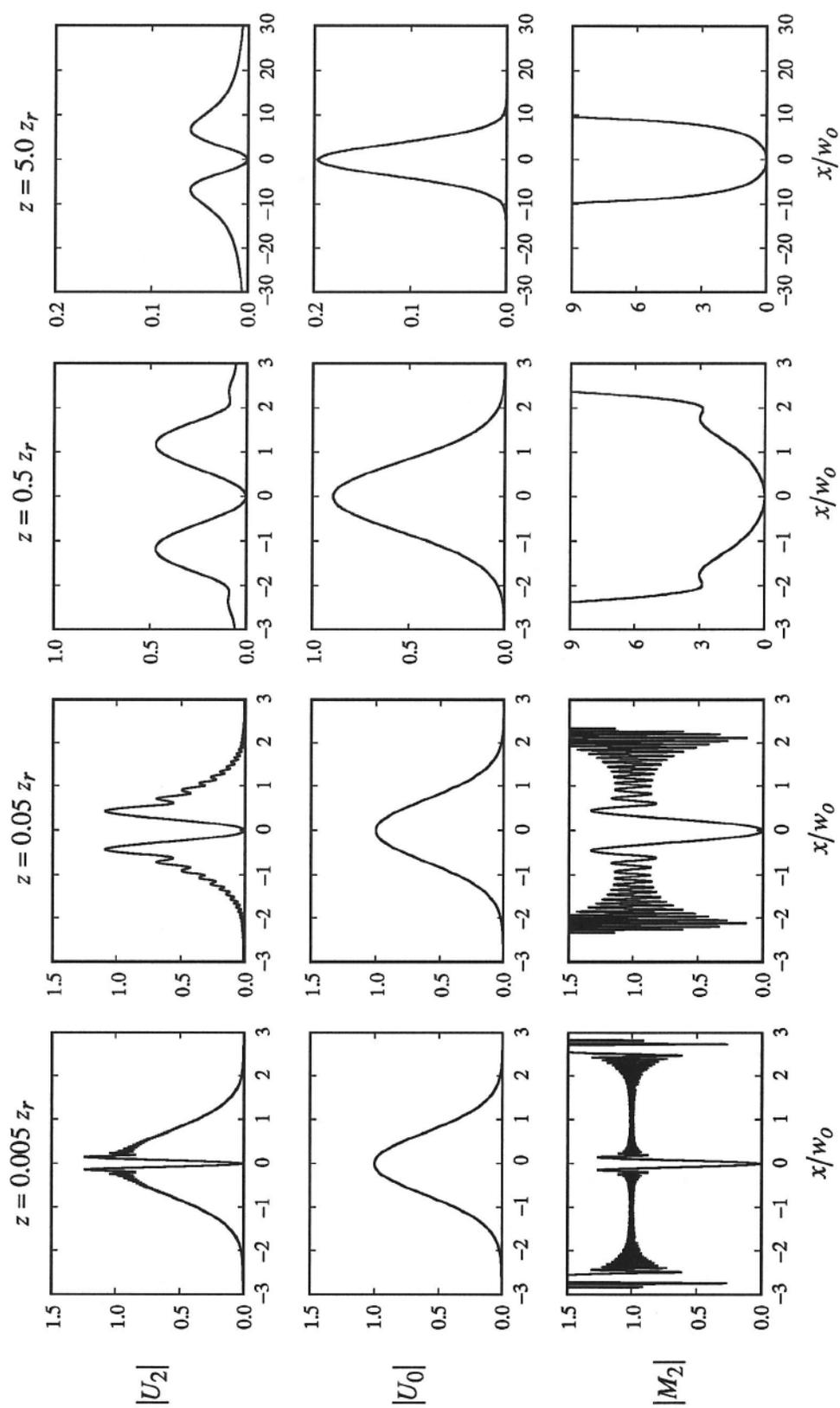

**Figure 6-5** (top) A Gaussian beam that has passed through a spiral phase plate with $m = 2$, (middle) an ordinary Gaussian beam in free space, and (bottom) the modifying function for $m = 2$.

There are two somewhat surprising aspects of Figs. 6-4 and 6-5: (i) that for propagation distances much smaller than the Rayleigh range the amplitudes of $U_1(\rho,\varphi,z)$ and of $U_2(\rho,\varphi,z)$ do change considerably on propagation, even though the amplitude of an ordinary Gaussian beam is nearly invariant over such distances and (ii) that there are high frequency radial oscillations present in $|U_1(\rho,\varphi,z)|$ and $|U_2(\rho,\varphi,z)|$ for $z = 0.005z_r$ and $z = 0.05z_r$. These oscillations become even more pronounced for larger $m$, but they disappear almost completely once the propagation distance is of the order of the Rayleigh range. Although the analysis here applies to the specific case of a spiral phase plate illuminated by a Gaussian beam, we would expect to observe similar effects for illumination by an arbitrary vortex-free coherent beam.

# CHAPTER 7

# SUMMARY

The primary objective of this thesis was to study the effects of diffraction of light in the near zone of apertures in thin opaque screens, with dimensions comparable to the wavelength. For this purpose, in addition to making use of standard techniques, several new analytical and numerical tools were developed specifically for examining the near field.

In Chapter 2, it was shown that for a scalar field near a point in space where the amplitude has a local extremum the surfaces of constant phase are relatively flat. Furthermore, if the extremum is a maximum, these surfaces are spaced further apart than the corresponding surfaces for a plane wave, whereas if it is a minimum, they are spaced closer together than those for a plane wave. In Chapter 4, this type of phase behavior was demonstrated explicitly for the near-field of a slit in a perfectly conducting plane. However, these properties apply to any scalar field in free space and, therefore, they can applied to understand various phase phenomena that occur in the vicinity of extrema of the amplitude, such as the phase anomaly present near focus.

In order to understand the propagation of a field in the near zone, the contributions of homogeneous and of evanescent waves to the total field were examined in Chapter 3 for the case of a two-dimensional scalar field. Several exact techniques for calculating these two contributions were discussed and approximate relations were derived for the near field. The concepts of total homogeneous intensity and total evanescent intensity were introduced as convenient measures of the relative importance of the two contributions. As an example, the diffraction of a plane wave by a slit was examined with the use of the approximate boundary conditions of the Rayleigh-Sommerfeld theory of the first kind. Even though these boundary conditions provide a poor approximation to the exact near field, they can be used to obtain insight into the behavior of the homogeneous and the evanescent contributions on propagation. As one would expect, near-zone modifications of field are dominated by the decay of the evanescent contribution, with only relatively small changes of the homogeneous contribution.

In Chapter 4, the finite-difference time-domain (FD-TD) method was used to reexamine the slit diffraction problem rigorously for the case of a thin perfectly conducting screen, without making use of any approximate boundary conditions. The near-fields for both E and H-polarizations were considered in detail. A comparison of the boundary fields obtained from the FD-TD method with the predictions of the Rayleigh-Sommerfeld theory of the first kind and of the Rayleigh-Sommerfeld theory of the second kind showed that neither approximate theory gives an accurate description of the near field. The discrepancy between the exact solution and the predictions of the approximate theories decreases, however, with increasing propagation distance. Futhermore, it was shown that, by modifying the boundary values of the Rayleigh-Sommerfeld theories, it is possible to obtain two new theories of diffraction that provide a better approximation to the near field. One of these new theories yields a reasonable approximation to the electric field for E-polarization and the other to the magnetic field for H-polarization. Generalizations of the new approximate theories to arbitrary aperture shapes were discussed in Chapter 5. Because the new theories are more difficult to employ than the usual approximate theories, further work is still needed to determine whether the additional computation efforts required by the new theories are worthwhile.

An iterative Fourier-based algorithm for treating diffraction by an aperture of arbitrary shape in a thin opaque screen was proposed in chapter 5. This new algorithm, which is based on the angular spectrum representation, can be implemented with the use of fast Fourier transforms for scalar fields incident upon apertures in Dirichlet-type or Neumann-type screens and for electromagnetic fields incident upon apertures in perfectly conducting screens. For these situations, the algorithm appears to be much more efficient than the FD-TD method. Preliminary results for the case of slit diffraction indicate that the algorithm converges to the exact solution very rapidly.

In Chapter 6, in order to show that an optical vortex can dramatically affect the diffraction of a three-dimensional field over small propagation distances, the field emerging from a spiral phase plate illuminated by a Gaussian beam was examined. For this example, a closed-form solution for the field behind the phase plate was found in terms of modified Bessel functions with complex argument. It was shown that there are appreciably changes to the field for propagation distances much smaller than the Rayleigh range. Hence, even though, strictly speaking, the propagation distances in question are not in the near zone because they are larger than the wavelength, they are considerably smaller than minimum distance usually associated with observable diffraction effects in beams. One would expect to see similar behavior in other situations in which an optical vortex is imposed on a previously vortex-free field.

In conclusion, it should be stressed that, in general, for apertures with dimensions comparable to the wavelength or smaller, near-field diffraction is difficult to analyze because the usual approximations used in physical optics for the propagator (the Green's function) and for the boundary values of the field are not applicable. Both the boundary value problem, which requires the determination of the boundary value of the field immediately behind the screen, and the propagation problem, which requires the determination of the field at some distance from the screen from this boundary field, must be treated carefully.

APPENDIX A

# SOME REMARKS ON THE POYNTING
# VECTOR AND ENERGY DENSITY

In Section 2.1.7, it was shown that, for a complex scalar field $U = A\,e^{i\phi}$, the (real) energy flux vector $\mathbf{F}$ is related to the amplitude $A$ and the phase $\phi$ by the simple expression

$$\mathbf{F} \;=\; 2\omega\alpha\,A^2\nabla\phi \;,\qquad\qquad (A.1)$$

where $\alpha$ is a real, positive constant. In this Appendix, we derive the analogous relation for an electromagnetic field and show that it reduces to the form of Eq. (A.1) when the field is linearly polarized.

The (real) time-averaged Poynting vector associated with a monochromatic electromagnetic field with electric field $\mathbf{E}$ and magnetic field $\mathbf{H}$ is given by

$$
\begin{aligned}
\mathbf{S} \;&=\; \frac{c}{8\pi}\;\operatorname{Re}\{\mathbf{E}\times\mathbf{H}^{*}\}\\[4pt]
&=\; -\frac{c}{8\pi k}\;\operatorname{Im}\{\mathbf{E}\times(\nabla\times\mathbf{E}^{*})\}\\[4pt]
&=\; -\frac{c}{8\pi k}\;\operatorname{Im}\{\mathbf{H}\times(\nabla\times\mathbf{H}^{*})\}
\end{aligned}
\qquad (A.2)
$$

and satisfies the conservation law

$$\nabla\cdot\mathbf{S} \;=\; 0 \;.\qquad\qquad (A.3)$$

Let us express the electric field $\mathbf{E}$ in terms of its magnitude $E$ and a (complex) polarization unit vector $\mathbf{u}_e$,

$$\mathbf{E} = E\mathbf{u}_e \, , \tag{A.4}$$

$$E = |\mathbf{E}| \, , \quad \mathbf{u}_e \cdot \mathbf{u}_e^* = 1 \, . \tag{A.5}$$

Then,

$$
\begin{aligned}
\mathbf{E} \times (\nabla \times \mathbf{E}^*) &= E\mathbf{u}_e \times [\nabla \times (E\mathbf{u}_e^*)] \\
&= E\mathbf{u}_e \times (E\nabla \times \mathbf{u}_e^* + \nabla E \times \mathbf{u}_e^*) \\
&= E^2 \mathbf{u}_e \times (\nabla \times \mathbf{u}_e^*) + E[(\mathbf{u}_e \cdot \mathbf{u}_e^*)\nabla E - (\mathbf{u}_e \cdot \nabla E)\mathbf{u}_e^*] \\
&= E^2 \mathbf{u}_e \times (\nabla \times \mathbf{u}_e^*) + E\nabla E - E\mathbf{u}_e^*(\mathbf{u}_e \cdot \nabla E)
\end{aligned}
\tag{A.6}
$$

and, because $\nabla \cdot \mathbf{E}(\mathbf{r}) = 0$,

$$\mathbf{u}_e \cdot \nabla E = -E\nabla \cdot \mathbf{u}_e \, . \tag{A.7}$$

By making use of Eqs. (A.2), (A.6) and (A.7), we readily obtain the desired expression for the Poynting vector,

$$\mathbf{S} = -\frac{c}{k} \, w_e \, \mathrm{Im}\{\mathbf{u}_e \times (\nabla \times \mathbf{u}_e^*) + \mathbf{u}_e^*(\nabla \cdot \mathbf{u}_e)\} \, , \tag{A.8}$$

where

$$w_e = \frac{1}{16\pi} \, |\mathbf{E}|^2 \tag{A.9}$$

is the electric energy density. Equation (A.8) should be compared with Eq. (A.1). Similarly to Eq. (A.1), where the direction of energy flux vector $\mathbf{F}$ depends only on the phase $\phi$, in Eq. (A.8) the direction of the Poynting vector $\mathbf{S}$ depends only on the polarization unit vector $\mathbf{u}_e$.

We next consider the case when the electric field is linearly polarized. Although this case is discussed extensively in Ref. 1, we reexamine it here to show that Eq. (A.8) does yield the correct result.

For a linearly polarized electric field, the complex unit polarization vector $\mathbf{u}_e$ may be written in terms of a real polarization unit vector $\mathbf{n}_e$ and a real phase $\phi_e$ as

$$\mathbf{u}_e = \mathbf{n}_e \, e^{i\phi_e} \ . \tag{A.10}$$

Hence

$$
\begin{aligned}
\mathbf{u}_e \times (\nabla \times \mathbf{u}_e^*) &= e^{i\phi_e} \, \mathbf{n}_e \times [\nabla \times (\mathbf{n}_e \, e^{-i\phi_e})] \\
&= \mathbf{n}_e \times (\nabla \times \mathbf{n}_e) \, - \, i\mathbf{n}_e \times (\nabla\phi_e \times \mathbf{n}_e) \\
&= \mathbf{n}_e \times (\nabla \times \mathbf{n}_e) \, - \, i\nabla\phi_e \, + \, i\mathbf{n}_e(\mathbf{n}_e \cdot \nabla\phi_e) \ ,
\end{aligned}
\tag{A.11}
$$

$$
\begin{aligned}
\mathbf{u}_e^*(\nabla \cdot \mathbf{u}_e) &= \mathbf{n}_e \, e^{-i\phi_e}[\nabla \cdot (\mathbf{n}_e \, e^{i\phi_e})] \\
&= \mathbf{n}_e(\nabla \cdot \mathbf{n}_e) \, + \, i\mathbf{n}_e(\mathbf{n}_e \cdot \nabla\phi_e)
\end{aligned}
\tag{A.12}
$$

and Eq. (A.8) becomes

$$\mathbf{S} = \frac{c}{k} \, w_e \, [\nabla\phi_e \, - \, 2\,\mathbf{n}_e(\mathbf{n}_e \cdot \nabla\phi_e)] \ . \tag{A.13}$$

If we now substitute from Eq. (A.10) into Eq. (A.7), we obtain the relation

$$\mathbf{n}_e \cdot \nabla E = - E\nabla \cdot \mathbf{n}_e \, - \, iE\mathbf{n}_e \cdot \nabla\phi_e \ . \tag{A.14}$$

By taking the imaginary part of this expression,

$$\mathbf{n}_e \cdot \nabla\phi_e = 0 , \tag{A.15}$$

we find that the polarization unit vector $\mathbf{n}_e$ is perpendicular to the surfaces of constant $\phi_e$. Equation (A.13) then reduces to the form of Eq. (A.1),

$$\mathbf{S} = \frac{c}{k} w_e \nabla \phi_e \; , \tag{A.16}$$

in agreement with Eq. (3.22) of Ref. 1.

It should be mentioned that, by writing the magnetic field $\mathbf{H}$ in a form analogous to Eq. (A.4),

$$\mathbf{H} = H \mathbf{u}_m \; , \quad H = |\mathbf{H}| \; , \quad \mathbf{u}_m \cdot \mathbf{u}_m^* = 1, \tag{A.17}$$

we can obtain an expression for the Poynting vector that is equivalent to Eq. (A.8), but which is in terms of the magnetic energy density $w_m$ and the (complex) polarization unit vector $\mathbf{u}_m$:

$$\mathbf{S} = -\frac{c}{k} w_m \; \text{Im}\{\mathbf{u}_m \times (\nabla \times \mathbf{u}_m^*) + \mathbf{u}_m^*(\nabla \cdot \mathbf{u}_m)\} \; , \tag{A.18}$$

$$w_m = \frac{1}{16\pi} |\mathbf{H}|^2 \; . \tag{A.19}$$

# APPENDIX B

# DERIVATION OF EQ. (3.3.3)

In this Appendix, we derive the series expansion (3.3.3) of the integral (3.3.2), i.e., of

$$T(x,z) = \int_0^\pi e^{ikx\cos\psi}\, e^{ikz\sin\psi}\, d\psi \; . \tag{B.1}$$

By making use of Eqs. (9.1.42)-(9.1.45) of Ref. 1, we can write the two exponentials in the integrand of Eq. (B.1) in terms of a series of Bessel functions of the first kind and integer order as

$$e^{ikz\sin\psi} = J_0(kz) + 2\sum_{m=1}^{\infty} J_{2m}(kz)\, \cos(2m\psi)$$
$$+ 2i\sum_{m=0}^{\infty} J_{2m+1}(kz)\, \sin[(2m+1)\psi] \; , \tag{B.2}$$

$$e^{ikx\cos\psi} = J_0(kx) + 2\sum_{n=1}^{\infty} (-1)^n J_{2n}(kx)\, \cos(2n\psi)$$
$$+ 2i\sum_{n=0}^{\infty} (-1)^n J_{2n+1}(kx)\, \cos[(2n+1)\psi] \; . \tag{B.3}$$

When these two expansions are substituted into Eq. (B.1), there are 9 different integrals that need to be evaluated, all of which are very simple. The results are

$$\int_0^\pi d\psi \;=\; \pi \;,$$

(B.4a)

$$\int_0^\pi \cos(2m\psi)\, d\psi \;=\; \int_0^\pi \cos(2n\psi)\, d\psi \;=\; 0\,, \quad (m \geq 1,\; n \geq 1)\,,$$

(B.4b)

$$\int_0^\pi \sin[(2m+1)\psi]\, d\psi \;=\; \frac{2}{2m+1} \;, \quad (m \geq 0)\,,$$

(B.4c)

$$\int_0^\pi \cos[(2m+1)\psi]\, d\psi \;=\; 0\,, \quad (n \geq 0)\,,$$

(B.4d)

$$\int_0^\pi \cos(2m\psi)\, \cos(2n\psi)\, d\psi \;=\; \frac{\pi}{2}\,\delta_{nm}\,, \quad (m \geq 1,\; n \geq 1)\,,$$

(B.4e)

$$\int_0^\pi \cos(2m\psi)\, \cos[(2n+1)\psi]\, d\psi \;=\; 0\,, \quad (m \geq 1,\; n \geq 0)\,,$$

(B.4f)

$$\int_0^\pi \sin[(2m+1)\psi]\, \cos(2n\psi)\, d\psi \;=\; \frac{2\,(2m+1)}{(2m+1)^2 - 4n^2}\;, \quad (m \geq 0,\; n \geq 1)\,,$$

(B.4g)

$$\int_0^\pi \sin[(2m+1)\psi]\, \cos[(2n+1)\psi]\, d\psi \;=\; 0\,, \quad (m \geq 0,\; n \geq 0)\,.$$

(B.4h)

In Eq. (B.4e) $\delta_{nm}$ denotes the Kronecker delta symbol. The resulting expression for $T(x,z)$ is then

$$T(x,z) = \pi J_0(kx) J_0(kz) + 4i J_0(kx) \sum_{m=0}^{\infty} \frac{1}{2m+1} J_{2m+1}(kz)$$

$$+ 2\pi \sum_{m=1}^{\infty} (-1)^m J_{2m}(kx) J_{2m}(kz)$$

$$+ 8i \sum_{m=0}^{\infty} \sum_{n=1}^{\infty} (-1)^n \frac{2m+1}{(2m+1)^2 - 4n^2} J_{2n}(kx) J_{2m+1}(kz).$$

$$(B.5)$$

We can simplify the first three terms of Eq. (B.5) with the help of the relations[2,3]

$$J_0\left(k\sqrt{x^2 + z^2}\right) = J_0(kx) J_0(kz) + 2\sum_{m=1}^{\infty} (-1)^m J_{2m}(kx) J_{2m}(kz)$$

$$(B.6)$$

and

$$\mathbf{H}_0(kz) = \sum_{m=0}^{\infty} \frac{1}{2m+1} J_{2m+1}(kz),$$

$$(B.7)$$

where $\mathbf{H}_0$ is a Struve function of zeroth order. Equation (B.5) then becomes

$$T(x,z) = \pi J_0\left(k\sqrt{x^2 + z^2}\right) + \pi i J_0(kx) \mathbf{H}_0(kz)$$

$$+ 8i \sum_{m=0}^{\infty} \sum_{n=1}^{\infty} (-1)^n \frac{2m+1}{(2m+1)^2 - 4n^2} J_{2n}(kx) J_{2m+1}(kz),$$

$$(B.8)$$

which is Eq. (3.3.3) of Chapter 3.

# APPENDIX C

# DERIVATION OF EQ. (3.3.14a)

Here we derive Eq. (3.3.14a) of Chapter 3 from the expression [see Eqs. (3.3.12) and (3.3.13)]

$$H_h(x,z) = \frac{k}{2\pi} \sum_{n=0}^{\infty} \frac{i^n}{n!} (kz)^n \int_{-1}^{1} e^{iku_x x} (1-u_x^2)^{n/2} du_x . \qquad \text{(C.1)}$$

We can evaluate the integrations with respect to $u_x$ using the following integral representation for Bessel functions of the first kind:[1]

$$J_\nu(\beta) = \frac{(\beta/2)^\nu}{\Gamma(\nu+\frac{1}{2})\Gamma(\frac{1}{2})} \int_{-1}^{1} e^{i\beta t} \left(1-t^2\right)^{\nu-\frac{1}{2}} dt , \quad \left(\text{Re } \nu > -\frac{1}{2}\right),$$
$$\text{(C.2)}$$

where $\Gamma$ is the gamma function. Equation (C.1) then becomes

$$H_h(x,z) = \frac{k}{2\pi} \sum_{n=0}^{\infty} \frac{i^n}{n!} (kz)^n \Gamma\left(\frac{n}{2}+1\right)\Gamma\left(\frac{1}{2}\right) \frac{J_{(n+1)/2}(kx)}{(kx/2)^{(n+1)/2}} . \qquad \text{(C.3)}$$

Let us examine the sum over even $n$ and odd $n$ separately and express the summation symbolically as

$$\sum_{n=0}^{\infty} = \sum_{\substack{n=0 \\ (even)}}^{\infty} + \sum_{\substack{n=1 \\ (odd)}}^{\infty} . \tag{C.4}$$

For the summation over odd $n$, if we use the relations

$$\Gamma\left(\tfrac{1}{2}\right) = \sqrt{\pi} , \tag{C.5}$$

$$\Gamma\left(\tfrac{n}{2}+1\right) = \frac{n!!}{2^{n/2}\, 2^{1/2}}\, \Gamma\left(\tfrac{1}{2}\right), \tag{C.6}$$

where $n!! = n(n-2)(n-4).....(5)(3)(1)$, and make the change of summation indices $n = 2m-1$, we obtain

$$\sum_{\substack{n=1 \\ (odd)}}^{\infty} = -i\pi \sum_{m=1}^{\infty} \frac{(-1)^m}{(m-1)!\, 2^{m-1}}\, (kz)^{2m-1}\, \frac{J_m(kx)}{(kx)^m} . \tag{C.7}$$

We can resum this expression explicitly using the multiplication theorem[2]

$$J_1(s\beta) = s \sum_{p=0}^{\infty} (-1)^p\, \frac{(s^2-1)^p\, (\beta/2)^p}{p!}\, J_{p+1}(s\beta) , \tag{C.8}$$

with $\beta = kx$, $s = \sqrt{1 + \left(z^2/x^2\right)}$ and $p = m-1$. Equation (C.7) then reduces to

$$\sum_{\substack{n=1 \\ (odd)}}^{\infty} = \frac{i\pi z}{\sqrt{x^2 + z^2}}\, J_1\!\left(k\sqrt{x^2 + z^2}\right) . \tag{C.9}$$

On the other hand, for even $n$, if we use Eqs. (C.5) and

$$\Gamma\left(\frac{n}{2}+1\right) = (n/2)! \tag{C.10}$$

and make the change of indices $n = 2m$, we have

$$\sum_{\substack{n=0 \\ (even)}}^{\infty} = \sqrt{\frac{2\pi}{kx}} \sum_{m=0}^{\infty} (-1)^m \frac{2^m\, m!}{(2m)!} (kz)^{2m} \frac{J_{m+1/2}(kx)}{(kx)^m} \tag{C.11}$$

or, in terms of spherical Bessel functions $j_m(\beta) \equiv \sqrt{\pi/(2\beta)}\, J_{m+1/2}(\beta)$,

$$\sum_{\substack{n=0 \\ (even)}}^{\infty} = 2 \sum_{m=0}^{\infty} (-1)^m \frac{2^m\, m!}{(2m)!} (kz)^{2m} \frac{j_m(kx)}{(kx)^m} \ . \tag{C.12}$$

After substituting from Eqs. (C.9) and (C.12) into Eq. (C.3), we obtain Eq. (3.3.14a) of Chapter 3 for the kernel $\mathrm{H}_h(x,z)$,

$$\mathrm{H}_h(x,z) = \frac{ikz}{2\sqrt{x^2+z^2}} J_1\!\left(k\sqrt{x^2+z^2}\right) + \frac{k}{\pi} \sum_{m=0}^{\infty} (-1)^m \frac{2^m\, m!}{(2m)!} (kz)^m \frac{j_m(kx)}{(kx)^m} \ . \tag{C.13}$$

# APPENDIX D

# DERIVATION OF ASYMPTOTIC SERIES (3.5.9)
# FOR THE TOTAL EVANESCENT INTENSITY

To understand the decay of the total evanescent intensity $I_{\text{tot}}^{(i)}(z)$ for the example considered in Section 3.5, we need to derive the first few terms in the asymptotic expansion, for large values of $kz$, of Eq. (3.5.4b), i.e., of

$$I_{\text{tot}}^{(i)}(z) \;=\; \frac{4I_{\text{tot}}(0)}{\pi kd} \int\limits_{1}^{\infty} \frac{\sin^2(u_x kd/2)}{u_x^2}\, e^{-2kz\sqrt{u_x^2-1}}\, du_x \;. \qquad \text{(D.1)}$$

If we make the change of variables $s = \sqrt{u_x^2 - 1}$, we can rewrite this equation in the form

$$I_{\text{tot}}^{(i)}(z) \;=\; \frac{4I_{\text{tot}}(0)}{\pi kd} \int\limits_{0}^{\infty} \frac{\sin^2\!\left[kd(1+s^2)^{1/2}/2\right]}{(1+s^2)^{3/2}}\, e^{-2kzs}\, s\, ds \;. \qquad \text{(D.2)}$$

When $kz \gg 1$ there is an appreciable contribution to the above integral only for $s \ll 1$. We can then make the following approximations:

$$\frac{1}{(1+s^2)^{3/2}} \;\approx\; 1 - \frac{3}{2}\, s^2 + \frac{15}{8}\, s^4 \;, \qquad \text{(D.3)}$$

$$\sin^2\left[kd(1+s^2)^{1/2}/2\right] \approx \sin^2\left[kd\left(\frac{1}{2}+\frac{1}{4}s^2\right)\right]$$

$$\approx \frac{1}{2} - \frac{1}{2}\left\{\cos(kd)\,\cos(kds^2/2) - \sin(kd)\,\sin(kds^2/2)\right\}$$

$$\approx \frac{1}{2} - \frac{1}{2}\left\{\cos(kd)\left[1-\frac{(kds^2/2)^2}{2}\right] - \sin(kd)\left(\frac{kds^2}{2}\right)\right\}$$

$$\approx \sin^2(kd/2) + \frac{kd}{4}\,\sin(kd)\,s^2 + \left(\frac{kd}{4}\right)^2\cos(kd)\,s^4 ,$$

$$\text{(D.3)}$$

$$\frac{\sin^2\left[kd(s^2+1)^{1/2}/2\right]}{(s^2+1)^{3/2}} \approx \sin^2(kd/2) + \left[\frac{kd}{4}\,\sin(kd) - \frac{3}{2}\,\sin^2(kd/2)\right]s^2$$

$$+ \left[\left(\frac{kd}{4}\right)^2\cos(kd) - \frac{3kd}{8}\,\sin(kd) + \frac{15}{8}\,\sin^2(kd/2)\right]s^4 .$$

$$\text{(D.4)}$$

After substituting Eq. (D.4) into Eq. (D.2) and performing the integrations over $s$ using the relation

$$\int_0^\infty e^{-2kzs}\,s^q\,ds = \frac{q!}{(2kz)^{q+1}} ,$$

$$\text{(D.5)}$$

we obtain Eq. (3.5.9) of Chapter 3, viz.,

$$I_{\text{tot}}^{(i)}(z) \sim \frac{4I_{\text{tot}}(0)}{\pi kd}\left\{\frac{1!}{(2kz)^2}\sin^2(kd/2) + \frac{3!}{(2kz)^4}\left[\frac{kd}{4}\,\sin(kd) - \frac{3}{2}\,\sin^2(kd/2)\right]\right.$$

$$\left. + \frac{5!}{(2kz)^6}\left[\left(\frac{kd}{4}\right)^2\cos(kd) - \frac{3kd}{8}\,\sin(kd) + \frac{15}{8}\,\sin^2(kd/2)\right]\right\} ,$$

$$\text{(D.6)}$$

which contains the first three terms in the asymptotic series of $I_{\text{tot}}^{(i)}(z)$ for large $kz$.

# APPENDIX E

# DESCRIPTION OF THE FD-TD NUMERICAL METHOD

In this Appendix, we describe the finite-difference time-domain (FD-TD) technique that we implemented to obtain the numerical results presented in Chapter 4. Although the specific FD-TD method we used is rather unsophisticated, it does provide very accurate results. Because there exists extensive literature on FD-TD,[1-3] we shall keep our discussion rather brief.

Since both E and H-polarizations can be completely described by a scalar field that satisfies the free-space Helmholtz equation, in the time domain the two polarizations can be treated by applying finite differences to the free-space wave equation,

$$\left( \nabla^2 - \frac{1}{c^2} \frac{\partial^2}{\partial t^2} \right) V(x,z,t) = 0 \ . \tag{E.1}$$

For a rectangular grid of points (see Fig. E-1), the discritized wave equation for the scalar field $V(x,z,t)$ can then be expressed in the form

$$V(x,z,t+\Delta t) = c^2 (\Delta t)^2 \nabla^2 V(x,z,t) + 2V(x,z,t) - V(x,z,t-\Delta t) \ , \tag{E.2}$$

where the second-order partial derivatives with respect to $x$ and $z$ are given by the formulas

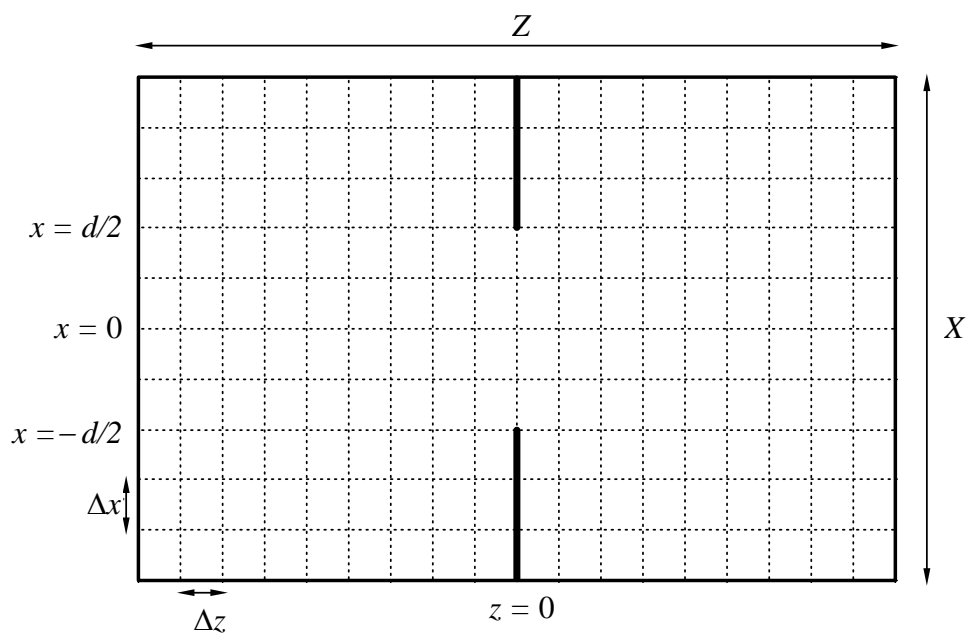

**Figure E-1** Numerical grid used in the FD-TD calculations for the diffraction of a plane wave by a slit. The grid points are at the intersections of the dotted lines.

$$\frac{\partial^2 V(x,z,t)}{\partial x^2} \;=\; \frac{V(x+\Delta x,z,t) \,-\, 2V(x,z,t) \,+\, V(x-\Delta x,z,t)}{(\Delta x)^2} \;, \quad \text{(E.3)}$$

$$\frac{\partial^2 V(x,z,t)}{\partial z^2} \;=\; \frac{V(x,z+\Delta z,t) \,-\, 2V(x,z,t) \,+\, V(x,z-\Delta z,t)}{(\Delta z)^2} \;. \quad \text{(E.4)}$$

Here $\Delta x \ll \lambda$ and $\Delta z \ll \lambda$ are small increments along $x$ and $z$ directions, respectively, and $\Delta t$ is a small increment of time. To obtain stable results, $\Delta t$ must satisfy the causality inequality

$$\Delta t \;<\; \frac{1}{c\sqrt{1/(\Delta x)^2 \,+\, 1/(\Delta z)^2}} \;. \quad \text{(E.5)}$$

From Eqs. (E.2)-(E.4), $V(x,z,t+\Delta t)$ can be computed from $V(x,z,t)$ and $V(x,z,t-\Delta t)$. This procedure can be invoked repeatedly to determine the field $V(x,z,T)$ at some time of interest $T = t_o + n(\Delta t)$, $n$ being an integer, from knowledge of $V(x,z,t_o)$ and $V(x,z,t_o - \Delta t)$ at some initial time $t_o$.

The difference between the treatment of the two polarizations is in the boundary conditions for $V(x,z,t)$ on the surface of the perfect conductor. For E-polarization, since $V$ represents the electric field $E_y$, $V$ must satisfy the condition

$$V(x,0,t) = 0 \qquad \text{for } |x| > \frac{d}{2}, \qquad (E.6)$$

and for H-polarization, since $V$ represents the magnetic field $H_y$, it must satisfy

$$V(x,-\Delta z,t) = V(x,0,t) \quad \text{and} \quad V(x,\Delta z,t) = V(x,2\Delta z,t) \quad \text{for } |x| > \frac{d}{2}.$$
$$(E.7)$$

Equations (E.6) and (E.7) ensure that the tangential electric field is zero on the conductor.

The above scheme was applied to the total field with the use of three two-dimensional arrays to store $V(x,z,t+\Delta t)$, $V(x,z,t)$ and $V(x,z,t-\Delta t)$ at the grid points. The slit plane was placed near the center of the grid and the incident plane wave was taken to propagate into the grid from the left edge. To enforce the radiation condition at the edges of grid in the half-space $z > 0$, 2nd order Engquist-Majda absorbing edge boundary conditions were used together with 1st order corner boundary conditions.[4,5,1] However, when such boundary conditions were applied in the half-space $z < 0$, they caused a significant disturbance to the incident plane wave. To eliminate this disturbance, the Neumann boundary condition $\partial V(x,z,t)/\partial x = 0$ was used in this half-space along the edges of the grid parallel to the *z*-axis, but it was then necessary to restrict the total computation time *T* so that no artifacts from grid-edge reflections would be present in the region of interest[*].   A more efficient approach would have been to use a scattered field FD-TD formulation to solve the complementary strip diffraction problem.  The results for slit diffraction could then have been obtained with the help of Babinet's principle.

Table E-1 contains the computation parameters that correspond to the numerical results presented in Chapter 4.  The accuracy of these results was checked by performing additional calculations with longer computation times and finer grids. Throughout the region  of interest, i.e., the region shown in the color images of Chapter 4, the more accurate results differed from those in Chapter 4 by at most 4% in amplitude and usually by 1 to 2%.

**Table E-1**  Values of Parameters Used in FD-TD Computations

| Width | Grid Size | | Grid Spacing | | Total Time | Time Spacing |
|:---:|:---:|:---:|:---:|:---:|:---:|:---:|
| *d* | *X* | *Z* | $\Delta x$ | $\Delta z$ | *T* | $\Delta t$ |
| ∞ (Half-Plane) | 6 $\lambda$ | 6 $\lambda$ | 0.013 $\lambda$ | 0.0029 $\lambda$ | 7 periods | 0.0018 periods |
| 0.2 $\lambda$ | 6 $\lambda$ | 6 $\lambda$ | 0.008 $\lambda$ | 0.0033 $\lambda$ | 7 periods | 0.002 periods |
| 0.5 $\lambda$ | 6 $\lambda$ | 6 $\lambda$ | 0.02 $\lambda$ | 0.0033 $\lambda$ | 7 periods | 0.002 periods |
| 1.0 $\lambda$ | 6 $\lambda$ | 6 $\lambda$ | 0.02 $\lambda$ | 0.0033 $\lambda$ | 7 periods | 0.002 periods |
| 1.5$\lambda$ | 8 $\lambda$ | 8 $\lambda$ | 0.02 $\lambda$ | 0.005 $\lambda$ | 9 periods | 0.003 periods |
| 2.0$\lambda$ | 8 $\lambda$ | 8 $\lambda$ | 0.02 $\lambda$ | 0.005 $\lambda$ | 9 periods | 0.0036 periods |
| 5.0$\lambda$ | 25 $\lambda$ | 20 $\lambda$ | 0.05 $\lambda$ | 0.0067 $\lambda$ | 24 periods | 0.0048 periods |

[*] The total time *T* was kept sufficiently large for a steady state to be reached.